\documentclass[prb,aps,amssymb,twocolumn,superscriptaddress,notitlepage]{revtex4-2}
\usepackage{amsmath}
\usepackage{amssymb}
\usepackage{amsthm}
\usepackage{wrapfig}
\usepackage{mathrsfs}
\usepackage{amsfonts}
\usepackage{listings}
\usepackage{physics}
\usepackage{ dsfont }
\usepackage{float}
\usepackage[title]{appendix}
\usepackage[makeroom]{cancel}
\usepackage{enumerate}
\usepackage{latexsym}
\usepackage{psfrag}
\usepackage{bm}
\usepackage{graphicx}
\usepackage[caption=false]{subfig}
\usepackage{blkarray}
\usepackage{array}
\usepackage[com pat=1.1.0]{tikz-feynman}
\usepackage{slashed}
\usepackage{wrapfig}
\usepackage{color}
\usepackage[normalem]{ulem}
\usepackage{hyperref}

\newcommand{\subsubsubsection}[1]{\mbox{}\paragraph{#1}\mbox{}\\}
\begin{document}
\title{Optical Response from Charge-Density Waves in Weyl Semimetals}
\author{Robert C. McKay}
\email[]{rcmckay2@illinois.edu}
\affiliation{Department of Physics and Institute for Condensed Matter Theory,
University of Illinois at Urbana-Champaign, Urbana, IL, 61801-3080, USA}

\author{Barry Bradlyn}
\email[]{bbradlyn@illinois.edu}
\affiliation{Department of Physics and Institute for Condensed Matter Theory,
University of Illinois at Urbana-Champaign, Urbana, IL, 61801-3080, USA}

\date{\today}

\begin{abstract}
    The study of charge-density wave (CDW) distortions in Weyl semimetals has recently returned to the forefront, inspired by experimental interest in materials such as $\text{(TaSe}_4\text{)}_2 \text{I}$.  
    However, the interplay between collective phonon excitations and charge transport in Weyl-CDW systems has not been systematically studied.  
    In this paper, we examine the longitudinal electromagnetic response due to collective modes in a Weyl semimetal gapped by a quasi-one-dimensional  charge-density wave order, using both continuum and lattice regularized models. 
    We systematically compute the contributions of the collective modes to the linear and nonlinear optical conductivity of our models, both with and without tilting of the Weyl cones. 
    We discover that, unlike in a single-band CDW, the gapless CDW collective mode does not contribute to the conductivity unless the Weyl cones are tilted. 
    Going further, we show that the lowest nontrivial collective mode contribution to charge transport with untilted Weyl cones comes in the third-order conductivity, and is mediated by the gapped amplitude mode.
    We show that this leads to a sharply peaked third harmonic response at frequencies below the single-particle energy gap.
    We discuss the implications of our findings for transport experiments in Weyl-CDW systems. 
\end{abstract}

\maketitle

\tableofcontents

\section{Introduction}

The interplay between electronic band topology and symmetry breaking order is at the forefront of modern condensed matter physics. 
Recent progress in the search for topological materials has revealed topological insulators and semimetals with local magnetic order, as well as highlighted the link between phonon-driven structural distortions and changes in electronic topology. 
One particularly illustrative material class is the bismuth halides Bi$_4$X$_4$\cite{liu2016weak,li2019pressure,zhou2015topological,huang2016quasi,noguchi2021evidence}, where a structural phase transition drives a band inversion between trivial and weak or higher-order topological insulating phases. 
Additionally, several topological semimetals such as (TaSe$_4$)$_2$I\cite{wang1983charge,tournier2013electronic,Bernevig_CDW_in_tase42I,perfetti_spectroscopic_data_old_TaSe42I, schaefer_dynamics_2013_TaSe42I,kim_CDW_in_TaSe42I_like_materials}, and ZrTe$_5$\cite{tang2019three} in a magnetic field, TaTe$_4$\cite{BinghaiCDWDDP}, and GdSbTe\cite{lei2020charge} undergo charge-density wave (CDW) distortions that can drastically change the electronic structure and band topology. 
It is then a compelling theoretical and experimental question as to how the CDW distortion in these systems influences dynamical properties, such as electronic transport.

Due to the spontaneous symmetry breaking, the low-energy excitations in a CDW consist not only of single-particle electronic degrees of freedom, but also involve collective excitations of the mean field order parameter. 
In the case of a CDW distortion, these are the gapped amplitude mode, and the nominally gapless phase mode\cite{gruner1988dynamics}. 
As their names imply, excitations of the amplitude mode create spatiotemporal variations of the amplitude of the density wave distortion; excitations of the phase mode create spatiotemporal variations of the phase of the density wave modulation. 
The zero-wavevector phase excitation, which corresponds to a uniform shift of the CDW phase, is conventionally referred to as the CDW sliding mode. 
Because they directly modulate the charge density, the collective modes in a CDW can impact low-frequency (subgap) 

charge transport in these systems. 
In a classic result\cite{FrohlichSuperconductivityCDW}, Fr\"{o}hlich argued that the sliding mode in a one-dimensional (1D) CDW, if not pinned, could carry a one-dimensional supercurrent. 
Further work on CDW transport focused on the contribution of the sliding mode to the low-frequency conductivity of quasi-1D CDWs, both with and without disorder\cite{RLAChargeDensityWaves,FentonCDW,RiceElectronPhononInteractions,allender1974,Rice_Supplemental_notes_on_orignal_RLA,RiceandCrossCDW,lee1979electric,rice1975theory,rice1976weakly,bardeen1979theory,rice1982elementary}. 

Until recently, this work on CDW transport was focused on quasi-1D systems, where the single-particle gap could be viewed as originating from nesting of a single-band Fermi surface. 
However, 
the discovery of Weyl semimetals\cite{Wan11,Weng15,Huang15,Xu15,Lv15,Xu15a,Lv15a,Huang2015,Hirschberger2016} and their anomalous transport properties\cite{BurkovWeylSemimetalReview, Burkov_fractional_QHE_WSM_2020, BradlynAxionicWeylCDW, xiao_berry_WSM, BradlynMultifoldChiralResponse_2018, Burkov_anomalous_WSM_response_2014, Landsteiner_WSM_anomaly_diagram_notes_2016, HaltermanWeylWaveguide, HaltermanWeylAbsorption} has sparked interest in the phenomena that emerges when the CDW order opens a gap between Weyl fermions of opposite chirality\cite{osterhoudt_CDW_in_WP2, Burkov_CDW,BradlynAxionicWeylCDW,Hughes_CDW_topological_response_WSM,WangAndZhangAxionic,Redell_resonant_axion_WSM,zyuzin2012weyl,wilczekaxion}. 
Theoretically, it has been shown that for a magnetic Weyl-CDW, the CDW phase couples to the electromagnetic field as a dynamical electromagnetic theta angle. 
This implies, for instance, that the sliding mode in a magnetic Weyl-CDW can induce a chiral magnetic effect and modulate the anomalous Hall conductance. 
Signatures of similar coupling of the sliding mode to magnetoelectric response has recently been seen in the non-magnetic Weyl-CDW (TaSe$_4$)$_2$I, where magnetoconduction due to the sliding mode was responsible for an observed negative quadratic magnetoresistance\cite{gooth_axionic_in_TaSe42I}. 
Further study of this material has provided increasing evidence that the CDW state emerges from a high-temperature Weyl semimetallic phase\cite{mu_CDW_SC_phase,zhang_first-principles_TaSe42I,  Bernevig_CDW_in_tase42I}.

A full understanding of these experiments requires an analysis of the collective mode contribution to \emph{longitudinal} conductivity in Weyl-CDW systems. 
Additionally, the recently discovered large nonlinear optical response of chiral Weyl semimetals may leave an imprint on the response functions in chiral CDWs such as (TaSe$_4$)$_2$I\cite{de_juan_topological_nonlinear_2019, wu_giant_nonlinear_resonse_experimental_2017, ahn_low-frequency_nonlinear_topological_response_2020, de_juan_quantized_galvanic_effect_2017, rees_quantized_multifold_experimental_2019, sodemann_quantum_weyl_nonlinear_response_2015, morimoto_semiclassical_nonlinear_2016, morimoto_topological_2016}. 
To that end, we study for the first time the collective mode contributions to the linear and nonlinear conductivity in simple minimal models of three-dimensional (3D) Weyl-CDWs. 
We focus on models with low-energy Weyl fermions that are gapped by a CDW distortion driven by electron-phonon coupling. 
Consistent with recent interest in magnetic topological materials\cite{MTQC,AshvinMagnetic,MTQCmaterials,MagneticNewFermion,yang2021symmetry,AxionExp1,AxionExp2,AxionZahid1,OtherAxion3,OtherAxion4,MagneticWeylZahid,MagneticWeylYulin,MagneticWeylHaim}, we analyze a minimal magnetic Weyl-CDW with two Weyl points in the normal state.
Extending the formalism of Lee, Rice, and Anderson (LRA)\cite{RLAChargeDensityWaves}, we first derive expressions for the zero wavevector amplitude and phase mode propagator in 3D. 
Using these, we are able to extend the diagrammatic technique for computing the nonlinear optical conductivity from Ref.~\cite{MooreDiagrammatic} to include contributions from the CDW collective modes. 
Crucially, we find that the involvement of multiple bands in Weyl-CDWs fundamentally changes the way collective modes enter into the conductivity calculation when compared with the simple single band formalism of Refs.~\cite{RLAChargeDensityWaves,FentonCDW,RiceElectronPhononInteractions,allender1974,Rice_Supplemental_notes_on_orignal_RLA}. In particular, we show that the massless collective mode only contributes to the linear conductivity when the Weyl cones are tilted. 
For an untilted Weyl semimetal, we do not find any zero-frequency (linear or nonlinear) conductivity due to the sliding mode. 
On the contrary, we find that the lowest-order collective mode contribution to optical response in the untilted case comes at third order from the massive collective mode. 
This leads to an enhancement of the third harmonic response at half the resonant frequency of the massive mode, which is well below the single-particle band edge.

The structure of this paper is as follows. 
First, we start by introducing our minimal model for a Weyl-CDW system in Sec.~\ref{model_section}. 
We consider both a full tight-binding model as well as a low-energy $\mathbf{k}\cdot\mathbf{p}$ approximation, both with and without tilting the Weyl cones.  
In Section \ref{propagator_calcs_section}
we calculate the mean field gap equation, electron propagators, and phonon propagators in the CDW phase. 
We also introduce the Feynman rules for diagrammatic perturbation theory in the CDW. 
In Sec.~\ref{conductivity_section} we introduce our diagrammatic scheme for calculating the contributions of the collective phonon contributions to the conductivity through third-order. 
We pay particular attention to the regularization of contact, or diamagnetic, contributions to the conductivity. 
We employ both a minimal subtraction regularization as well as a lattice completion.  
We show the first nonzero collective conductivity for an untilted Weyl semimetal comes from third-order processes. 
We further show that tilting the Weyl nodes leads to a nonzero DC linear conductivity mediated by the sliding mode of the CDW.  
We conclude in Section \ref{conclusion_section} with a summary of our results and an outlook towards future experimental applications. 
We include several appendices containing details of the calculations and a review of the calculation of the conductivity for 1D CDWs.

\section{Models}
\label{model_section}

\subsection{Lattice Model}
\begin{figure*}[t]
      \centering
\centering
\includegraphics[width=0.9\hsize]{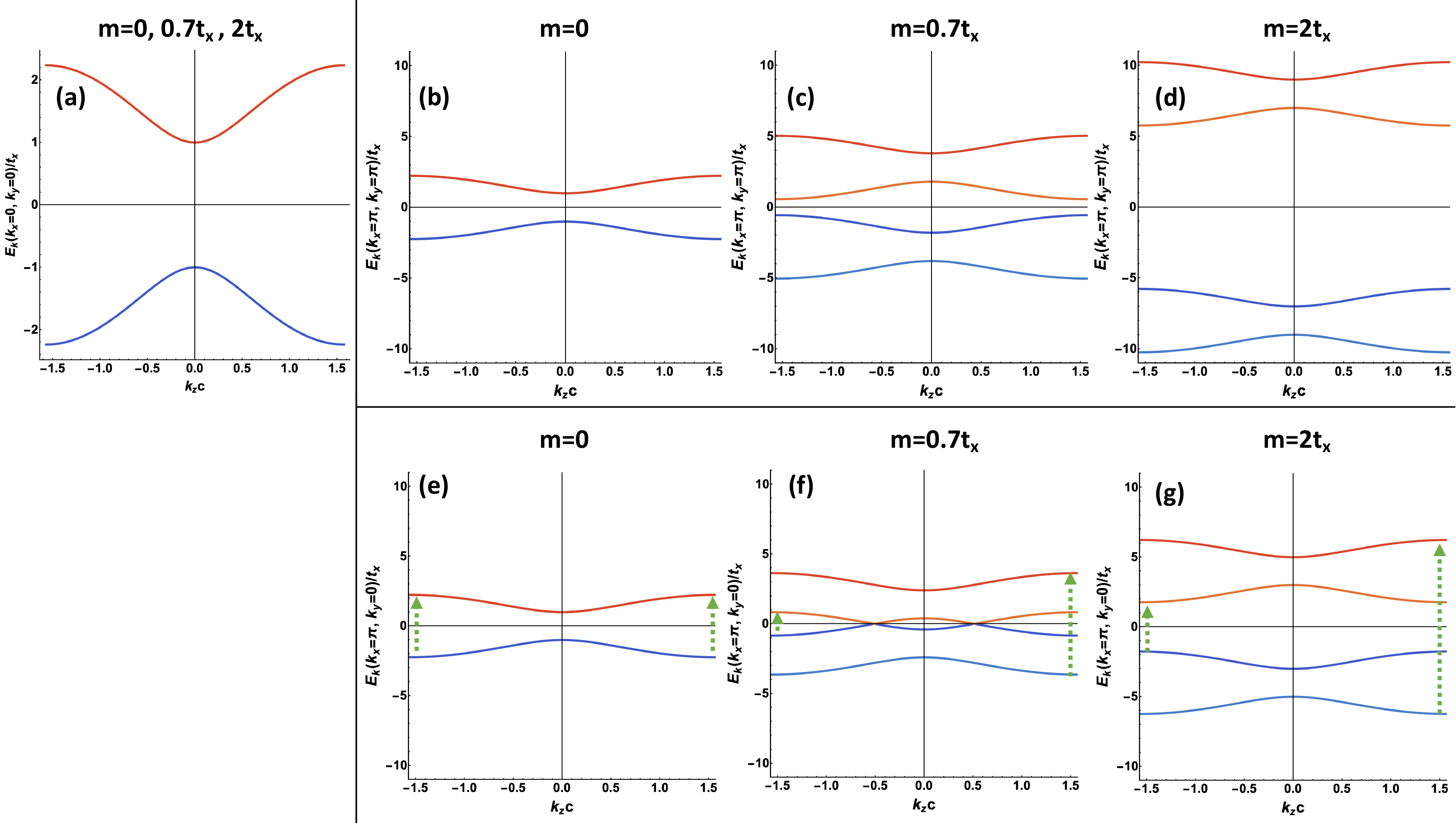}
\caption{The spectra for the lattice model with mean field electron-phonon interactions in the shifted Brillouin zone scheme, with $t_x = t_y = t_z$ and $\Delta = 0.5 t_x$.  
(a) is a cut through $k_x = k_y = 0$, which shows the gapped Weyl nodes at $k_z c=0$. 
(b-d) are cuts through $(k_x, k_y)=(\pi, \pi)$, illustrating that the parameter $m$ lifts the band degeneracy near the Brillouin zone edge.  
At $m=0$, three additional pairs of Weyl points exist at the Brillouin zone edge, which are gapped by the electron-phonon interaction as can be seen in (b) at $k_z c= 0$.  
Plots (e-g) show a different slice of the spectra at $(k_x a, k_y b)=(\pi, 0)$, and the green arrows signify excitations described in Section \ref{harmonicSection}.}
\label{UntiltedLatticeSpectrumPlot}
\end{figure*}

Our starting point is a simple (spinless) two-band model for a Weyl semimetal adopted from the time-reversal-breaking model of Refs.~\cite{McCormickMinimalModel, RanTRBreakingModel, Delplace_TRBreaking_WSM_model_2012}.  
The electronic part of the tight-binding Hamiltonian is
\begin{equation}
\label{H0Ham}
\begin{split}
H_0=& \sum_{\mathbf{k}} \vec{c}^{\dagger}_{\mathbf{k}}\left[ 2(-t_{x} \sigma_x \sin (k_{x}a) - t_{y}\sigma_y \sin( k_{y} b)
\right.\\& \left.
+ t_{z}(\cos (k_{z} c) -\cos (k_f c) ) \sigma_{z})
\right.\\& \left.
-m(2-\cos( k_{x} a) - \cos( k_{y} b))\sigma_z \right] \vec{c}_{\mathbf{k}},
\end{split}
\end{equation}
where $\sigma_x$, $\sigma_y$, $\sigma_z$ are Pauli matrices acting in the orbital space, and $m$, $t_x$, $t_y$, and $t_z$ are energy parameters of the model.  
The parameter $m$ acts to lift additional nodal degeneracies at the edges of the Brillouin zone.  
Also, $a$, $b$, $c$ are characteristic lengths between neighboring sites.  
For the remainder of this work, we take $k_x$, $k_y$, and $k_z$ in units of $1/a$, $1/b$, and $1/c$ respectively.
The model features two Weyl nodes at $(0,0,\pm k_f)$,  
and we denote the Weyl node separation vector as $\mathbf{Q} = (0, 0,2 k_f)$.

Following Ref.~\cite{BradlynAxionicWeylCDW}, we now consider electron-phonon coupling in this model. 
We assume the model to be quasi-one-dimensional along the direction of $\mathbf{Q}$, meaning the electron-phonon coupling strength is only non-negligible for phonons propagating along the $\hat{\mathbf{k}}_f=\hat{\mathbf{z}}$ direction. 

The terms in the Hamiltonion responsible for the free phonon dispersion and the electron-phonon interaction are respectively given by 

\begin{equation}
\label{gapAndPhonon}
\begin{split}
    H_1 =& \sum_{\mathbf{q}} \omega_{\mathbf{q}} b^{\dagger}_{\mathbf{q}} b_{\mathbf{q}}, 
    \\
    H_2 =& g\sum_{\mathbf{k}, \mathbf{q}} b_{\mathbf{q}}\vec{c}^{\dagger}_{\mathbf{k+q}}\sigma_z\vec{c}_{\mathbf{k}}+b^{\dagger}_{\mathbf{q}}\vec{c}^{\dagger}_{\mathbf{k}}\sigma_z\vec{c}_{\mathbf{k+q}}.
\end{split}
\end{equation}
As we will see, this interaction term can couple Weyl nodes of opposite chirality\cite{WangAndZhangAxionic, Hughes_CDW_topological_response_WSM, BradlynAxionicWeylCDW}.

To proceed further, we expand the electronic creation and annihilation operators around the two Weyl nodes as $\vec{c}^{\dagger}_{\mathbf{k} + \mathbf{Q} / 2}$ and $\vec{c}^{\dagger}_{\mathbf{k} - \mathbf{Q} / 2}$, where $\mathbf{Q} \equiv 2 \mathbf{k}_f$ \cite{FrohlichSuperconductivityCDW, RLAChargeDensityWaves, RiceandCrossCDW}. 
We will now examine a charge-density wave transition that couples the two Weyl nodes.  
We define a new two-component operator as $\vec{c}^{\: \prime}_{\mathbf{k}} \equiv (\vec{c}_{\mathbf{k} + \mathbf{Q}/2}, \vec{c}_{\mathbf{k} - \mathbf{Q}/2})$.  
The corresponding Pauli matrices in this $\pm \mathbf{Q}/2$ subspace are denoted by $\tau_{i \in \{x, y, z, 0\}}$.
When phonons of wavevector $\mathbf{Q}$ condense, the phonon annihilation operator acquires a translation symmetry-breaking expectation value defined by $g \langle b_{\mathbf{Q}} \rangle = \Delta  e^{-i \phi}$, which opens a gap in the single-particle spectrum\cite{FrohlichSuperconductivityCDW, RLAChargeDensityWaves, BurkovWeylSemimetalReview}.    
{Note that although our choice of coupling constant $g$ is consistent with Refs.~\cite{FentonCDW,RiceElectronPhononInteractions}, it differs from LRA by a factor of $i$. 
This only affects the mean field dynamics via a phase shift of $\phi + \pi / 2$, which, as we show in the following sections, does not impact our results.}  
This choice of symmetry-breaking  means momentum is only conserved modulo $\mathbf{Q}$ for the electronic part of the Hamiltonian. 
{Our focus will be on the effect of zero-momentum fluctuations 
\begin{equation}
b_\mathbf{Q} = (\Delta /g + \delta b)e^{i(\delta\theta-\phi)}\label{eq:mfansatz}
\end{equation}
 on the response of this system to electric fields. 
 Here $\delta b$ represents the amplitude fluctuations of the CDW, and $\delta\theta$ represents the phase fluctuations. }

Our goal will be to analyze the collective modes and conductivity of an incommensurate CDW, i.e.~ when $\mathbf{Q}$ is not a rational multiple of a reciprocal lattice vector. 
While there is no lattice periodicity in this case, it will be convenient to start from a model with a bounded spectrum in order to eliminate spurious divergences in our subsequent transport calculations. 
To do so, we will introduce a convenient lattice regularization of an incommensurate Weyl-CDW, taking inspiration from the lattice model Eq.~\eqref{FullLatticeModel}.  
We use a convenient shifted Brillouin zone scheme, where the Weyl fermions are recentered at the origin of the shifted zone. 
We give the details of how this is implemented in Appendix \ref{shiftedZoneSchemeDerivation}. 
When $\mathbf{Q}=\pi$, we can derive a simple four-band model that is suitable for a lattice regularized starting point for further analysis.  
We first apply our mean field ansatz Eq.~(\ref{eq:mfansatz}) to the interaction Hamiltonian (\ref{gapAndPhonon}) to obtain the mean field Hamiltonian $\overline{H}_2$. For $\mathbf{Q}=\pi$, we have $\overline{H}_2 = \vec{c}^{\dagger \: \prime}_{\mathbf{k}} \{ 2 |\Delta|\sigma_z \otimes (\cos \phi \tau_x) \}\vec{c}^{\: \prime}_{\mathbf{k}}$.    
Note that $\bar{H}_2$ would produce a strong pinning of the phase of the CDW, as the electronic band structure depends strongly on $\phi$. 
For an incommensurate CDW we expect there to be an unpinned--and therefore massless--dynamical sliding mode; this implies that the single-particle energy spectrum should be independent of the phase $\phi$ in Eq.~\eqref{eq:mfansatz}. 
To simulate an incommensurate modulation within the context of our lattice regularized four-band model, we can replace the mean field Hamiltonian $\bar{H}_2$ with
\begin{equation}
\label{meanFieldGapHam}
    H_2' = \sum_k \vec{c}^{\dag \: \prime}_{\mathbf{k}}\left[2|\Delta|\sigma_z\otimes\left(\cos \phi \tau_x +\sin\phi\tau_y\right)\right]\vec{c}^{\: \prime}_{\mathbf{k}},
\end{equation}
which is the form of the mean field electron-phonon interaction we would expect for the low-energy theory of an incommensurate Weyl-CDW\cite{Burkov_CDW,BradlynAxionicWeylCDW} (i.e., what we would obtain following our mean field decomposition with $\mathbf{Q}\neq\pi$). 
Since we are primarily interested in the properties of unpinned, incommensurate CDWs, we will use $H_2'$ as our mean field Hamiltonian throughout this paper.

From Eq.~\eqref{H0Ham}, moving to the shfited Brillouin zone with $\mathbf{Q}=\pi\hat{z}$ gives the Hamiltonian 
\begin{equation}
\label{FullLatticeModel}
\resizebox{1\hsize}{!}{$
\begin{aligned}
    H_{0, \text{lat}} + H'_{2} =& \sum_{\mathbf{k}} \vec{c}^{\dagger \: \prime}_{\mathbf{k}}\left\{  -2 (t_x \sin(k_x) \sigma_x + t_y \sin{k_y} \sigma_y ) \otimes \tau_0 
     \right.
    \\&
     \left.
    + 2 t_z\sin(k_z)\sigma_z \otimes \tau_z
    \right.
     \\& 
     \left.
     - m \left(2 - \cos(k_x) - \cos(k_y) \right) \sigma_z \otimes \tau_0
     \right.
     \\& 
     \left.+2 |\Delta|\sigma_z \otimes (\cos \phi \tau_x) +2 |\Delta|\sigma_z \otimes (\sin \phi \tau_y) \right\}\vec{c}^{\: \prime}_{\mathbf{k}}.
\end{aligned}
$}
\end{equation}
In this shifted Brillouin zone, the boundaries are $k_x \in [-\pi, \pi]$, $k_y \in [-\pi, \pi]$, and $k_z \in [-\pi/2, \pi/2]$.  
Several slices of the spectra in the shifted Brillouin zone scheme for different values of $m$ are given in Fig.~\ref{UntiltedLatticeSpectrumPlot}.

In this work, we will consider the low-energy physics not only for a lattice model, but also for an ideal model lineraized about the untilted Weyl cones, as well as a titled lattice Weyl Hamiltonian.  
In the remaining subsections, we give explicit forms for the Hamiltonians for both cases.

\subsection{Ideal Model}
\begin{figure}[ht]
      \centering
\begin{minipage}{0.8\hsize}
\centering
\includegraphics[width=0.8\hsize]{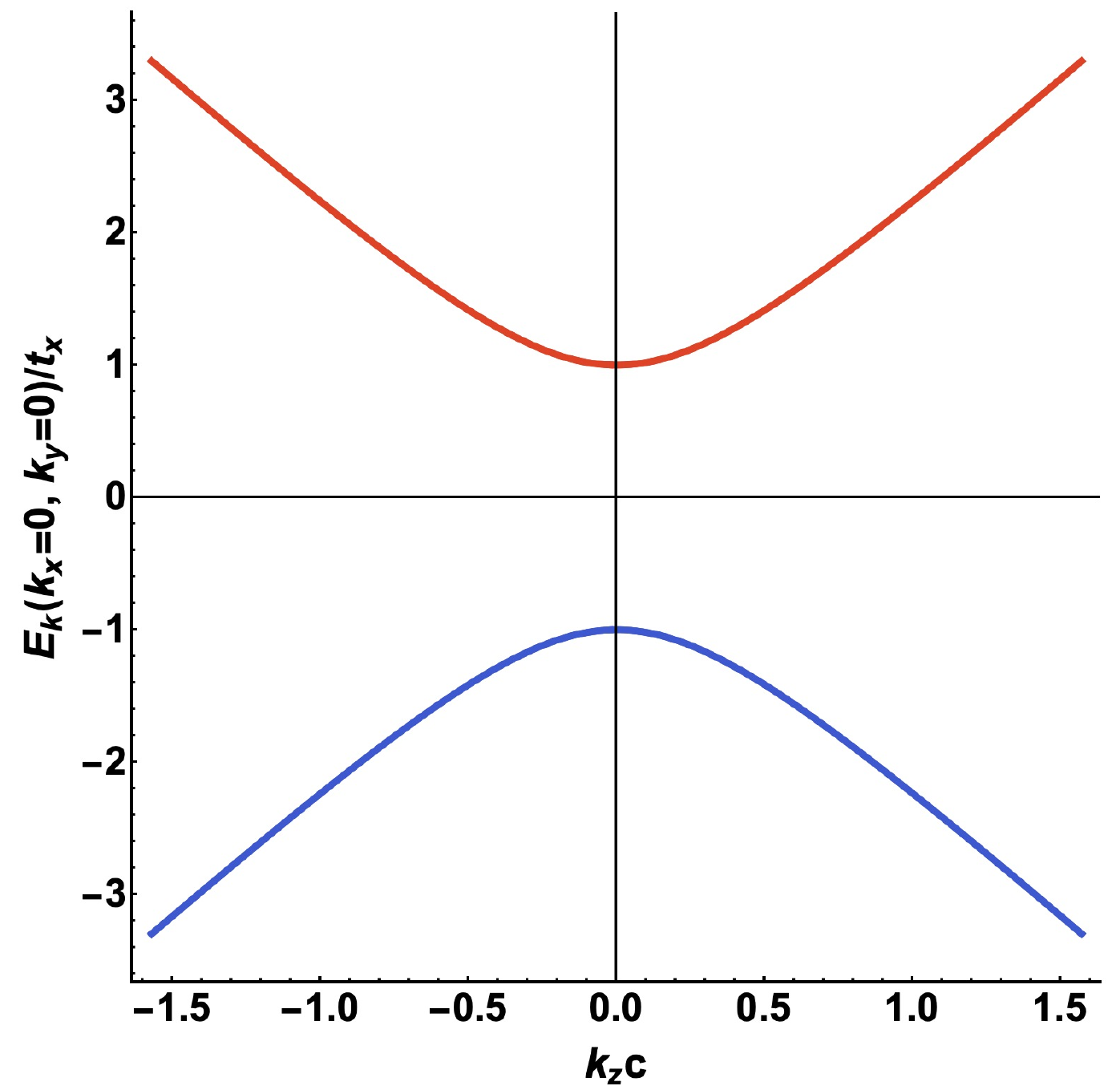}
\end{minipage}
\caption{Spectrum of the idealized Weyl semimetal model in a cut of the shifted Brillouin zone at $k_x = k_y = 0$.  The parameter values are $t_x = t_y = t_z$ and $\Delta = 0.5 t_x$.  The Weyl nodes, which are shifted onto each other, are gapped by the nonzero $\Delta$.}
\label{LinearSpectrumPlot}
\end{figure}

To obtain our ideal Weyl semimetal, we perform a $\mathbf{k}\cdot\mathbf{p}$ expansion near the Weyl nodes of the lattice model.  
Near these points, the idealized electronic Hamiltonian becomes
\newline
\newline
\begin{equation}
\label{ideal_ham}
\begin{split}
    H_{0, \text{ideal}} +  H_{2} \approx  \sum_{\mathbf{k}} & \vec{c}^{\dagger \: \prime}_{\mathbf{k}}\left[-(2 t_{x} k_{x} \sigma_x+2 t_{y} \sigma_y k_y)\otimes \tau_0 
     \right.\\&\left.
     +2t_{z} k_z    \sigma_z \otimes \tau_z
    \right.\\&\left.
    +2 |\Delta|\sigma_z \otimes (\cos \phi \tau_x + \sin \phi \tau_y)\right]\vec{c}^{\: \prime}_{\mathbf{k}}.
\end{split}
\end{equation}{}

It will also be convenient to define symbols for the two positive eigenvalues of $H_{0, \text{ideal}}$ and $H_{0,  \text{ideal}} + H_{2}$, 
\begin{equation*}
\begin{split}
   & \epsilon_{\mathbf{k}} \equiv \sqrt{ (2 t_x k_x)^2 + (2 t_y k_y)^2 + (2 t_z k_z )^2}, 
    \\&
    E_{\mathbf{k}} \equiv \sqrt{\epsilon^2_{\mathbf{k}} + 4 |\Delta|^2}.
\end{split}
\end{equation*}
A slice of $E_{\mathbf{k}}$ is shown in Fig.~\ref{LinearSpectrumPlot}.

\subsection{Tilted Lattice Model}
\begin{figure*}[t]
\centering
\includegraphics[width=0.9\hsize]{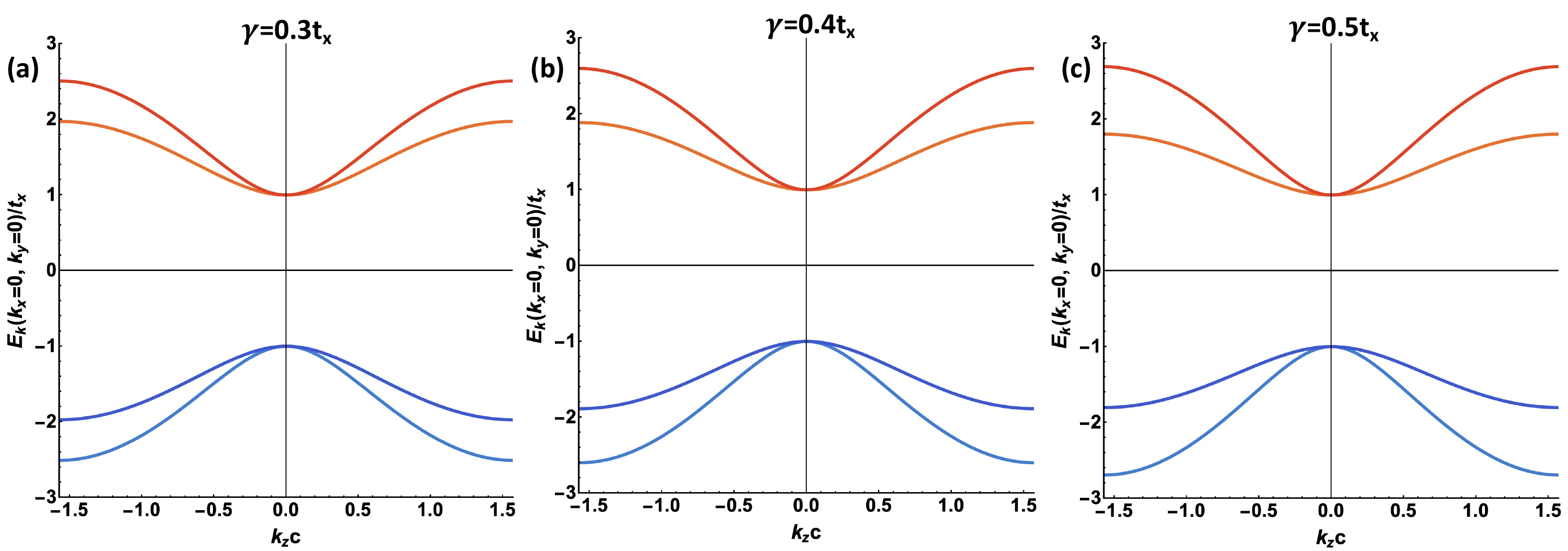}
\caption{Spectra for the tilted lattice model with mean field order in the shifted Brillouin zone scheme.  Each plot has $t_x = t_y = t_z$, $\Delta = 0.5 t_x$ and $m=0$.  Plot (a-c) show the line $k_x = k_y = 0$ for the tilts $\gamma=0.3 t_x$, $\gamma=0.4 t_x$, and $\gamma=0.5 t_x$ respectively.  Notice the bands are nondegenerate away from $k_z=0$. }
\label{TiltedLatticeSpectrumPlot}
\end{figure*}

Implementing a tilted Weyl model will be useful to explore the effects of inversion and particle-hole symmetry breaking.  
To induce a tilt in the unshifted Brillouin Zone\cite{McCormickMinimalModel}, we add $\sum_{\mathbf{k}} \vec{c}^{\dagger}_{\mathbf{k}} \gamma (\cos{k_z} - \cos{Q/2}) \sigma_0 \vec{c}_{\mathbf{k}}$ to the lattice Hamiltonian. 
The parameter $\gamma$ controls the degree of tilt. 
In the shifted Brillouin Zone scheme with $\mathbf{Q}=\pi\hat{z}$, this term can be written
\begin{equation}
    \label{tiltedLatticeTerm}
\begin{split}
    H_{\text{tilt}} =& \sum_{\mathbf{k}} \vec{c}^{\dagger \: \prime}_{\mathbf{k}} \left[ \gamma \sigma_0 \otimes  \tau_z \sin(k_z) \right]\vec{c}^{\: \prime}_{\mathbf{k}}.
\end{split}
\end{equation}
When we consider a tilt along with the mean field CDW Hamiltonian, large enough tilts can lead to the emergence of electron and hole Fermi surfaces. 
To avoid complications due to Fermi surface carriers, we restrict our attention in this work to $0\le \gamma \le 2$.  
A plot of the gapped CDW spectrum for different tilts is given is Fig.~\ref{TiltedLatticeSpectrumPlot}.

\section{Calculation of Propagators}
\label{propagator_calcs_section}
Next, we move on to consider the dynamical effects of phonon fluctuations in the ordered phase, with a particular focus on the computation of propagators and vertices that will enter into the conductivity calculations. 
To begin, we will derive the self-consistent gap equation relating the phonon frequency $\omega_\mathbf{Q}$ at the CDW wavevector to the amplitude $|\Delta|$ of the gap.
To do so we will first introduce the bare (unperturbed) electron and phonon propagators.

\subsection{Bare Propagators}
To perform the Kubo formula calculations, we will need the bosonic collective mode propagator and the fermionic electron propagator. 
We will calculate these in the imaginary time Matsubara formalism. 
The equation of motion for the Wick rotated Schr\"{o}dinger Green function equation is sufficient to calculate the bare electronic propagator: $(\delta_{n_1 \xi_1 n_2 \xi_2}\partial_{\mathscr{T}} - H_{0, n_1 \xi_1 n_2 \xi_2})G_{ n_2 \xi_2 n_3 \xi_3}(t, \mathbf{x}) = \delta(t) \delta(\mathbf{x}) \delta_{n_1 \xi_1 n_3 \xi_3}$.  
The Fourier transformed version of this equation will be more useful.  
For concreteness, for the ideal Weyl semimetal Eq.~(\ref{ideal_ham}), this yields
\begin{equation}
\label{greenFunctions}
\begin{split}
    G(i\nu, \mathbf{k}) =& \frac{ \left( \overline{k}_{\nu} \gamma^0 \gamma^{\nu} +  2 \Delta \cos{\phi} \gamma^0 \gamma_5 + i 2 \Delta \sin{\phi} \gamma^0 \right)  }{E^2 + \nu^2 }
    \\=&
    \frac{\gamma^0 \slashed{\overline{k}}  +  2 \Delta \cos{\phi} \gamma^0 \gamma_5 + i 2 \Delta \sin{\phi} \gamma^0  }{E^2 + \nu^2 }.
\end{split}
\end{equation}
Since $\mathscr{T}$ is defined in Euclidean space, $\nu$ is a fermion Matsubara frequency, given as $\nu = \pi (2 n + 1)/\beta$ for $n \in \mathds{Z}$.  
Also, $\overline{k}_{\nu}$ is defined to be the scaled version of the four-component momentum, $\overline{k} \equiv (i \nu, 2 t_x k_x, 2 t_y k_y, 2 t_z \sin(Q/2) k_z )$. \tabularnewline
The corresponding gamma matrices are 
\begin{align}
\gamma_0 &= i \sigma_z \otimes \tau_y,\;\; \gamma_1 =  \sigma_{y} \otimes \tau_y,\nonumber \\
\gamma_2 &= -\sigma_x \otimes \tau_{y},\;\; \gamma_3 = -\sigma_0 \otimes \tau_x, \nonumber \\
\gamma_5 &= i \gamma_0 \gamma_1 \gamma_2 \gamma_3.
\end{align}
The other bare propagator we will need is that of the phonon.  
This will be defined as $D_0(\mathscr{T}, \mathbf{q}) = \langle T_{\mathscr{T}}A_{\mathbf{q}}(\mathscr{T}) A^{\dagger}_{\mathbf{q}} \rangle$ where $A_{\mathbf{q}} \equiv b_{\mathbf{q}} + b^{\dagger}_{-\mathbf{q}}$ (Note that this is the operator which couples to $c^\dag_{k+q}\sigma_z c_k$ in Eq.~(\ref{gapAndPhonon})).  
In Matsubara space, this becomes 
\begin{equation}D_0(i\omega, \mathbf{q}) =- \frac{2 \omega_{\mathbf{q}}}{\omega^2 + \omega^2_{\mathbf{q}}}.
\end{equation} 
The full derivation is provided in Appendix \ref{FreePhononPropagator}.  

Since we will be interested in the conductivity at zero wavevector, we will only need to focus on the effects of phonons with wavevector $\mathbf{q}\approx \mathbf{Q}$. 
We can thus make the approximation $\omega_{\mathbf{\mathbf{Q}+\mathbf{\delta q}}} \approx \omega_{\mathbf{Q}}$ \cite{RLAChargeDensityWaves, Rice_Supplemental_notes_on_orignal_RLA}
yielding, in the $\delta\mathbf{q}\rightarrow 0$ limit,
\begin{equation}
    D_0(i\omega) = -\frac{2 \omega_{\mathbf{Q}}}{\omega^2 + \omega^2_{\mathbf{Q}}},
\end{equation}
which will enter into our subsequent calculations.  
This approximation will be valid at sufficiently low temperatures in the CDW phase, where we are interested in the effect of low-energy excitations at wavevector $\mathbf{Q}$.

\subsection{Gap Equation}
\label{gap}
The gap equation provides us with a self-consistent equation for the order parameter $\Delta$. 
We will see in the $T\rightarrow 0$ limit, the gap equation takes the form as a constraint between the phonon frequency $\omega_\mathbf{Q}$ and the electron-phonon coupling constant $g$.
The gap equation can be found by minimizing the free energy with respect to the CDW order parameter \cite{RLAChargeDensityWaves}.  
That is, from the free energy, $F = -\frac{1}{\beta} \ln {\Tr{e^{- \beta H}}}$, the variational derivative may be taken with respect to $\frac{\delta}{\delta \Delta}$ or $\frac{\delta}{\delta \Delta^*}$.  
To minimize the the free energy, these variational derivatives are set equal to zero.  
In the low temperature limit, $T \rightarrow 0$, where $\omega_{\mathbf{q}} \approx \omega_{\mathbf{Q}}$ and $\Delta(T) \approx \Delta$, the gap equation reduces to a statement on $\omega_{\mathbf{Q}}$.  
For a general Hamiltonian, which also applies to the previous tiltless and tilted lattice models, consider $H_{\text{gen}} = d_1(\mathbf{k}) \sigma_x \otimes \tau_0 + d_2(\mathbf{k}) \sigma_y \otimes \tau_0 + d_3(\mathbf{k}) \sigma_z \otimes \tau_z + d_4(\mathbf{k}) \sigma_z \otimes \tau_0 + d_5(\mathbf{k}) \sigma_0 \otimes \tau_z + d_6(\mathbf{k}) \sigma_0 \otimes \tau_0$.  
We consider this generalized Hamiltonian because it not only allows for the tilting term from Eq.~(\ref{tiltedLatticeTerm}), but it  is also general enough to extend beyond the Weyl semimetal model.  To find the value of $\Delta$, we minimize the  free energy 
\begin{equation}
F = -\frac{1}{\beta}  \ln{\sum_i e^{-\beta E_i}} + \frac{\Delta \Delta^{*}}{g^2} \omega_\mathbf{Q}
\end{equation}
with respect to $\Delta$. 
The easiest way to do this is to rewrite the Free energy in terms of a Euclidean effective action\cite{Casalbuoni2018lecturesOnSuperconductivity,altland2010condensed}
\begin{widetext}
\begin{equation}
S_{\text{eff}} = -\int [d k]\frac{i}{2}\text{Tr}\left[ \ln\left(  i \nu - H_{\text{gen}} - H_{2, \text{lat}}  \right) \right] - \int [d k] \frac{\Delta \Delta^*}{g^2}\omega_\mathbf{Q},
\end{equation}
where $[dk] = d(\nu)d\mathbf{k}$. 
Taking the variational derivative and minimizing this effective action results in
\begin{equation}
\begin{split}
    \Delta^{*} \omega_{Q} = -2 g^2 \int [d k] \text{Tr}\left\{\left[ \frac{\delta}{\delta \Delta}(i \nu - H_{\text{gen}} - H_{2, \text{lat}}) \right] \left[ i \nu - H_{\text{gen}} - H_{2, \text{lat}} \right]^{-1} \right \}.
\end{split}
\end{equation}
Upon integrating over the Matsubara frequency $\nu$, this generates a gap equation
\begin{equation}
\resizebox{0.95\hsize}{!}{$
\label{gapGeneral}
\begin{aligned}
    \omega_{Q} = & 4 g^2 \int [d \mathbf{k}]  \left( \frac{1}{E_{1, \mathbf{k}}} - \frac{1}{E_{2, \mathbf{k}}} \right) 
    \left[ \frac{(E_{1, \mathbf{k}} -d_6(\mathbf{k}))^2}{(E_{1, \mathbf{k}} + E_{3, \mathbf{k}})(E_{1, \mathbf{k}} + E_{4, \mathbf{k}})}
    -  \frac{(E_{2, \mathbf{k}} -d_6(\mathbf{k}))^2}{(E_{2, \mathbf{k}} + E_{3, \mathbf{k}})(E_{2, \mathbf{k}} + E_{4, \mathbf{k}})}
    \right. \\& \left.
     + \left[ (2 \Delta )^2 + d^2_1(\mathbf{k}) + d^2_2(\mathbf{k}) + d^2_3(\mathbf{k}) - d^2_4(\mathbf{k}) + d^2_5(\mathbf{k})\right] \left( \frac{-1}{(E_{1, \mathbf{k}} + E_{3, \mathbf{k}})(E_{1, \mathbf{k}} + E_{4, \mathbf{k}})} + \frac{1}{(E_{2, \mathbf{k}} + E_{3, \mathbf{k}})(E_{2, \mathbf{k}} + E_{4, \mathbf{k}})} \right)\right].
\end{aligned}
$}
\end{equation}
\end{widetext}
The quantities $E_{i, \mathbf{k}}$ are the eigenvalues lower eigenvalues $H_{\text{gen}} + H_{2, \text{lat}}$, ordered in terms of increasing energy.  
We also assumed that $\Delta$ is sufficiently large such that the CDW bands do not produce Fermi surfaces at the Fermi energy.  
The value of $\phi$ will switch the eigenstates corresponding to each of the $E_{i, \mathbf{k}}$.  
However, regardless of the choice of $\phi$, the final result will come out the same.

For the idealized model, we have that $d_4(\mathbf{k}) = d_5(\mathbf{k}) = d_6(\mathbf{k}) = 0$ and $(d_1(\mathbf{k}), d_2(\mathbf{k}), d_3(\mathbf{k})) = (2 t_x k_x, 2 t_y k_y, 2 t_z k_z)$. 
In this limit, the gap equation (\ref{gapGeneral}) reduces to
\begin{equation}
\label{idealGapEquation}
\resizebox{0.81\hsize}{!}{$
   \omega_\mathbf{Q} = \int [d \mathbf{k}] \frac{2 g^2}{\sqrt{(2 t_x k_x)^2 + (2 t_y k_y)^2 + (2 t_z k_z)^2 + (2 \Delta)^2}} ,
   $}
\end{equation}
which implcitly determines $|\Delta|$ in terms of the phonon frequency $\omega_\mathbf{Q}$ at $T=0$. 
\subsection{Feynman Rules}
\label{FynmanRulesSecion}

To calculate the {linear and nonlinear} conductivities, we will make use of diagrammatic perturbation theory, generalizing the formalism of Ref.~\onlinecite{MooreDiagrammatic} to include contributions from dynamical collective excitations.  
First, consider the phonon propagator expanded about the CDW wavevector $\mathbf{Q}$: $D_{d_1, d_2}(\mathbf{q}, \mathscr{T}) = i \left\langle T_{\mathscr{T}} b_{\mathbf{q} + d_1 \mathbf{Q } }(\mathscr{T}) b^{\dagger}_{ \mathbf{q} + d_2 \mathbf{Q}}(0) \right\rangle$, where $T_{\mathscr{T}}$ is the time-ordered operator for the Wick rotated imaginary time, $\mathscr{T}$, and $d_1, d_2$ takes either $(+)$ or $(-)$.  
The value of $D_{d_1, d_2}(\mathbf{q}, \mathscr{T})$ is determined by a set of coupled Dyson equations involving the bare ($g=0$) phonon propagator, and the electron propagator.  
We will focus solely on one-loop contributions to the collective phonon self-energy, since this is the leading contribution.

The Feynman rules that account for the interactions are:
\begin{enumerate}
  \item The free electron propagator is denoted by       \begin{tikzpicture}[baseline={(current bounding box.center)}]
    \begin{feynman}
    \vertex (a) ;
    \vertex [       left=of a] (b) ;
    \diagram* {
        (b) -- [fermion]  (a),
    };
    \end{feynman}
    \end{tikzpicture}.
  \item The free phonon propagator is denoted by       \begin{tikzpicture}[baseline={(current bounding box.center)}]
    \begin{feynman}
    \vertex (a) ;
    \vertex [       left=of a] (b) ;
    \diagram* {
        (b) -- [charged scalar]  (a),
    };
    \end{feynman}
    \end{tikzpicture}.
    \item The electromagnetic field is denoted by \begin{tikzpicture}[baseline={(current bounding box.center)}]
    \begin{feynman}
    \vertex (a) ;
    \vertex [       left=of a] (b) ;
    \diagram* {
        (b) -- [photon]  (a),
    };
    \end{feynman}
    \end{tikzpicture}.
    \item Each electron-phonon vertex provides a contribution of $ g$.
    \item Each electron-electromagnetic field vertex in the $\alpha$ direction provides a contribution of $\frac{i e}{\omega_{\alpha} \hbar}$ with an outgoing field denoted by \begin{tikzpicture}[baseline={(current bounding box.center)}]
    \begin{feynman}
    \vertex (a) ;
    \vertex [empty dot](b)[       left=1.5emof a]{}  ;
    \diagram* {
        (b) -- [photon]  (a),
    };
    \end{feynman}
    \end{tikzpicture} and an incoming field denoted by \begin{tikzpicture}[baseline={(current bounding box.center)}]
    \begin{feynman}
    \vertex (a) ;
    \vertex [dot](b)[       left=1.5emof a]{}  ;
    \diagram* {
        (b) -- [photon]  (a),
    };
    \end{feynman}
    \end{tikzpicture}.
    \item The momentum must also be indexed by $\xi = \pm 1$ for particle propagators, since momentum is only conserved modulo $\xi \mathbf{Q}$.  
    Similarly, $d = \pm 1$ indexes the phonon propagators with momentum modulo $d \mathbf{Q}$.  
    Each phonon vertex will specify $(+)$ or $(-)$.
    \item The orbital index will be denoted by $n$, which can take the values $1$ and $2$.
    \item To avoid double counting, only topologically unique diagrams should be considered.  
    Particularly, if the exchange of two or more frequencies is not unique, then an appropriate multiplicity factor must be included.
\end{enumerate}
It is useful to introducing a rank-3 tensor for the electron-phonon vertex.  
In the mean field approximation, the input vertex (one phonon to two electrons) extracted by applying the shifted zone scheme to Eq.~(\ref{gapAndPhonon}) is
\begin{equation}
\label{phononElectronVertexIn}
P_{n_1 \xi_1, n_2 \xi_2; d_1} = g \delta_{d_1, \xi_1} \left( \sigma_z \otimes \tau_x  \right)_{n_1 \xi_1, n_2 \xi_2}.
\end{equation}
Similarly, the output vertex (two electrons to one phonon) is
\begin{equation}
\label{phononElectronVertexOut}
P_{d_1; n_1 \xi_1, n_2 \xi_2} = g \delta_{d_1, \xi_2} \left( \sigma_z \otimes \tau_x \right)_{n_1 \xi_1, n_2 \xi_2}.
\end{equation}
These vertex tensors take a set of rank-2 tensors from particle Green functions, denoted by the combined indices $n$ and $\xi$ for the orbital and momentum bases respectively, and output a rank-1 phonon propagator of momentum basis, $d_1$.

\subsection{Collective Propagators}
\begin{figure}[ht]
      \centering
\fbox{
\begin{minipage}{0.93\hsize}
\centering
\includegraphics[width=0.93\hsize]{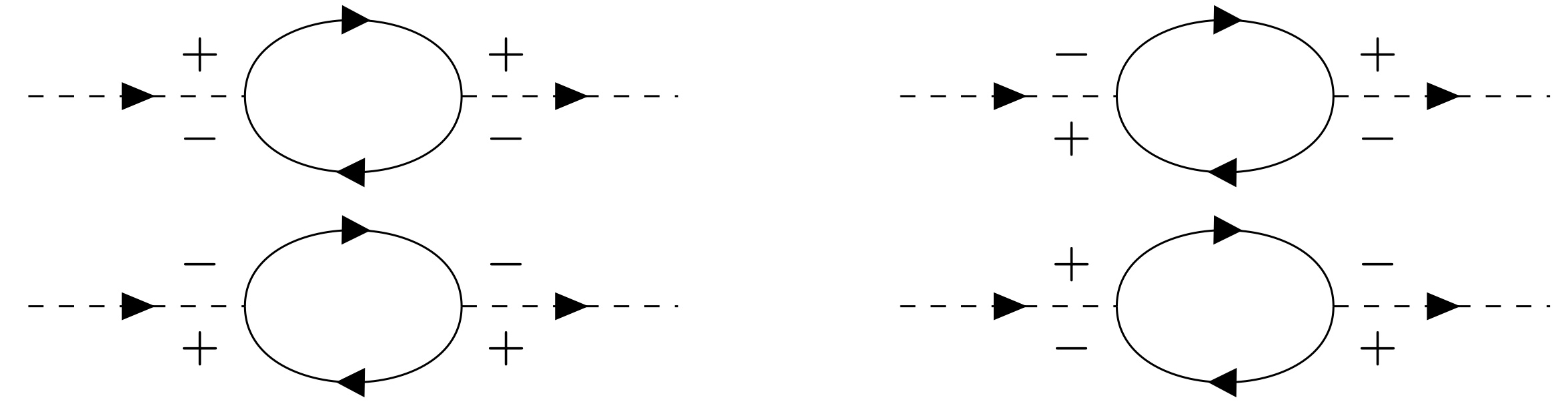}
\end{minipage}
}
\caption{The possible single-particle interactions for the interacting phonon propagator. }
\label{allPhononSingleParticleDiagrams}
\end{figure}
\begin{figure*}[t!]
      \centering
\fbox{
\begin{minipage}{13 cm}
\centering
\includegraphics[width=13cm]{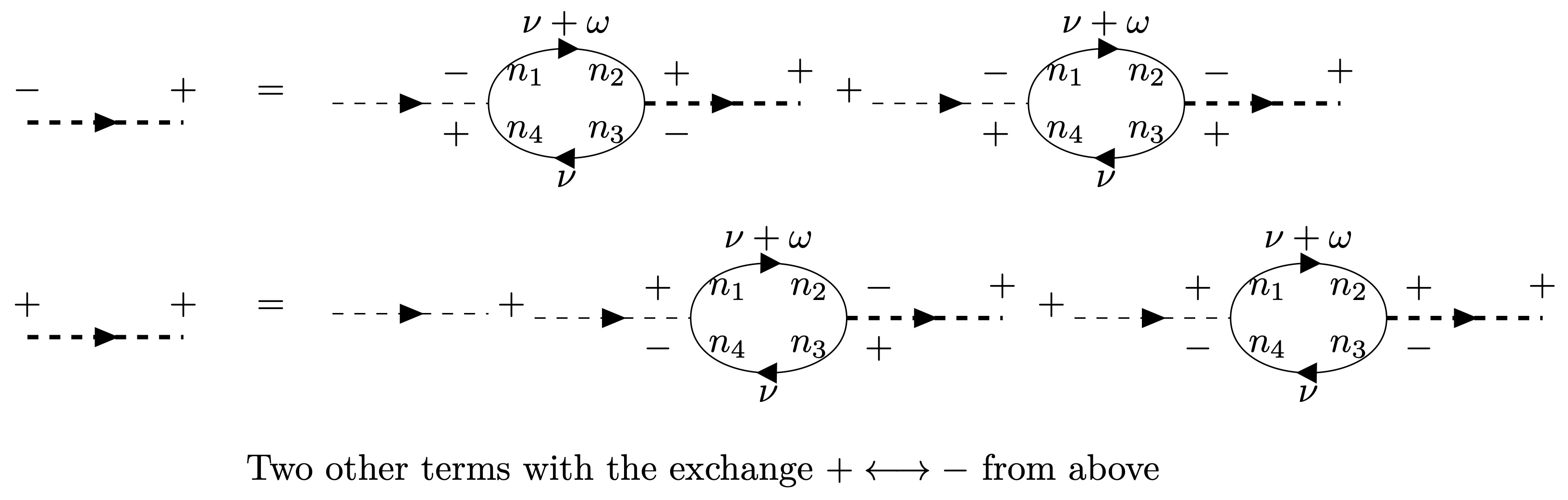}
\end{minipage}
}
\fbox{
\begin{minipage}{13 cm}
\centering
\includegraphics[width=10cm]{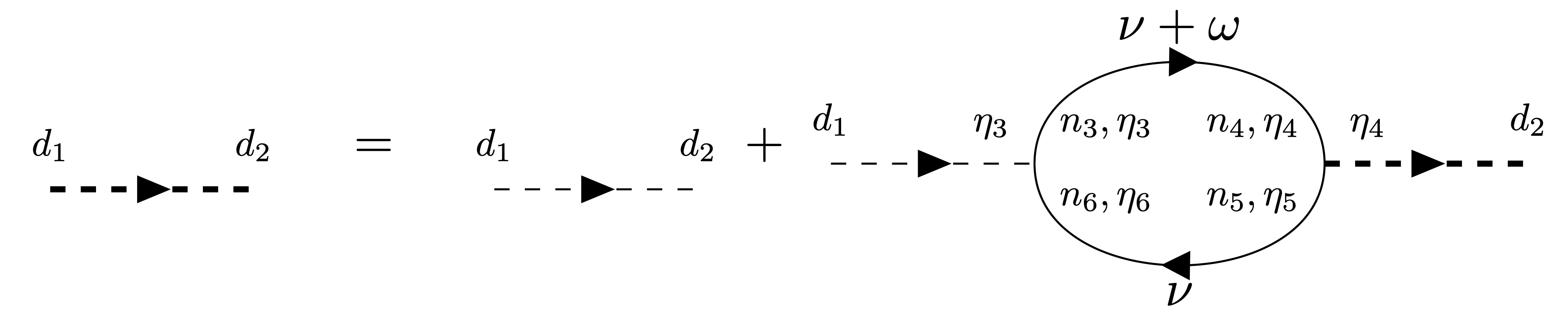}
\end{minipage}
}
\caption{The Dyson equations for the collective mode propagators.  The bold dotted line denotes the recursively defined collective phonon.  The bottom diagram generalizes the above diagrams \cite{SuzumuraCDWReflectivity, RLAChargeDensityWaves, FentonCDW, KuriharaPinningEffect}. }
\label{collectivePhononProgagatorDiagramEquation}
\end{figure*}

As phonons propagate through a material, they repeatedly interact with electron-hole pairs, and are renormalized by the electron-phonon interaction.  
Due to the presence of the charge-density wave condensate, the electron-phonon interaction can lead to a nonzero value of the off-diagonal propagator $D_{+-}(\mathbf{q}, \mathscr{T})$ \cite{RLAChargeDensityWaves, RiceElectronPhononInteractions}. 
Using our Feynman rules, we show in Fig.~\ref{allPhononSingleParticleDiagrams} the one-loop diagrams that contribute to the renormalized phonon propagator.  
We can represent the recursive effects of the electron-phonon interaction through a Dyson equation, as represented in Fig.~\ref{collectivePhononProgagatorDiagramEquation}.

Converting from diagrams to symbols, let $D_{d_1 d_2}(i\omega)$ represent the collective phonon propagator in Matsubara space.  
Then the Dyson equation reads
\begin{equation}
\resizebox{1\hsize}{!}{$
\label{GeneralD}
\begin{aligned}
    D_{d_1 d_2}(i \omega) =& \left(\delta_{d_1 d_6} 
    +\int [d k] \left( D_0(i\omega) \right)_{d_1 d_3} P_{d_3; n_7 \xi_7, n_3 \xi_3} G(i\omega + i\nu, \mathbf{k})_{n_3 \xi_3, n_4 \xi_4}
     \right.
    \\&
    \times
    P_{n_4 \xi_4, n_5 \xi_5; d_6} G(i\nu, \mathbf{k})_{n_5 \xi_5, n_7 \xi_7}\Big)^{-1} \left( D_0(i\omega) \right)_{d_6 d_2},
\end{aligned}
$}
\end{equation}
where we remind the reader we have specialized to the zero-momentum phonon propagator, which is sufficient for our forthcoming calculation of the conductivity. 
The computation of $D^{-1}(i\omega)$ applied to the ideal Weyl model reduces to
\begin{equation}
\resizebox{1\hsize}{!}{$
\begin{aligned}
    D^{-1}(i \omega) =& \begin{bmatrix}
D^{-1}_{++}(i\omega) & D^{-1}_{+-}(i\omega)\\
D^{-1}_{-+}(i\omega) & D^{-1}_{--}(i\omega) 
\end{bmatrix} 
\\=& 
D^{-1}_0(i \omega) \sigma_0 - 2 g^2\int[d \mathbf{k}] \begin{bmatrix}
\frac{-2 E^2_{\mathbf{k}} + (2 \Delta)^2}{E_{\mathbf{k}} (4 E^2_{\mathbf{k}} + \omega^2)} & \frac{ (2 \Delta)^2 e^{-2 i \phi} }{E_{\mathbf{k}} (4 E^2_{\mathbf{k}} + \omega^2)} \\
\frac{ (2 \Delta)^2 e^{2 i \phi} }{E_{\mathbf{k}} (4 E^2_{\mathbf{k}} + \omega^2)} & \frac{-2 E^2_{\mathbf{k}} + (2 \Delta)^2}{E_{\mathbf{k}} (4 E^2_{\mathbf{k}} + \omega^2)}
\end{bmatrix}.
\end{aligned}
$}
\end{equation}
The linear combinations $D_{++}(\omega) + e^{2 i \phi}D_{+-}(\omega) $ and $D_{++}(\omega) - e^{2 i \phi} D_{+-}(\omega)$ are more important to later calculations.  
These two linear combinations can be interpreted as the massive amplitude and massless phase collective excitations, respectively \cite{RLAChargeDensityWaves, Rice_Supplemental_notes_on_orignal_RLA, RiceandCrossCDW, TakanoDigrammaticalCDW, RiceElectronPhononInteractions, FentonCDW}.  
To see this, note that after a Wick rotation from Matsubara frequency space to real frequency space we have,
\begin{align}
    D_{++}(\omega) - e^{2 i \phi} D_{+-}(\omega) &= \frac{1}{\frac{ -\omega^2 + \omega^2_{\mathbf{Q}} }{2 \omega_{ \mathbf{Q} }}  - g^2\int[d \mathbf{k}] \frac{4 E_{\mathbf{k}} }{4 E_{\mathbf{k}}^2 -\omega^2}}\nonumber
    \\ 
    &=
    \frac{1}{-\frac{ \omega^2}{2 \omega_{\mathbf{Q}}}   + g^2\int[d \mathbf{k}] \frac{-\omega^2 }{E_{\mathbf{k}} (4 E_{\mathbf{k}}^2 -\omega^2)}},\label{MasslessMode}
    \\
    D_{++}(\omega) + e^{2 i \phi}D_{+-}(\omega) &= \frac{1}{\frac{ -\omega^2 + \omega^2_{\mathbf{Q}} }{2 \omega_{ \mathbf{Q} }}  -g^2\int[d \mathbf{k}] \frac{4 (E^2_{\mathbf{k}} - (2 \Delta)^2) }{E_{\mathbf{k}}(4 E_{\mathbf{k}}^2 -\omega^2)}}\nonumber
    \\ 
    &=
    \frac{1}{-\frac{ \omega^2}{2 \omega_{\mathbf{Q}}}   + g^2\int[d \mathbf{k}] \frac{ 4 (2 \Delta)^2 -\omega^2 }{E_{\mathbf{k}} (4 E_{\mathbf{k}}^2 -\omega^2)}}.\label{MassiveMode}
\end{align}
The frequency $\omega_{\mathbf{Q}}$ is related to $\Delta$ through the gap equation. 
Substituting the mean field ansatz \eqref{eq:mfansatz} into the definition of $D(\omega)$ reveals that Eq.~(\ref{MasslessMode}) is proportional to the propagator for $\delta\theta$, while Eq.~\eqref{MassiveMode} is proportional to the propagator for $\delta b$. Using the ideal gap equation \eqref{idealGapEquation}, we see that the massless mode propagator \eqref{MasslessMode} becomes singular as $ \omega \rightarrow 0$ while the massive mode propagator \eqref{MassiveMode} is singular at a nonzero resonant frequency $\omega_{\text{res}}$, approximated in Appendix \ref{massiveResonanceSection}. 
Note that in the weak coupling limit, we will generally have {$\omega_{\text{res}} < |\Delta|$.} 

\section{Conductivity}
\label{conductivity_section}
Having constructed the propagators for the collective mode and electronic excitations in the CDW phase, we can now compute the electromagnetic response for our models, focusing on the role of the collective modes. 
We will follow the diagrammatic approach to nonlinear optical conductivity outlined in Ref.~\onlinecite{MooreDiagrammatic}. 
To begin, recall that from the Kubo formula, the $n$-th order optical conductivity can be expressed in terms of a correlation function of $n$ velocity operators. 
To define the velocity operator, we can use the ``velocity gauge'' minimal coupling
$\mathbf{k}\rightarrow \mathbf{k} - q \mathbf{A}(t)$, where $\mathbf{E}(t) = E_0 e^{i \omega t} = - i \omega \mathbf{A}(t) $.  
Any Hamiltonian can na\"{i}vely be expressed as an expansion in the vector potential: $H(\mathbf{k} +
\frac{e}{\hbar} \mathbf{A}(t)) = H(\mathbf{k}) + \sum^{\infty}_{n = 1} \frac{1}{n!} \Pi_i\Big[ \frac{e}{\hbar}A^{\alpha_i}\partial_k^{\alpha_i} H(\mathbf{k}) \Big]$.  
However, instead of applying the ordinary derivative in this expansion, the covariant derivative should be used to ensure the result is gauge invariant with respect to $\mathbf{k}$-dependent changes of basis for the occupied Bloch functions. 
This amounts to defining the velocity as the time-derivative of the position operator, expressed in terms of the Berry connection\cite{MooreDiagrammatic}.  
Therefore, an arbitrary operator's derivative should be replaced by the following covariant version: $\partial_{k^{\alpha^i}} \hat{\mathcal{O}}_{a b} \rightarrow \partial_{k^{\alpha^i}}\hat{\mathcal{O}}_{a b} - i [\mathcal{A}^{\alpha^i}, \hat{\mathcal{O}}]_{a b} \equiv \hat{D}^{\alpha^i} \mathcal{O}_{ab}$, where $\mathbf{\mathcal{A}}$ is the Berry connection.  
{The derivation of how the Berry connection shows up in the velocity operator is given in Appendix \ref{VelocityOperatorDerivation}.}  
We can then expand the Hamiltonian as $H(\mathbf{k} + \frac{e}{\hbar} \mathbf{A})=H(\mathbf{k})+\hat{V}_E(\mathbf{k})$, where $\hat{V}_E(\mathbf{k})$ includes all (Berry) gauge invariant coupling terms to the vector potential. 
In general,
\begin{align}
    \hat{V}_E(\mathbf{k}) &= \frac{e}{\hbar} A^{\alpha} [\hat{D}^{\alpha}, H(\mathbf{k})] + \frac{1}{2}\frac{e^2}{\hbar^2} A^{\alpha} A^{\beta} [\hat{D}^{\beta}, [\hat{D}^{\alpha}, H(\mathbf{k})]]\nonumber \\
    &+ \frac{1}{6}\frac{e^3}{\hbar^3} A^{\alpha} A^{\beta} A^{\gamma} [\hat{D}^{\gamma}, [\hat{D}^{\beta}, [\hat{D}^{\alpha}, H(\mathbf{k})]]] + \dotsb.\label{eq:VEint}
\end{align}
Note that the quadratic term in this expansion gives the diamagnetic current.
We will see that this diamagnetic contribution is critical for regularizing the singularities from the ideal Weyl model \cite{BradlynDiamagnetic}. 
More generally, {we will use this definition of $\hat{V}_E(\mathbf{k})$ to define the diagrammatic electronic velocity vertices, $h^{\alpha_1 \alpha_2 \cdot \cdot \cdot \alpha_n} = \frac{\delta^n}{\delta A^{\alpha_1} \delta A^{\alpha_2} \delta A^{\alpha_n} } \hat{V}_E(\mathbf{k})$.}

To apply this formalism to our present case, we must carefully treat $ \frac{d \hat{x}}{d t} $, in the presence of the electron-phonon interaction.  
Consider the current operator, $\hat{j}_{\mathbf{q}}$ obeying the continuity equation, $e[ H, \sum_{p} \vec{c}^{\dagger}_{\mathbf{q}} \vec{c}_{\mathbf{p}+\mathbf{q}^{\prime}}] =  \mathbf{q}^{\prime} \hat{j}_{q^{\prime}}$.  
Expanding this current operator provides a consistent definition of the velocity operator, which is derived in Appendix \ref{VelocityOperatorDerivation}.  
Notably, there are no tricky contributions coming from the chosen basis using $c_{\mathbf{k} \pm \mathbf{Q}}$.  
Thus, the relation, $\langle n_1, \xi_1 |  \frac{d \hat{x}^{\mu}}{d t} | n_2, \xi_2 \rangle = \hat{v}^{\mu}_{E, n_1 \xi_1, n_2 \xi_2} = \left[ \hat{D}^{\mu} H_0(\mathbf{k} + \frac{e}{\hbar} \mathbf{A}(t)) \right]_{n_1 \xi_1, n_2 \xi_2}$, above still holds.  

Having worked out the appropriate velocity operator, we can now diagrammatically compute the $n-$th order conductivity using a generating function approach \cite{MooreDiagrammatic}.  
The expectation value of the electric current is given by
\begin{equation}
\label{masterEqnforJ}
    \langle \hat{J}^{\mu} \rangle = \frac{1}{Z} \text{Tr} \left[ T_{t} e \hat{v}^{\mu}_{E} e^{-i \int d t^{\prime}\hat{H_0}(t^{\prime}) + \hat{H_1}(t^{\prime}) +\hat{H_2}(t^{\prime})} \right],
\end{equation}
where $T_{t}$ denotes time-ordering of the exponential. 
The definition of the nonlinear conductivity includes contributions from multiple electric field sources, and so is specified as
\begin{align}
\langle \hat{J}^{\mu}(t) \rangle &= \sum^{\infty}_{n = 1}\int(\prod_{j=1}^n dt_j)\sigma^{\mu \alpha_1 \dotsb \alpha_n}(t,t_1,t_2,\dots,t_n) \nonumber \\\
&\times \prod^{n}_{i = 1} E^{\alpha_i}(t_i).
\end{align}
Therefore, taking functional derivatives and Fourier transforming gives the $(n)$-th order conductivity as\cite{sipe_second-order_formalism_2000, sipe_third_order_optical_response_2020, peterson_formal_nonlinear_theory_1967}  
\begin{equation}
\label{masterEqnforSigma}
\begin{split}
    \sigma^{\mu \alpha_1 \dotsb \alpha_n}(\omega, \omega_1, \dotsb, \omega_n) = \left. \int [dt] e^{i \omega t} \prod^{n}_{i= 1}\int [d t_i]e^{i \omega_i t_i}
    \right.&
    \\
    \left.
    \times \frac{\delta}{\delta E^{\alpha_i}(t_i)} \langle \hat{J}^{\mu} \rangle \right |_{\mathbf{E} = 0}.&
\end{split}
\end{equation}
With this functional form, the current may be computed to all orders in the electric field.  
Note that this method produces time-ordered, rather than the causal response functions that are more conventional in condensed matter physics.  
To connect these two, care must be taken in how we extend Eq.~(\ref{masterEqnforSigma}) into the complex frequency domain. 
While this is a textbook problem for the linear conductivity\cite{BruusAndFlensberg}, it is not necessarily trivial for the nonlinear collective response.  
In our framework, we will follow the procedure outlined in Refs.~\cite{MooreDiagrammatic,passos2018nonlinear} and shift all frequencies $\omega_\alpha$ into the complex plane in an identical way  (i.e. $\omega_\alpha + \omega_\beta \rightarrow \omega_{\alpha} + \omega_{\beta} + 2 i \eta$ for the small complex part, $i \eta$).  
Since this issue does not affect our central results as to the relative contributions of the collective contributions to the conductivity, we will defer a systematic examination to future work.

As we compute the conductivity below, it will be convenient to separate the contributions into two categories: those with and without contributions from the collective mode.  
The collective contribution is understood to mean the contributions to the conductivity mediated by the collective mode propagator via the electron-phonon interaction.  
We compute the conductivity to order $\mathcal{O}(g^2)$ in the electron-phonon coupling. 
Furthermore, in the spirit of Migdal's theorem, we neglect vertex corrections\cite{FentonCDW}.  
In this way, we will explore to what extent the massless and massive collective modes in a 3D Weyl-CDW carry an electric current, generalizing the classical Fr\"{o}hlich result\cite{FrohlichSuperconductivityCDW,RLAChargeDensityWaves}. 
To make the calculation more convenient, we will work in the orbital basis of the Hamiltonian rather than converting to the block diagonalized band basis.  
The main reason for this is to avoid messy conversions of the electron-phonon vertex.  
A secondary reason is that the orbital basis basis leads to the absence of the Berry connection in the covariant derivative.  
That is, the Berry curvature contributions to the current, which are invariant, are absorbed into definitions of the Green function and particle velocity vertices.  
A tertiary reason is that the orbital basis is generally more computationally friendly than the energy band basis. 
However, the final results for the conductivities are basis independent.

In the subsequent subsections, we compute the conductivity for the Weyl-CDW using two parallel approaches.  
First, we use the low-energy, ideal Weyl model directly, where regularization of divergences in frequency is necessary.  
To obtain finite results, we implement a minimal subtraction regularization scheme.  
Second, we compute the conductivity for a lattice completion of the Weyl-CDW.  
The full Weyl lattice model is used, where the current vertices are defined as variational derivatives of the generating function.  
The lattice completion method acts to approximate the incommensurate lattice, and directly parallels the phenomenology of of the low-energy regularization method.  
We will start by applying these methods to compute the linear conductivity in Subsection~\ref{subsec:linear}, followed by the second-order conductivity in Subsection~\ref{subsec:second}.
Finally, we examine the third-order conductivity in Subsection~\ref{subsec:third}.

\subsection{Linear Conductivity}\label{subsec:linear}

We start by examining the linear conductivity. 
Since (as we will see) the collective mode contributions only manifest in the longitudinal conductivity $\sigma^{zz}(\omega)$, we will focus our attention primarily on these components\footnote{{It should be noted there will be an additional anomalous Hall response given as $ \mathbf{j}_{\text{anom}} = \frac{1}{2 \pi^2} \int [d \mathbf{k}] \left( \mathbf{k} \phi + \frac{1}{2} \overline{\mathbf{Q}} \right) \times \mathbf{E}$.  However, the derivation of this in the field-theoretic description relies on topological triangle diagrams described in References ~\onlinecite{fujikawa_genesis_2020_1}, ~\onlinecite{fujikawa_regularization_2020_2}, and ~\onlinecite{fujikawa_path-integral}.  It is understood such responses are already accounted for in the generating functional description of the current in Section \ref{conductivity_section}.  In other words, the current response described in this paper is non-topological.}}. 

\subsubsection{Linear Non-collective Contributions ($\sigma_{\text{NC}}^{zz}(\omega_{\beta})$)}
\label{LinearNonCollectiveSection}

\begin{figure}[ht]
      \centering
\fbox{
\begin{minipage}{0.93\hsize}
\includegraphics[width=0.93\hsize]{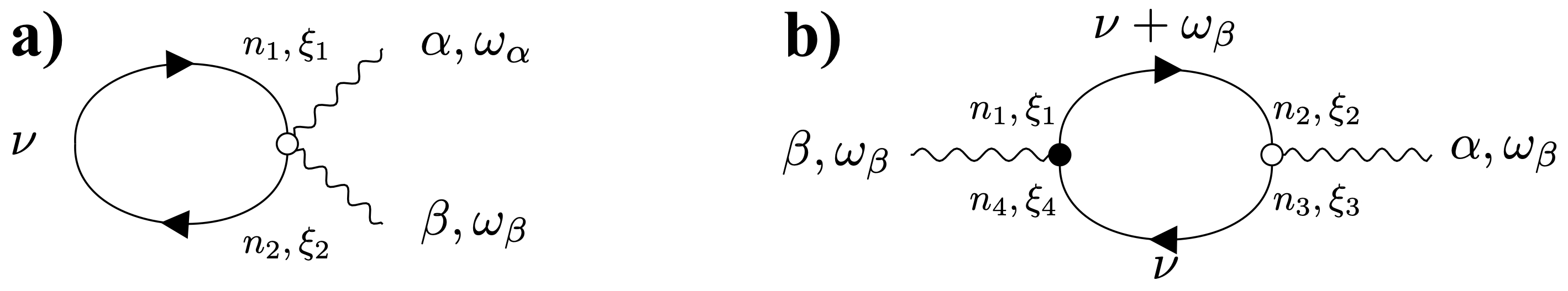}
\end{minipage}
}
\caption{Feynman diagrams for the non-collective contributions to the linear conductivity. a) is the diamagnetic term. b) is the single-particle term. }
\label{LinearNonCollectiveDiagrams}
\end{figure}

We start by considering the ordinary single-particle (non-collective) contribution to the linear optical conductivity.
The sum of contributions from Fig.~\ref{LinearNonCollectiveDiagrams} and the application of the Feynman Rules from Section \ref{FynmanRulesSecion} gives the linear order contribution.  
The non-collective conductivity, denoted by subscript ``NC", is
\begin{equation}
\label{linearNonCollectiveEquation}
\begin{split}
    \sigma^{\alpha \beta}_{\text{NC}}(i\omega_{\beta}) =& \frac{i e^2}{\hbar i \omega_{\beta}}\int [d k] \left[ G_{n_1 \xi_1, n_2 \xi_2}(i\nu, \mathbf{k}) h^{\alpha \beta}_{n_2 \xi_2, n_1 \xi_1}
    \right.
    \\&
    \left.
    + G_{n_1 \xi_1, n_2 \xi_2}(i\nu + i\omega_{\beta}, \mathbf{k}) h^{\beta}_{n_2 \xi_2, n_3 \xi_3} 
    \right.
    \\&
    \left.
     \times G_{n_3 \xi_3, n_4 \xi_4}(i\nu, \mathbf{k}) h^{\alpha}_{n_4 \xi_4, n_1 \xi_1} \right].
\end{split}
\end{equation}
Since our system of interest is a single-particle band insulator, we expect $\sigma_{NC}^{zz}(i\omega)$ to vanish in the limit of low frequency. 
Examining this in detail will shed light on the role of the two-photon vertex $h^{\alpha\beta}$.

\subsubsubsection{Ideal Weyl Model}
We start by considering the linearized ideal gapped Weyl model.  
{Using the definition of the covariant derivative, the one-photon velocity vertex is 
\begin{equation}
\label{linearizedVelocityVertexWithOneIndex}
    h^{\alpha} = \left[ -2(t_x \sigma_x\delta^{\alpha}_x + t_y \sigma_y\delta^{\alpha}_y) \otimes \tau_0+ 2\delta^{\alpha}_z t_z \sigma_z \otimes \tau_z \right].
\end{equation}}  
For this system, it turns out that $h^{\alpha \beta}_{n_1 \xi_1, n_2 \xi_2}$ is zero, since the Hamiltonian is linear in $\mathbf{k}$.  
However, this is clearly not the case for any lattice completion of the model (which must describe an insulator).  
In the ideal gapped Weyl linear response, a singularity exists in the non-collective response as $\omega \rightarrow 0$, which is not present in the response from the lattice model, which is a band insulator with $\sigma^{\alpha\alpha}_{NC}(\omega\rightarrow 0)=0$.  
Furthermore, we expect the diagmagnetic contribution from Fig.~\ref{LinearNonCollectiveDiagrams}a) to ensure the single particle part from Fig.~\ref{LinearNonCollectiveDiagrams}b) does not diverge as $ \omega \rightarrow0$.  
To accomplish this, it must be true that the integral from Eq.~\eqref{linearNonCollectiveEquation} conspires to be zero at $i \omega_{\beta} = 0$.  
To recover this physics in the linearized model, we introduce a regularizing velocity vertex, $\overline{h}^{\alpha \beta}_{n_1 \xi_1, n_2 \xi_2}$ with the demand that
\begin{equation}
\begin{split}
\label{linearRegularizationEquation}
    0=&\int [d \nu] \left[ G_{n_1 \xi_1, n_2 \xi_2}(i\nu, \mathbf{k}) \overline{h}^{\alpha \beta}_{n_2 \xi_2, n_1 \xi_1}
    \right.
    \\&
    \left.
    + G_{n_1 \xi_1, n_2 \xi_2}(i\nu + i\omega_{\beta}, \mathbf{k}) h^{\beta}_{n_2 \xi_2, n_3 \xi_3} G_{n_3 \xi_3, n_4 \xi_4}(i\nu, \mathbf{k})
    \right.
    \\&
    \left.
    \times h^{\alpha}_{n_4 \xi_4, n_1 \xi_1} \right]_{i \omega_{\beta} = 0} .
\end{split}
\end{equation}
Note that since the integral here is only over frequency $\nu$, this is a stronger condition than the demand $\sigma^{\alpha\alpha}_{NC}(\omega\rightarrow 0)=0$; nevertheless, we find this allows for the simplest solution for $\overline{h}^{\alpha \beta}_{n_1 \xi_1, n_2 \xi_2}$. 
Some constraints on $\overline{h}^{\alpha \beta}_{n_1 \xi_1, n_2 \xi_2}$ derived from the lattice model are: it should be block diagonal, not depend on the integration frequency $\nu$, and be Hermitian.  
Additionally, there are integration parity constraints on this vertex.  
For example, the lattice model vertex $h^{zz}$ is odd in $k_z$, so the regularized $\overline{h}^{zz}$ should also be odd in $k_z$.  
Upon solving Eq.~\eqref{linearRegularizationEquation}, we find one possible solution 
\begin{equation}
    \label{completeRegularizedHzz}
    \sum_{n, \xi} \text{sgn}(\xi) (-1)^n \overline{h}^{zz}_{n \xi, n \xi} = \frac{8 t_z (E^2_{\mathbf{k}} -(2 k_z)^2)}{( k_z) E^2_{\mathbf{k}}}.
\end{equation}
This choice of $\overline{h}^{\alpha \beta}_{n_1 \xi_1, n_2 \xi_2}$ is not unique.  
There will always exist a family of transformations, $\overline{h}^{\alpha \beta}_{n_1 \xi_1, n_2 \xi_2} \rightarrow \overline{h}^{\alpha \beta}_{n_1 \xi_1, n_2 \xi_2} + \overline{X}^{\alpha \beta}_{n_1 \xi_1, n_2 \xi_2}$ each of which will satisfy Eq.~(\ref{linearRegularizationEquation}) provided that $\int [d k] \text{Tr}(G(\nu, \mathbf{k}) \overline{X}^{\alpha \beta} ) = 0$.  
However, the ambiguity due to $\overline{X}^{\alpha \beta}$ will not affect the final answers in either Sections \ref{linCondCollIdealWeyl} or \ref{secNCCondIdealModel}, since taking traces only require Eq.~(\ref{completeRegularizedHzz}). 

\subsubsubsection{Linear Collective Contributions ($\sigma^{zz}_{\text{coll}}(\omega_{\beta})$)}
\label{LinearCollectiveSection}
\begin{figure}[h]
      \centering
\fbox{
\begin{minipage}{0.93\hsize}
\centering
\includegraphics[width=0.93\hsize]{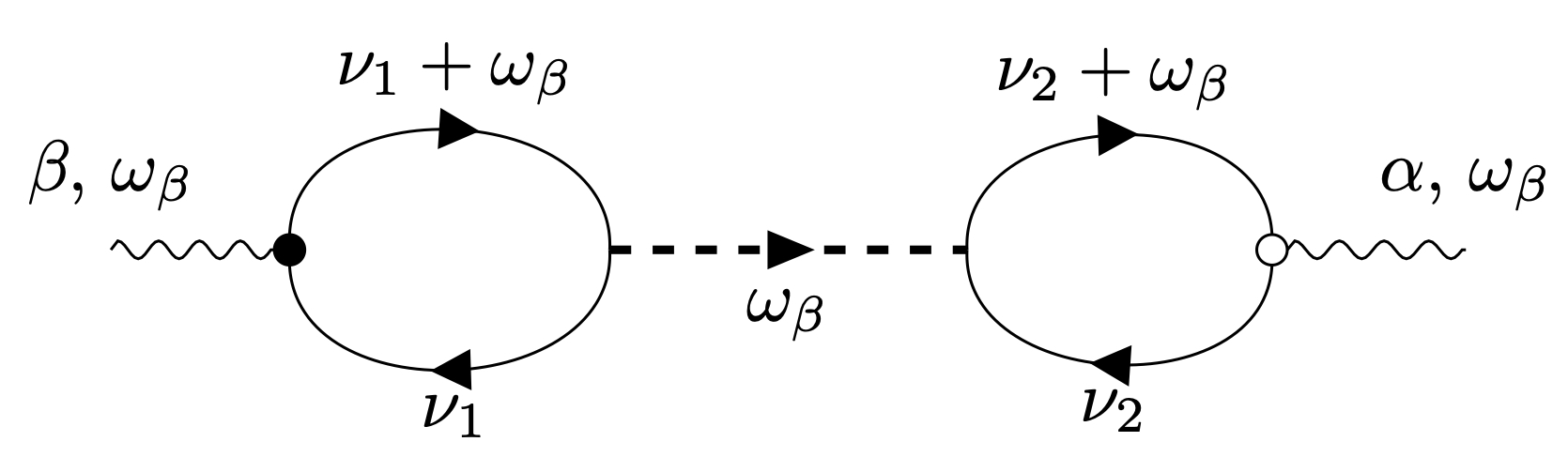}
\end{minipage}
}
\caption{Diagram for the collective mode contribution to the linear conductivity \cite{RLAChargeDensityWaves, RiceandCrossCDW, FentonCDW}.  To alleviate crowding in the diagrams, the indices to be summed over ($n$, $\xi$, and $d$) will be left implicit in the diagram and in all the diagrams that follow.}
\label{LinearCollectiveDiagrams}
\end{figure}

Next we calculate the contribution to the conductivity from the collective modes, denoted by a subscript ``coll".  
Converting from the diagram of Fig.~\ref{LinearCollectiveDiagrams} to a mathematical expression, we find:
\begin{equation}
\label{LinearCollectiveFullCalc}
\begin{split}
    \sigma^{\alpha \beta}_{\text{coll}}(i\omega_{\beta}) =& \frac{i e^2}{\hbar i\omega_{\beta}}\sum \int[d k_1] [d k_2] G_{n_1, \xi_1, n_2; \xi_2}(i\nu_1, \mathbf{k}_1) 
    \\
    &
    \times h^{\beta}_{n_2, \xi_2; n_3, \xi_3} G_{n_3, \xi_3; n_4, \xi_4}(i\nu_1 + i\omega_{\beta}, \mathbf{k}_1)
     \\
    &
    \times P_{n_4, \xi_4, n_1, \xi_1; d_1} D_{d_1; d_2}(i\omega_{\beta})
    \\ &
    \times  P_{ d_2;  n_8, \xi_8, n_5, \xi_5} G_{n_5, \xi_5; n_6, \xi_6}(i\nu_2 + i\omega_{\beta}, \mathbf{k}_2) 
    \\ &
    \times
    h^{\alpha}_{n_6, \xi_6; n_7, \xi_6} G_{n_7, \xi_7; n_8, \xi_8}(i\nu_2, \mathbf{k}_2).
\end{split}
\end{equation}
We will now evaluate this expression for the ideal linearized model, the lattice model, and the tilted lattice model.
\subsubsubsection{Ideal Weyl Model}
\label{linCondCollIdealWeyl}
To analyze the collective conductivity in the linearized model, we plug the Green's functions from Eq.~\eqref{greenFunctions} and the velocity vertex for the linearized model into Eq.~\eqref{LinearCollectiveFullCalc}.  
The (not yet Wick-rotated) resulting equation is
\begin{widetext}
\begin{equation}
\label{idealLinearCollectiveCondutivityEquation}
    \sigma^{\alpha \beta}_{\text{coll}}(i\omega_{\beta}) = \frac{ i e^2}{\hbar i \omega_{\beta}} \int [d k_1] [d k_2] \frac{32 g^2 \left( D_{++}( i\omega_{\beta}) + e^{2 i \phi} D_{+-}(i\omega_{\beta})  \right) \left( \overline{k}_1^{\beta} \overline{v}_1^{\beta} \right) \left( \overline{k}_2^{\alpha}  \overline{v}_2^{\alpha} \right) (2 \Delta)^2  }{(E^2_{\mathbf{k}_1} + \nu^2_1)(E^2_{\mathbf{k}_1} + (\nu_1 + \omega_{\beta})^2)(E^2_{\mathbf{k}_2} + \nu^2_2)(E^2_{\mathbf{k}_2} + (\nu_2 + \omega_{\beta})^2)},
\end{equation}
\end{widetext}
where $\overline{v} = (0, 2 t_x, 2 t_y, -2 t_z \sin Q / 2)$ is the scaled velocity vector.  
Since this result in linear in $\overline{k}_1$ and $\overline{k}_2$, the integration will force $\sigma^{\alpha \beta}_{\text{coll}}(\omega_{\beta})$ to vanish.  
This vanishing is expected as the massive mode corresponds to oscillations in the amplitude of the charge-density wave modulation, which should be unable to carry an electric charge in an inversion- and particle-hole symmetric system at linear order \cite{RLAChargeDensityWaves}. 

Importantly, note that the vanishing of the collective contribution holds for all loops involving a single electric field vertex, $h^{\mu}_{n_1 \xi_1, n_2 \xi_2}$ in the collective conductivity.  
Therefore, diagrams with loops having a single electric field vertex, as in Fig.~\ref{LinearCollectiveDiagrams}, will be ignored in higher orders of the conductivity when we examine the untilted Weyl models.

Let us briefly comment on the dependence of our result on the choice of Fermi energy.  
Because the integrand in Eq.~(\ref{idealLinearCollectiveCondutivityEquation}) is an odd function, the collective contribution to the conductivity will vanish for any Fermi energy in the gap.  
For temperatures approaching zero, there is an implied Heaviside step function in Eq.~\eqref{idealLinearCollectiveCondutivityEquation}, $\Theta(\pm E_{\mathbf{k}} - \epsilon_F)$, which comes about from integrating the Matsubara frequency.  
Hence, for Fermi energies in the gap, the integrand remains odd and vanishes.  
Conversely, for Fermi energies above or below the gap, the argument of the Heaviside function will disrupt the parity of the integrand.  
However, the collective contribution to the conductivity will be small for small Fermi surface volumes, and will generally be overshadowed by the non-collective contribution.  
A study of the response of systems with energies above the gap is outside the scope of this paper.

Note crucially that unlike in the case of a 1D CDW, the massless mode does not contribute to the linear conductivity in the untilted Weyl model.  
The formative paper by Lee, Rice, and Anderson (LRA) provided a one-dimensional calculation showing why the massless mode propagator contributes for a single, quadraticly dispersing band \cite{RLAChargeDensityWaves, RiceandCrossCDW}, which has one degree of freedom at each Fermi point.    
Here, however, we have two degrees of freedom per Fermi point in each one-dimensional $z$-directed slice of the Brillouin zone. Furthermore, in the untilted model, each of these degrees of freedom has the opposite velocity. 
The ideal Weyl model can thus be thought of as an LRA model with velocity $v_z=\pm v_f$ at each Fermi point, combined with a second ``anti-LRA'' model with velocity $v_z=\mp v_f$ at each Fermi point.  
Because the two velocities are opposite on either Weyl node, the two contributions cancel.  
A more detailed analysis is provided in Appendix \ref{MasslessModeVanishingDerivation}. 

To get some physical intution for this, consider our ideal Weyl model in the undistorted phase, with two small Fermi pockets around each Weyl node, separated by wavevector $\mathbf{Q}$. 
When the CDW order parameter is nonzero, we condense particle-hole pairs consisting of a hole in one Fermi pocket, and an electron separated by momentum $\mathbf{Q}$. 
In our ideal Weyl model, these electron and hole states exhibit the \emph{same} velocity, unlike for a 1D CDW. Because of this, exciting the phase mode does not produce a net velocity in the untilted 3D case; in the semiclassical picture of Bardeen\cite{allender1974,gruner1988dynamics}, every left moving state that is depopulated at one pocket is compensated by an extra left moving state at the opposite pocket (and vice-versa for right moving states). 
Thus, in order for the sliding mode to contribute to the DC conductivity in a Weyl-CDW, we must have an asymmetry between the velocities in different Weyl pockets.  

There are a several symmetry-breaking terms that may allow the collective modes to contribute to the conductivity at linear order. 
For instance, our ideal model has inversion, mirror, and particle-hole symmetry.  
If either inversion symmetry or mirror symmetry $M_z: z\rightarrow-z$ is broken, then the integral in Eq.~(\ref{idealLinearCollectiveCondutivityEquation}) will not be forced to vanish due to integration parity considerations.  
Furthermore, particle-hole symmetry breaking can be introduced by including a term proportional to $\sigma_0$ in the Hamiltonian.  
The $\sigma_0$ term modifies the Pauli matrix structure {and} the traces in Eq.~(\ref{LinearCollectiveFullCalc}) (from Appendix \ref{MasslessModeVanishingDerivation}) to allow for a nonvanishing contribution of the massless mode to the conductivity.  
This can be accomplished by shifting the nodes in energy or inducing a tilt to the Weyl nodes, which we explore in the next sections.

\subsubsubsection{Lattice Weyl Model Without Tilt}

The same arguments from the ideal Weyl model still apply to the lattice model.  
Namely, there is no collective contribution to the conductivity in the absence of tilt.  
This is due to massive mode propagator involving an integrand that is odd in $k_z$, as in Eq.~\eqref{idealLinearCollectiveCondutivityEquation}, and the vanishing of diagrams involving the massless mode propagator due to the trace identities outlined in Appendix \ref{MasslessModeVanishingDerivation}.  
The conclusions are exactly the same as Section \ref{linCondCollIdealWeyl}.

\subsubsubsection{Lattice Weyl Model With Tilt}
\label{SectionLinearResponseWithTilt}

\begin{figure*}[ht]
      \centering
\centering
\includegraphics[width=0.95\hsize]{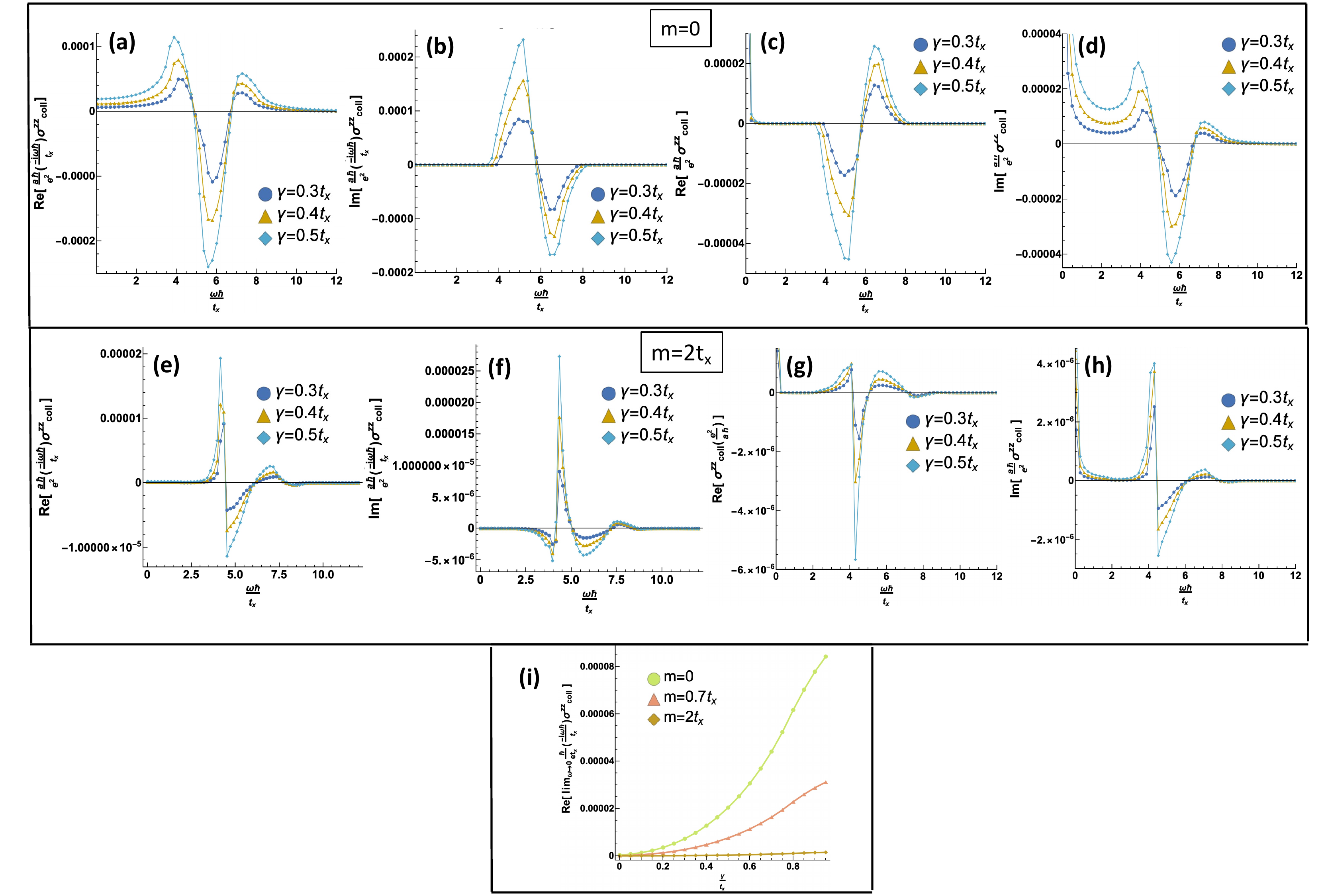}
\caption{Linear collective conductivity for the lattice model ($m=0$ and $m=2 t_x$ respectively). We plot both $-i\omega \sigma^{zz}_{\text{coll}}(\omega)$ [in (a), (b), (e), and (f)] and  $\sigma^{zz}_{\text{coll}}(\omega)$ [in (c), (d), (g), and (h)].  (a), (c), (e), and (g) are the real parts of the function, while (b), (d), (f), and (h) are the imaginary parts.  In all cases $\Delta = 0.5 t_x$.  (i) gives the real part of $-i\omega \sigma^{zz}_{\text{coll}}(\omega)$ as $\omega \rightarrow 0$ as a function of the tilt, $\gamma$.}
\label{linearCollectiveConductivityPlotWithTilt}
\end{figure*}

To avoid the exact cancellation of the massless mode contribution to the conductivity, we can introduce a tilt of the Weyl nodes via the Hamiltonian from Eq.~\eqref{tiltedLatticeTerm}.  
This ensures the velocity matrix at each node has nonzero trace, and so the LRA contribution and anti-LRA contribution described in Section \ref{linCondCollIdealWeyl} do not cancel.  
The result is the massless sliding mode acquires a nonzero effective charge, and carries a current.  
The inclusion of the tilting term from Eq.~\eqref{tiltedLatticeTerm} to the lattice model from Eq.~\eqref{FullLatticeModel} can be numerically evaluated to find $\sigma^{\alpha \beta}_{\text{coll}}(\omega_{\beta})$, shown in Fig.~\ref{linearCollectiveConductivityPlotWithTilt}.

There are several characteristics from the energy dispersion that contribute to the features observed in $\sigma^{zz}_\text{coll}(\omega)$:
\begin{enumerate}
        \item Only the massless phonon propagators contribute to the linear collective conductivity in the tilted model. 
        The non-zero intercepts at $\omega = 0$ in Fig.~\ref{linearCollectiveConductivityPlotWithTilt}(a), (e), and (i) illustrate this. 
        These reflect a divergence in the DC conductivity characteristic of the unpinned charge-density wave sliding mode.
        \item We expect to see a change of slope in the conductivity at $\omega = 4 \Delta$, when the electric field can excite an electron across the gap.  
        However this feature is imperceptible for this choice of $m=0$.  At $m = 2 t_x$, there is a slight change in slope, discernible in Fig.~\ref{linearCollectiveConductivityPlotWithTilt}(f) and (g). 
        Note that when we consider the total conductivity, including single-particle contributions, we expect this effect to be small compared to single-particle contributions to the charge response coming from direct excitation of electron-hole pairs.
        \item The next distinguishable feature is when  a photon is able to excite an electron past the highest point of the first valence band.  
        Then, only the second valence band would be accessible to excited states. This occurs when
        \begin{equation*}
        \resizebox{0.9\hsize}{!}{$
        \hbar \omega = 2 \sqrt{\gamma^2 -4 \sqrt{((t_x)^2 +(t_y)^2 +(t_z)^2) \gamma^2} + 4 ((t_x)^2 +(t_y)^2 +(t_z)^2 + \Delta^2)}$}
        \end{equation*}
        for $m=0$.
        \item Finally, we expect a sharp feature when the frequency is high enough to excite electrons across the highest point of the second band.  
        Since there are no available states to excite to beyond here, this quenches the conductivity.  
        At $m=0$, the frequency at which this occurs is 
        \begin{equation*}
        \resizebox{0.9\hsize}{!}{$\hbar \omega = 2 \sqrt{\gamma^2 +4 \sqrt{((t_x)^2 +(t_y)^2 +(t_z)^2) \gamma^2} + 4 ((t_x)^2 +(t_y)^2 +(t_z)^2 + \Delta^2)}.
        $}
        \end{equation*}
\end{enumerate}

The results of Fig.~\ref{linearCollectiveConductivityPlotWithTilt}(a-h) may be compared to the results of the 1D cosine and quadratic dispersion models given in Appendix \ref{cosineSection} with Fig.~\ref{cosineFullPlot} and Fig.~\ref{quadradicPlots} respectively.
Especially at $m=0$, many of the qualitative features are similar.
With $m\neq 0$, the bands are non-degenerate at the edges of the Brillouin zone, causing additional spectral transitions for each new band in that frequency region.

Since the massless contribution to the collective conductivity diverges at low frequency, the physically relevant observable is the residue of the pole at $\omega=0$, which we can extract from $\lim_{\omega\rightarrow 0}-i\omega \sigma^{zz}_{\text{coll}}(\omega)$.  
This is a property of the massless mode propagator, $D_{++}(\omega)- e^{2 i \phi} D_{+-}(\omega) $, which peaks at $\omega \rightarrow 0$. 
We provide the details of this computation in Appendix \ref{zeroFreqIntercept}. 
{This intercept plays the role of the superfluid density in the Fr\"{o}hlich superfluid model of CDW conduction.\cite{tinkham2004introduction}} 
At $m = 2 t_x$, the intercept value is small compared to the $m \neq 2 t_x$ intercepts, but still is nonzero.  

In typical commensurate CDWs, the residual discrete translational symmetry pins the sliding mode, and so the massless propagator acquires a mass. 
As our four-band model comes from a regularization of an incommensurate CDW, we do not see this pinning effect here. 
We can trace this back to the simple $e^{i\phi}$ phase dependence in the mean field Hamiltonian. 
If we were instead to apply the shifted-zone scheme to our Hamiltonian for generic commensurate $\mathbf{Q}=2\pi M/N$, we would find that the energy gap varies as a function of $\phi$, containing harmonics of the form $\cos n\phi$ for $n<N$\cite{RLAChargeDensityWaves} [recall also our discussion preceding Eq.~(\ref{meanFieldGapHam})]. 
In all cases except $N=2$, this will lead to pinning of the CDW. For $N=2$, however, the phase mode and the amplitude mode become inexorably linked, since the mean field Hamiltonian contains only $\Delta\cos\phi = \Delta'$. 
In this case there is only one collective mode. 

Now, we will move on to examine the contributions of the collective mode to the second- and third-order conductivities. 
For numerical simplicity, we will focus primarily on the untilted case.

\subsection{Second-Order Conductivity}\label{subsec:second}

To examine the second-order collective conductivity, we will first review the calculation of the non-collective second-order conductivity. 
This will serve to introduce the diagrams that will appear in the calculation of the collective conductivity at both second- and third-order.

\subsubsection{Second-Order Non-collective Contributions ($\sigma^{zzz}_{\text{NC}}(\omega_{\beta \gamma}; \omega_{\beta}, \omega_{\gamma})$)}
\begin{figure}[h]
      \centering
\fbox{
\begin{minipage}{0.93\hsize}
\centering
\includegraphics[width=0.93\hsize]{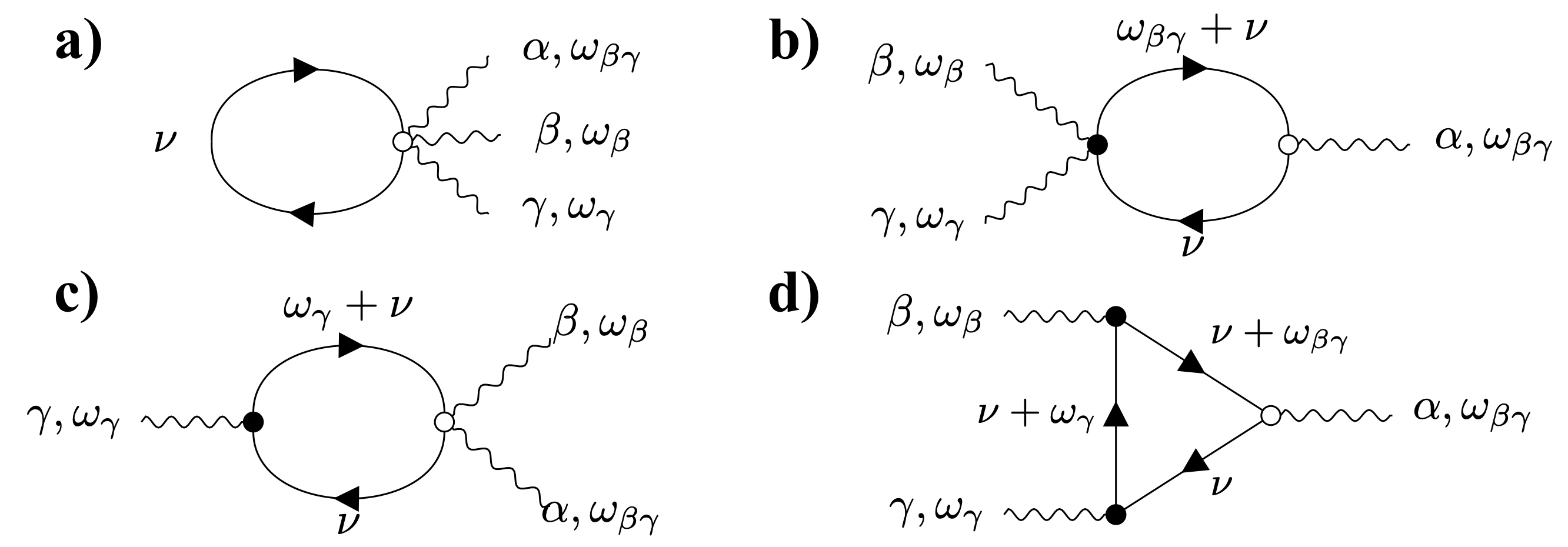}
\end{minipage}
}
\caption{Feynman diagrams for the non-collective second-order conductivity\cite{MooreDiagrammatic}.  Diagram d) is the only diagram that does not require regularization, since all the vertices are linear in the electric field.  
Diagrams a), b), and c) exist to regulate diagram d) as $\omega_{\gamma}$ and $ \omega_{\beta}$ hit various singularities.}
\label{SecondOrderNonCollectiveDiagrams}
\end{figure}
The Feynman diagrams contributing to the single-particle conductivity at second-order are provided in Fig.~\ref{SecondOrderNonCollectiveDiagrams}.  
These diagrams, in respective order, translate to
\begin{widetext}
\begin{equation}
\label{secondOrderNoncollective}
\resizebox{0.93\hsize}{!}{$
\begin{aligned}
    \sigma^{\alpha \beta \gamma}_{\text{NC}}(i\omega_{\beta \gamma}; i\omega_{\beta}, i\omega_{\gamma}) =& \frac{e (i e)^2}{\hbar^2 (i\omega_{\beta}) (i\omega_{\gamma})}\int [d k] \left [ \frac{1}{2} \left( G_{n_1 \xi_1, n_2 \xi_2}(i\nu, \mathbf{k}) h^{\gamma \beta \alpha}_{n_2 \xi_2, n_1 \xi_1} \right) \right.
    \\&
     \left. + \frac{1}{2}\left( h^{\gamma \beta}_{n_1 \xi_1, n_2 \xi_2}G_{n_2 \xi_2, n_3 \xi_3}(i\nu + i\omega_{\beta \gamma}, \mathbf{k}) h^{\alpha}_{n_3 \xi_3, n_4 \xi_4} G_{n_4 \xi_4, n_1 \xi_1}(i\nu, \mathbf{k}) \right) \right.
     \\&
     \left. +   \left( h^{\gamma}_{n_1 \xi_1, n_2 \xi_2}G_{n_2 \xi_2, n_3 \xi_3}(i\nu + i\omega_{\gamma}, \mathbf{k}) h^{\beta \alpha}_{n_3 \xi_3, n_4 \xi_4} G_{n_4 \xi_4, n_1 \xi_1}(i\nu, \mathbf{k}) \right) \right.
     \\&
     \left. +   \left( h^{\gamma}_{n_1 \xi_1, n_2 \xi_2}G_{n_2 \xi_2, n_3 \xi_3}(i\nu + i\omega_{\gamma}, \mathbf{k}) h^{\beta}_{n_3 \xi_3, n_4 \xi_4} G_{n_4 \xi_4, n_5 \xi_5}(i\nu+i\omega_{\beta \gamma }, \mathbf{k}) h^{\alpha}_{n_5 \xi_5, n_6 \xi_6} G_{n_6 \xi_6, n_1 \xi_1}(i\nu, \mathbf{k}) \right) \right]
     \\&
     + (\beta, i\omega_{\beta}) \longleftrightarrow (\gamma, i\omega_{\gamma}).
\end{aligned}
$}
\end{equation}
\end{widetext}
The last term symmetrizes over the two possible input frequencies, $i\omega_{\beta}$ and $i\omega_{\gamma}$.

\subsubsubsection{Ideal Weyl Model}
\label{secNCCondIdealModel}

As in the linear case from Section \ref{linCondCollIdealWeyl}, both the two-photon vertex and three-photon vertex are zero for the ideal Weyl model, without tilt.  
Thus, both $\overline{h}^{\alpha \beta}$ and $\overline{h}^{\alpha \beta \gamma}$ need to be fixed, and $\overline{h}^{\alpha \beta}$ needs to be in a gauge that respects the regularized solution in Section \ref{LinearNonCollectiveSection}.  
With regard to the previous process, $\overline{h}^{\alpha \beta \gamma}$ takes care of the singularity as $\omega_{\beta}=\omega_{\gamma} = 0$ since the integrand in Fig.~\ref{SecondOrderNonCollectiveDiagrams}a) is independent of frequency.  Furthermore, Fig.~\ref{SecondOrderNonCollectiveDiagrams}b) and c) reduce to the linear conductivity diagrams of Fig.~\ref{LinearNonCollectiveDiagrams} as either $ \omega_{\beta} \rightarrow0$ or $\omega_{\gamma} \rightarrow0$.  
Since $\overline{h}^{\alpha \beta \gamma}$ is not necessary for the collective conductivity, we will not compute its mathematical form.

Although a concrete calculation of the non-collective conductivities is outside the scope of this work, we remark here that for our ideal Weyl model, we have that inversion symmetry guarantees by integration parity that all terms in Eq.~\eqref{secondOrderNoncollective} are zero, and hence $\sigma_{NC}^{\alpha\beta\gamma}=0$. 
{This statement can be generalized: any Hamiltonian that preserves inversion symmetry necessarily has vanishing second-order non-collective conductivity.}

\subsubsection{Second-Order Collective Contributions ($\sigma^{zzz}_{\text{coll}}(\omega_{\beta \gamma}; \omega_{\beta}, \omega_{\gamma})$)}
\begin{figure}[ht]
      \centering
\fbox{
\begin{minipage}{0.93\hsize}
\centering
\includegraphics[width=0.93\hsize]{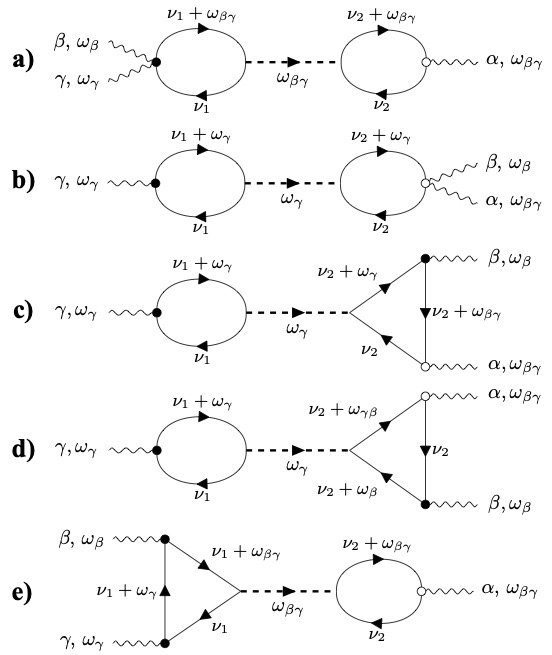}
\end{minipage}
}
\caption{Feynman diagrams for collective mode contributions to the second-order conductivity.  Each diagram has a single-photon vertex loop contribution.}
\label{SecondOrderCollectiveDiagrams}
\end{figure}
\subsubsubsection{Ideal and Lattice Weyl Model Without Tilt}

As mentioned in Section \ref{LinearCollectiveSection}, any loop containing only a single electric field line and a single phonon line sums to zero for the untilted lattice and ideal Weyl models.  
There are no diagrams to order $\mathcal{O}(g^2)$ that can be drawn where the loop integrates to a nonzero value in the second-order optical response. 
Thus, there is no collective mode contribution to the second-order conductivity in the untilted models.

\subsubsubsection{Lattice Weyl Model With Tilt}

Initially, the tilted model might seem promising for yielding a nonzero collective second-order response, given its nonzero contribution from the linear result.  
However, the second-order collective response is zero for this model as well.  
This time, the diagrams in Fig.~\ref{SecondOrderCollectiveDiagrams} show that each single photon vertex loop is accompanied by either a double photon vertex loop or a single photon vertex triangle.  
To codify this, consider the simplified expression:
\begin{equation}
\resizebox{1\hsize}{!}{$
\begin{aligned}
&\sigma^{\alpha \beta \gamma}_{\text{coll}} (i \omega_{\alpha \beta \gamma}; i \omega_{\beta}, i \omega_{\gamma}) =  \frac{e (i e)^2}{\hbar^2 (i \omega_{\beta})(i \omega_{\gamma})} \times
\\&
\left((G^{\text{double}})^{\beta \gamma}_{d_1}(\omega_{\beta \gamma}) + (G^{\text{tri}})^{\beta \gamma}_{d_1}(\omega_{\beta}, \omega_{\beta \gamma}) \right) D_{d_1 d_2}(\omega_{\beta \gamma}) (G^{\text{single}})_{d_2}^{\alpha} (\omega_{\beta \gamma})
\\& + 
(G^{\text{single}})^{ \gamma}_{d_1}(\omega_{ \gamma}) D_{d_1 d_2}(\omega_{\gamma}) \left( (G^{\text{double}})_{d_2}^{\alpha \beta } (\omega_{\beta \gamma}) + (G^{\text{tri}})_{d_2}^{ \alpha \beta } (\omega_{ \gamma}, \omega_{\beta \gamma})  \right),
\end{aligned}
$}
\end{equation}
where the superscript ``single", ``double", and ``tri" denote the single photon vertex loop, the double photon vertex loop, and the single photon vertex triangle as illustrated in Fig.~\ref{SecondOrderCollectiveDiagrams}. 
Here, recall that $d_{1, 2} = \pm$ denotes components of the $\pm \mathbf{Q} / 2$ subspace.

Notice that $(G^{\text{double}})^{\beta \gamma}_{1}(\omega_{\beta \gamma})  = (G^{\text{double}})^{\beta \gamma}_{2}(\omega_{\beta \gamma})$ and $ (G^{\text{tri}})^{\beta \gamma}_{1}(\omega_{\beta}, \omega_{\beta \gamma}) =  (G^{\text{tri}})^{\beta \gamma}_{2}(\omega_{\beta}, \omega_{\beta \gamma})$, meaning that the double photon vertex loop and single photon vertex triangle are symmetric under $d_1\leftrightarrow d_2$.  
On the other hand, the single photon vertex loop is antisymmetric, i.e. $(G^{\text{single}})^{ \gamma}_{1}(\omega_{ \gamma}) = -(G^{\text{single}})^{ \gamma}_{2}(\omega_{ \gamma})$.  
This parity argument stems from the parity of the velocity vertex, which comes from the tilt contribution in the Hamiltonian.  
Since momentum conservation is broken, this allows for exact cancellations between contributions from the conductivity in different valleys, even in the absence of inversion symmetry. 

In Section \ref{SectionLinearResponseWithTilt}, the linear response depended on the massless collective mode, which was the result of the single photon vertex loop being component-wise odd.  
In the next section, we will see that at third-order, the massive mode contributes to the collective conductivity when both left and right hand sides of the phonon propagator can be contracted against component-wise even diagrams.  
However, when the left and right hand diagrams have different component-wise parity, as in all the diagrams of Fig.~\ref{SecondOrderCollectiveDiagrams}, then the resulting contribution will always cancel when contracted against the collective phonon diagram.
Thus, we see that our single tilting term is not sufficient to allow for a collective contribution to the second-order conductivity. 
In Sec.~\ref{conclusion_section}, we remark on the effect of more general symmetry breaking terms.

\subsection{Third-Order Conductivity}\label{subsec:third}
\subsubsection{Third-Order Non-collective Contributions ($\sigma^{zzzz}_{\text{NC}}(\omega_{\beta \gamma \delta}; \omega_{\beta}, \omega_{\gamma}, \omega_{\delta})$)}
\begin{figure}[ht]
      \centering
\fbox{
\begin{minipage}{0.93\hsize}
\centering
\includegraphics[width=0.93\hsize]{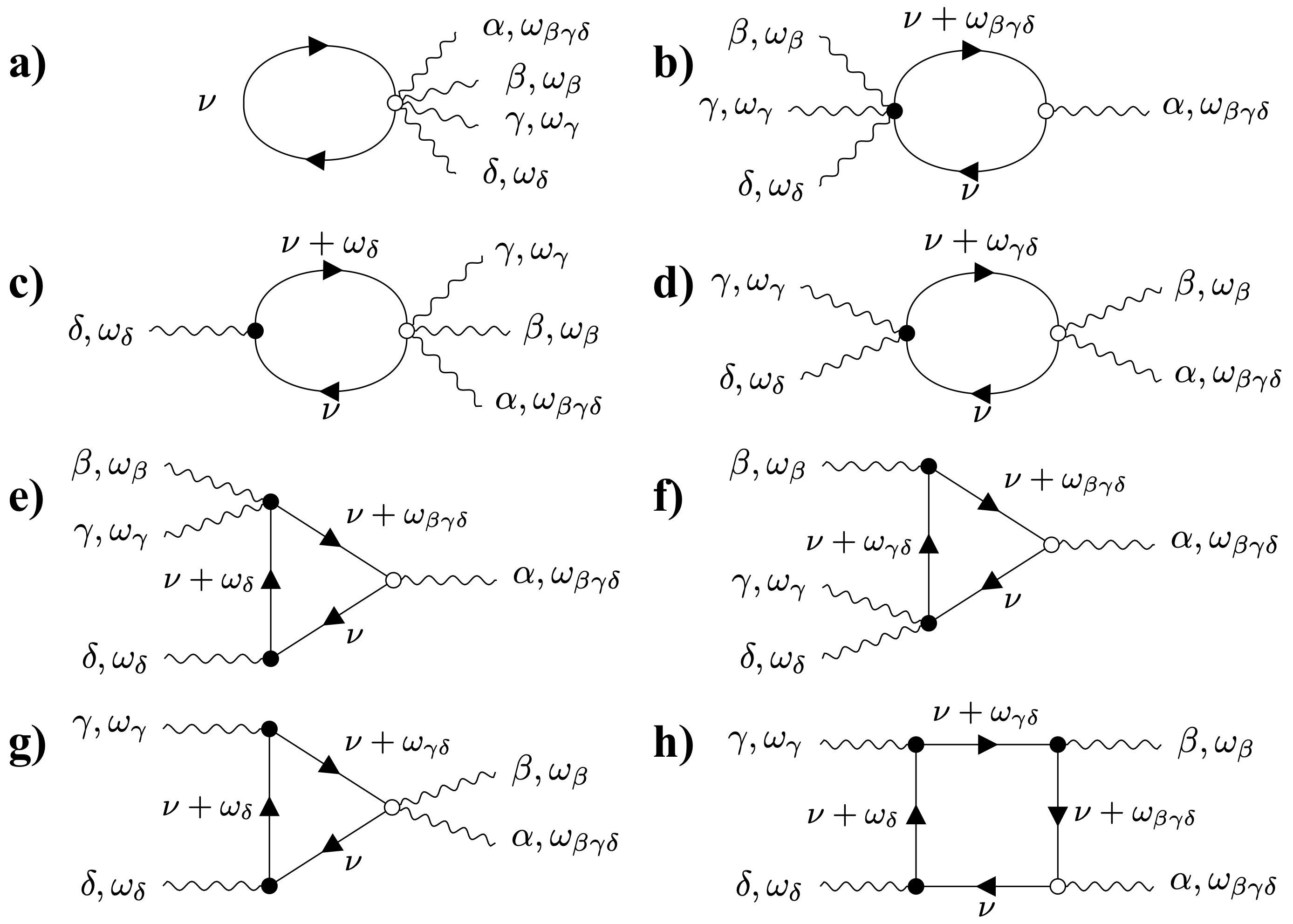}
\end{minipage}
}
\caption{Feynman diagrams for the non-collective contributions to the third-order conductivity\cite{MooreDiagrammatic}.  Diagram h) is the only diagram that does not require regularization.  Diagrams a), b), c), d), e), f), and g) exist to regulate diagram h) as $\omega_{\delta}$, $ \omega_{\gamma}$, and $ \omega_{\beta}$ hit various singularities.}.
\label{ThirdOrderNonCollectiveDiagrams}
\end{figure}
At third-order, we encounter both diagrams containing previously defined vertices, and a new four-photon vertex, $h^{\alpha \beta \gamma \delta}$.  With brevity in mind, the orbital and $\pm \mathbf{Q}/2$ indices will be left implicit in the following explicit expressions.  The set of diagrams from Fig.~\ref{ThirdOrderNonCollectiveDiagrams} gives a non-collective conductivity\cite{sipe_third_order_optical_response_2020} of
\begin{widetext}
\begin{equation}
\begin{split}
     \sigma^{\alpha \beta \gamma \delta}_{\text{NC}}(i\omega_{\beta \gamma \delta}; i\omega_{\beta}, i\omega_{\gamma}, i\omega_{\delta}) = & \frac{e (i e)^3}{\hbar^3 (i\omega_{\beta}) (i\omega_{\gamma})(i \omega_{\delta})} \int [d k] \left [ \frac{1}{6}\left( G(i\nu, \mathbf{k})  h^{\delta \gamma \beta \alpha} \right) + \frac{1}{6}\left( h^{\delta \gamma \beta}G(i\nu + i\omega_{\beta \gamma \delta}, \mathbf{k})  h^{\alpha} G(i\nu, \mathbf{k}) \right) \right.
     \\ &\left.
     + \frac{1}{2}\left( h^{\delta} G(i\nu + i\omega_{\delta}, \mathbf{k}) h^{\gamma \beta \alpha} G(i\nu, \mathbf{k}) \right) + \frac{1}{2}\left( h^{\delta \gamma} G(i\nu + i\omega_{\gamma \delta},\mathbf{k}) h^{\beta \alpha} G(i\nu, \mathbf{k}) \right) \right.
     \\ &\left.
     + \frac{1}{2}\left( h^{\delta} G(i\nu + i\omega_{\delta}, \mathbf{k}) h^{\gamma \beta} G(i\nu + i\omega_{\beta \gamma \delta}, \mathbf{k}) h^{\alpha} G(i\nu, \mathbf{k}) \right) \right.
     \\ &\left.
     + \frac{1}{2}\left( h^{\delta \gamma} G(i\nu + i\omega_{ \gamma \delta}, \mathbf{k}) h^{ \beta} G(i\nu + i\omega_{\beta \gamma \delta}, \mathbf{k}) h^{\alpha} G(i\nu, \mathbf{k}) \right) \right.
      \\ &\left.
     + \left( h^{\delta} G(i\nu + \omega_{ \delta}, \mathbf{k}) h^{ \gamma} G(i\nu + i\omega_{ \gamma \delta}, \mathbf{k}) h^{\beta \alpha} G(i\nu, \mathbf{k}) \right) \right.
     \\ &\left.
     + \left( h^{\delta} G(i\nu + i\omega_{ \delta}, \mathbf{k}) h^{ \gamma} G(i\nu + i\omega_{ \gamma \delta}, \mathbf{k}) h^{\beta} G(i\nu + i\omega_{\beta \gamma \delta}) h^{ \alpha} G(i\nu, \mathbf{k}) \right)
      \right] 
      \\&
      + \text{5 permutations of $(i\omega_{\beta}, \beta)$, $(i\omega_{\gamma}, \gamma)$, and $(i\omega_{\delta}, \delta)$}.
\end{split}
\end{equation}
\end{widetext}
\subsubsubsection{Ideal Weyl Model}

To regularize this expression, we may use the same bootstrapping methods from Sections \ref{linCondCollIdealWeyl} and \ref{secNCCondIdealModel} to find the regularized vertex $\overline{h}^{\alpha \beta \gamma \delta}$ for the idealized Weyl model.  The condition that must be obeyed is that this conductivity does not diverge as $\omega_i \rightarrow 0$ for all $i \in \{\beta, \gamma, \delta\}$.  However, since the collective conductivity is devoid of diagrams that would require $\overline{h}^{\alpha \beta \gamma \delta}$, we leave the detailed study of this expression for future work, and instead move on to examine the contributions of the collective mode to the third-order conductivity.

\subsubsection{Third-Order Collective Contributions ($\sigma^{zzzz}_{\text{coll}}(\omega_{\beta \gamma \delta} ; \omega_{\beta}, \omega_{\gamma}, \omega_{\delta})$)}
\begin{figure}[ht]
      \centering
\fbox{
\begin{minipage}{0.93\hsize}
\centering
\includegraphics[width=0.93\hsize]{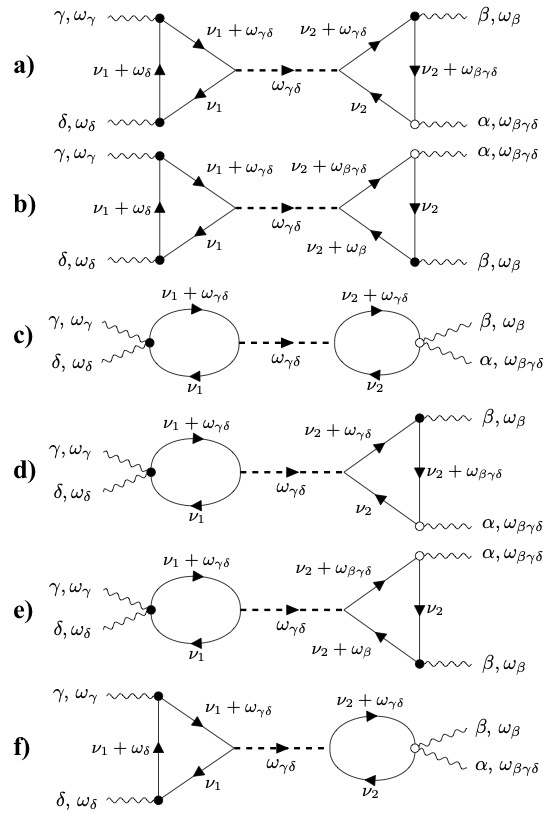}
\end{minipage}
}
\caption{Feynman diagrams for the non-vanishing contributions to the third-order collective conductivity in the untilted models.  The circular diagrams serve to regularize the singularities in the triangle diagrams.}
\label{ThirdOrderCollectiveDiagrams}
\end{figure}
This is the lowest order in the electric field where the collective conductivity is non-vanishing for the untilted models up to order $\mathcal{O}(g^2)$. 
The diagrams contributing to this response are illustrated in Fig.~\ref{ThirdOrderCollectiveDiagrams}.  
The main diagrams are the triangle diagrams Fig.~\ref{ThirdOrderCollectiveDiagrams}(a), (c) and (d), whereas the loop diagrams will regulate the triangle diagram.  
There is a parallel to this process in high energy physics, where the triangle diagram (see Fig.~\ref{ThirdOrderCollectiveDiagrams}a), c), and d)) is reminiscent of the bosonic contribution to the gluon-gluon scattering matrix arising via the Higgs mechanism\cite{HiggsGluonVasquez, HiggsGluonDelDuca, HiggsGluonDicus, HiggsGluonKniehl, HiggsGluonMaltoni, StreicherHigssBoson}.  
The electric fields take up the role of the gluons and the outgoing collective phonon takes up the role of the Higgs boson.

For the sake of convenience, the triangle and loop diagrams can be separately integrated.  
We start by defining
\begin{equation}
\resizebox{1\hsize}{!}{
$
\begin{aligned}
\label{TriangleTerms}
    \left( G^{\text{tri in}} \right)^{\delta \gamma}_{d_1}(i\omega_1, i\omega_{12}) =&
    \int [d k] h^{\delta}_{n_1 \xi_1, n_2 \xi_2} G(i\nu + i\omega_{1}, \mathbf{k})_{n_2 \xi_2, n_3 \xi_3} 
    \\
    & 
    \times h^{\gamma}_{n_3 \xi_3, n_4 \xi_4}  G(i\nu + i\omega_{12} , \mathbf{k})_{n_4 \xi_4, n_5 \xi_5}
     \\
    & 
    \times P_{n_5 \xi_5, n_6 \xi_6; d_1} G(i\nu, \mathbf{k})_{n_6 \xi_6, n_1 \xi_1},
    \\
   \left(G^{\text{ tri out}} \right)^{\beta \alpha}_{ d_2}(i\omega_{12}, i\omega_{123}) =&
   \int [d k] P_{d_2; n_7 \xi_7, n_8 \xi_8} G(i\nu + i\omega_{12}, \mathbf{k})_{n_8 \xi_8, n_9 \xi_9} 
     \\
    & 
    \times h^{\beta}_{n_9 \xi_9, n_{10} \xi_{10}} G(i\nu +i \omega_{123}, \mathbf{k})_{n_{10} \xi_{10}, n_{11} \xi_{11}}
    \\
    &
    \times h^{\alpha}_{n_{11} \xi_{11}, n_{12} \xi_{12}} G(i\nu , \mathbf{k})_{n_{12} \xi_{12}, n_{7} \xi_{y}}.
\end{aligned}$
}
\end{equation}
{Although there are two types of topologically distinct output triangles, shown in Fig.~\ref{ThirdOrderCollectiveDiagrams} (a), (b) as well as (d),(e), they are numerically equal after integrating over Matsubara frequencies.  
Therefore, we will denote the single output triangle as $G^{\text{ tri out}}$ and count it twice in the collective conductivity. 
Similarly, we define}
\begin{equation}
\resizebox{1\hsize}{!}{$
\begin{aligned}
\label{LoopTerms}
    \left( G^{\text{loop in}} \right)^{\delta \gamma}_{d_1}(i\omega_{12}) =& \int [d k] h^{\delta \gamma}_{n_1 \xi_1, n_2 \xi_2} G(i\nu + i\omega_{12}, \mathbf{k})_{n_2 \xi_2, n_3 \xi_3}
    \\&
    \times P_{n_3 \xi_3, n_4 \xi_4; d_1} G(i\nu , \mathbf{k})_{n_4 \xi_4, n_1 \xi_1},
    \\
    \left(G^{\text{loop out}} \right)^{\beta \alpha}_{ d_2}(i\omega_{12}) =& \frac{1}{2}\int [d k] P_{d_2; n_5 \xi_5, n_6 \xi_6} G(i\nu + i\omega_{12}, \mathbf{k})_{n_6 \xi_6, n_7 \xi_7} 
    \\&
    \times h^{\beta \alpha}_{n_7 \xi_7, n_{8} \xi_{8}} G(i\nu, \mathbf{k})_{n_{8} \xi_{8}, n_{5} \xi_{5}}.
\end{aligned}
$
 }
\end{equation}
The collective conductivity may be compactly written as
\begin{equation}
\resizebox{1\hsize}{!}{$
\begin{aligned}
    \sigma^{\alpha \beta \gamma \delta}_{\text{coll}}&(i\omega_{\alpha \beta \gamma \delta}; i\omega_{\beta}, i\omega{\gamma}, i\omega_{\delta}) =  \frac{e (i e)^3}{ \hbar^3 (i \omega_{\delta}) (i \omega_{\gamma}) (i \omega_{\beta})}
    \\&
    \times \left[ \left( G^{\text{tri in}} \right)^{\delta \gamma}_{d_1}(i\omega_{\delta}, i\omega_{\gamma \delta}) + \frac{1}{2}\left( G^{\text{loop in}} \right)^{\delta \gamma}_{d_1}( i\omega_{\gamma \delta}) \right]  D_{d_1 d_2}(i\omega_{\gamma \delta})
    \\&
    \times \left[ 2 \left( G^{\text{tri out}} \right)^{\beta \alpha}_{d_2}(i\omega_{\gamma \delta}, i\omega_{\beta \gamma \delta }) +  \left( G^{\text{loop out}} \right)^{\beta \alpha}_{d_2}(i\omega_{\gamma \delta}) \right] 
    \\& +  \text{5 permutations of $(i\omega_{\beta}, \beta)$, $(i\omega_{\gamma}, \gamma)$, and $(i\omega_{\delta}, \delta)$}.
\end{aligned}
$}
\label{CompactifiedCollectiveThirdOrderConductivity}
\end{equation}
For this calculation, only $\alpha = \beta = \gamma = \delta = z$ is non-zero.  
This makes sense since the CDW distortion and hence the collective mode is oriented along the $\hat{z}$ direction; this is mathematically guaranteed by the parity of the $\mathbf{k}$ integration.

\subsubsubsection{Ideal Weyl Model}

For the ideal model, we can compute the third-order collective conductivity analytically. 
For the diagrams defined in Eq.~(\ref{TriangleTerms}--\ref{LoopTerms}), and using Eq. ~(\ref{completeRegularizedHzz}) for regularization, we find
\begin{widetext}
\begin{equation}
\label{analyticLoops}
     \left( G^{\text{loop in}} \right)^{zz}( \omega_{12}) = \left[ \left( G^{\text{loop out}} \right)^{zz}( \omega_{12}) \right]^* = 16 t_z g \Delta \int [d \mathbf{k}]  \begin{bmatrix}
    e^{i \phi} \\
    e^{-i \phi}
    \end{bmatrix} \frac{ E^2_{\mathbf{k}} - (2 k_z)^2 }{E^{3}_{\mathbf{k}}(4 E^2_{\mathbf{k}} - \omega^2_{12})}
\end{equation}

\begin{equation}
\resizebox{1\hsize}{!}{$
\label{analyticTriIn}
\begin{aligned}
     \left( G^{\text{tri in}} \right)^{zz}(\omega_1, \omega_1 + \omega_2) =&  -8 t_z g \Delta \int [d \mathbf{k}] \begin{bmatrix}
    e^{i \phi} \\
    e^{-i \phi}
    \end{bmatrix}
    \\&
    \times \frac{ 16 (E^2_{\mathbf{k}} - 3 (2 k_z)^2)E^2_{\mathbf{k}} - 4 \omega^2_1(E^2_{\mathbf{k}} - (2 k_z)^2)  - \omega_1 \omega_2 (12 E^2_{\mathbf{k}} - 4 (2 k_z)^2_z - \omega^2_2) - 2 \omega^2_2 (2 E^2_{\mathbf{k}} - 2(2 k_z)^2 - \omega^2_1)}{ E_{\mathbf{k}} \left( 4 E^2_{\mathbf{k}} -\omega^2_1 \right) \left( 4 E^2_{\mathbf{k}} -\omega^2_2 \right) \left( 4 E^2_{\mathbf{k}} -\omega^2_{12} \right) } 
\end{aligned}
$}
\end{equation}

\begin{equation}
\label{analyticTriOut}
     \left( G^{\text{tri out}} \right)^{zz}(\omega_{12}, \omega_{123}) =\left[ \left( G^{\text{tri in}} \right)^{\delta \gamma}_{d_1}( \omega_{12}, \omega_3) \right]^*  - 8 t_z g  \int [d \mathbf{k}] \begin{bmatrix}
    e^{-i \phi} \\
    e^{i \phi}
    \end{bmatrix} \frac{\omega_3(2 \omega_{12} +\omega_3)(8 E^2_{\mathbf{k}} -\omega_{12} (\omega_{123}))}{E_{\mathbf{k}} \left( 4 E^2_{\mathbf{k}} -\omega^2_{12} \right) \left( 4 E^2_{\mathbf{k}} -\omega^2_3 \right) \left( 4 E^2_{\mathbf{k}} -\omega^2_{123} \right)},
\end{equation}
\end{widetext}
where the entries of the column vector correspond to $d=\pm 1$.
\begin{figure*}[t]
      \centering
\centering
\includegraphics[width=0.7\hsize]{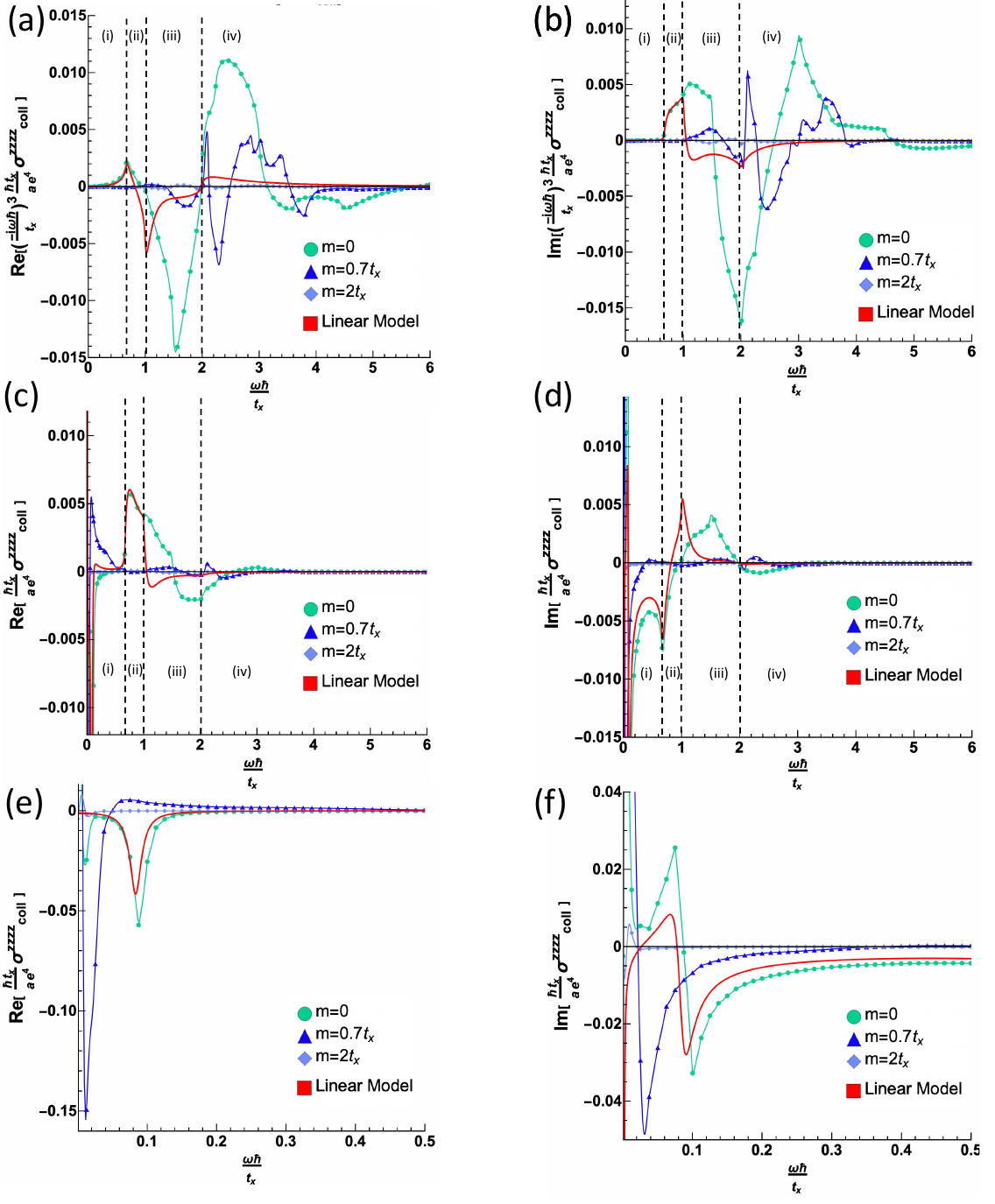}
\caption{Plots of the collective mode contribution to third harmonic generation.  The real and imaginary parts are labeled in (a, c, e) and (b, d, f) respectively.  Plots (a) and (b) are of $(-i \omega)^3 \sigma^{zzzz}_{\text{coll}}$, (c) and (d) are of $ \sigma^{zzzz}_{\text{coll}}$, and (e) and (f) are zoomed windows of (c) and (d).  The parameters $g=t_x$, $t_x = t_y = t_z$, and $\Delta = 0.5 t_x$ were used.  A cutoff of $\Lambda = \pi$ was also used in the ideal model, since this is the scale where the bands are approximately linear in the $m=0$ lattice model.  The ideal model conductivity is scaled up by a factor of four to illustrate the similarity with the $m=0$ case, which has four times as many Weyl nodes.  Region (i) indicates the region where three photons are not yet able to excite electrons across the gap.  Region (ii) marks where three photons can excite electrons across the gap, but two photons cannot.  Region (iii) indicates where two photons can excite electrons across the gap, but one photon cannot. Region (iv) is where one photon is sufficient to excite electrons across the gap.}
\label{harmonicFullPlot}
\end{figure*}

Notice, however, that the regularized Eq.~\eqref{analyticLoops}, along with Eqs.~\eqref{analyticTriIn} and \eqref{analyticTriOut}, does not force Eq.~\eqref{CompactifiedCollectiveThirdOrderConductivity} to converge as all frequencies are taken to zero.  
Although the choice of $\overline{h}^{\alpha \beta}$ works to regularize the non-collective first-order conductivity in the same frequency limit, the switching from electron-photon vertices to electron-phonon vertices changes the effect of this regularization.  
Even though $\overline{h}^{\alpha \beta}$ produces decent results away from $\omega_{\delta} = \omega_{\gamma} = \omega_{\beta} = 0$, a different choice for the two-photon vertex is necessary to get the third-order collective conductivity to converge at zero frequency.
More generally, as pointed out in Refs.~\cite{MooreDiagrammatic,passos2018nonlinear}, this is due to the fact that the minimal subtraction regularization using $\overline{h}^{\alpha\beta}$ requires the ordinary conductivity sum rule to be satisfied, but does not properly satisfy higher-order sum rules. 

One method to circumvent this and obtain convergent results for the third-order conductivity is to use an alternative scheme guaranteed to satisfy these higher-order sum rules (possibly at the expense of the linear conductivity sum rule). 
To this end, we can take Eqs.~\eqref{analyticTriIn} and \eqref{analyticTriOut} in Eq.~\eqref{CompactifiedCollectiveThirdOrderConductivity}, and solve for $\overline{h}^{\alpha \beta}$ as defined in Eqs.~\eqref{LoopTerms} in order to force the third-order conductivity to vanish as all frequencies are taken to zero.  
This generates a slightly different choice of regularization, denoted by $\left(\overline{h}_{\text{coll}} \right)^{\alpha \beta}$:
\begin{equation}
    \label{collectiveRegularizedHzz}
    \sum_{n, \xi} \text{sgn}(\xi) (-1)^n \left( \overline{h}_{\text{coll}} \right)^{zz}_{n \xi, n \xi} = \frac{8 t_z (E^2_{\mathbf{k}} -3 (2 k_z)^2)}{( k_z) E^2_{\mathbf{k}}}.
\end{equation}

An alternative method to regularize the nonlinear conductivity is to compute it using our lattice completion of the ideal linearized model. 
In the lattice model, both the first-order vertex $h^\alpha$ and the second-order vertex $h^{\alpha\beta}$ are modified at order $\mathcal{O}(k^2)$ and higher in order to ensure that the nonlinear conductivity is vanishing as all frequencies go to zero (as is appropriate for an insulator).
That is, the lattice model regularizes the ideal Weyl model in a way that guarantees all sum rules are satisfied. 
We will consider both the collective minimal subtraction and lattice regularization in what follows below.

We will next consider two experimentally relevant frequency regimes: harmonic generation and self-focusing.
We will show that the collective mode contributes to the third-order conductivity, yielding distinct features below the energy scale set by the single particle gap.

\subsubsubsection{Harmonic Generation: Comparing the Tiltless Ideal Case to the Lattice Model}
\label{harmonicSection}
Harmonic generation\cite{MooreDiagrammatic, SipeSecondOrderConductivity} is the case where $\omega_{\delta}=\omega_{\gamma} = \omega_{\beta} = \omega$.
The harmonic generation response results in a current at three times the incident photon frequency.  
The harmonic generation will show resonances for one-photon, two-photon, and three-photon processes.  
Additionally, the massive collective mode propagator will contribute a single resonance when $\omega$ is equal to the mass of the mode.  
Since the massless mode does not contribute to the lattice collective conductivity, $\lim_{\omega \rightarrow 0} d^n/d\omega^n[\omega^3\sigma^{zzz}_{\text{coll}}(3 \omega; \omega, \omega, \omega)]$, must be zero for our insulating system for $0\le n \le3$.  
This ensures the collective conductivity is zero in the $\omega\rightarrow 0$ limit. 
Although the numerical treatment for generating Fig.~\ref{harmonicFullPlot} appears to obscure this fact, a more complete analysis is provided in Appendix \ref{collectiveInterceptVsSelfEnergy}.

We can numerically compute the third-order conductivity $\sigma^{zzz}_{\text{coll}}(3 \omega; \omega, \omega, \omega)$ from the full lattice model Eq.~\eqref{FullLatticeModel} and compare this with our analytical results for the ideal model from Eq.~\eqref{analyticLoops}, \eqref{analyticTriIn}, \eqref{analyticTriOut}.  
However, in the lattice model, when $m=0$, four additional Weyl points exist along the edges of the Brillouin zone.  
Therefore, compared with the ideal model, we expect the results in the $m=0$ lattice model to be approximately four times as large.  
Additionally, a cutoff momentum of  $\Lambda = \pi$ is chosen, as this is approximately where quadratic corrections to the lattice model dispersion become important. 
We show our results in Fig.~\ref{harmonicFullPlot}. There are several distinct processes to explain this plot:  
\begin{enumerate}

	\item Since the massive mode propagator diverges at a resonant frequency $\omega_{\text{res}}$, there exists a peak in the conductivity at $2 \omega = \omega_{\text{res}}$, which is most evident in ideal and $m=0$ models in Figs.~\ref{harmonicFullPlot} (e) and (f).

    \item When  $3 \omega = 2 (2 \Delta)$, three photons can excite an electron across the gap between the valence and conduction bands.  
    This process is only allowed in the triangle diagrams and constitutes two-photon interactions converting into a phonon before a third photon adds enough energy to cross the gap (when the diagrams are read left-to-right).  
    This is most evident in Fig.~\ref{harmonicFullPlot}(c) and Fig.~\ref{harmonicFullPlot}(d), where the real and imaginary parts of the conductivity have extrema at the boundary between regions (i) and (ii).
    
    \item At $2 \omega = 2 ( 2 \Delta)$, two photons are able to excite an electron across the gap. 
    We again see sharp features in the real and imaginary parts of the conductivity in Fig.~\ref{harmonicFullPlot}(a, c) and (b, d), this time at the boundary between regions (ii) and (iii). Above this frequency, both the two-photon loops and the two-photon triangle processes contribute to the conductivity.
     
    \item A single photon is able to excite an electron-hole pair when $\omega = 2 (2 \Delta)$.  
    The real part of [$(-i\omega)^3$ times] the conductivity is nearly zero and the imaginary part has a minimum in Fig.~\ref{harmonicFullPlot}(a) and (b) respectively, which we see at the boundary between regions (iii) and (iv) in Fig.~\ref{harmonicFullPlot}.
     
    \item Three photons are able to excite electrons across the gap at the (shifted) Brillouin zone boundary when $3 \omega = 4| \sqrt{(t_z)^2 + (\Delta)^2} \pm m |$.  
    {This transition is illustrated by the green arrows in Fig.~\ref{UntiltedLatticeSpectrumPlot}(e-g).}  
    The real part of [$(-i\omega)^3$ times] the conductivity has a minimum, and the imaginary part transitions between two peaks, such as in Fig.~\ref{harmonicFullPlot}(a) and (b). 
    For example, this occurs near $\omega \approx 1.49 t_z$ for the $m = 0$ plot.
     
     \item Two-photon processes are able to excite electrons across the gap at the zone boundary when  $2 \omega = 4 |  \sqrt{(t_z)^2 + (\Delta)^2} \pm m |$.  
     This is demarcated by a change in slope in the real part of [$(-i\omega)^3$ times] the conductivity, as observed in Fig.~\ref{harmonicFullPlot}(a)  at $m=0$, which occurs near $\omega \approx 2.23 t_x$.
     
     \item One-photon processes are able to excite electrons across the gap at the zone boundary when  $\omega = 4|\sqrt{(t_z)^2 + (\Delta)^2} \pm m|$.  
     A smaller local minimum characterizes this in plots of the real part.  
     An example of this transition at $m=0$ in Fig.~\ref{harmonicFullPlot}(a) occurs at $\omega \approx 4.47 t_x$.
     
     \item Points 2, 3, and 4 also generalize to the case when when $m \neq 0$. 
     However, for nonzero $m$ the gaps at the Brillouin zone boundary are given by $2 (m \pm \Delta)$.  
     Hence, interband transitions become active when $n\omega=4(m \pm \Delta)$ for $n=1,2,3$.  
     For example, there is a clear local minimum in Fig.~\ref{harmonicFullPlot}(b) at $\omega = 2.4 t_x$ for $m = 0.7 t_x$. This corresponds to the resonant frequency of the two-photon transition [i.e. $2 \omega = 4 (m + \Delta)$].
\end{enumerate}

We expect that the ideal Weyl model will yield a similar conductivity to the lattice model, provided we focus on frequency regimes where the Brillouin zone boundaries can be neglected.  
The ideal and lattice conductivities becomes dissimilar once transitions at the Brillouin Zone edge are involved or multi-band transitions become appreciable.
Additionally, once the parameter $m$ becomes comparable to $t_x$, quadratic terms in the dispersion start becoming important, making $m=0$ the best fit for the ideal model.  
In particular, responses with $\omega\lesssim \Delta/3$, which are dominated by the effects of the collective mode, are qualitatively insensitive to the details of the lattice regularization.
As we expect no single-particle response at these subgap frequencies (consistent with a band insulator), this means that the low frequency harmonic generation response will be dominated by the effect of the massive collective mode illustrated in Fig.~\ref{harmonicFullPlot}  (e) and (f). 
Fig.~\ref{harmonicFullPlot} corroborates these statements.

\subsubsubsection{Self-Focusing: Comparing the Tiltless Ideal Case to the Lattice Model}
\begin{figure*}[t]
      \centering
\centering
\includegraphics[width=0.7\hsize]{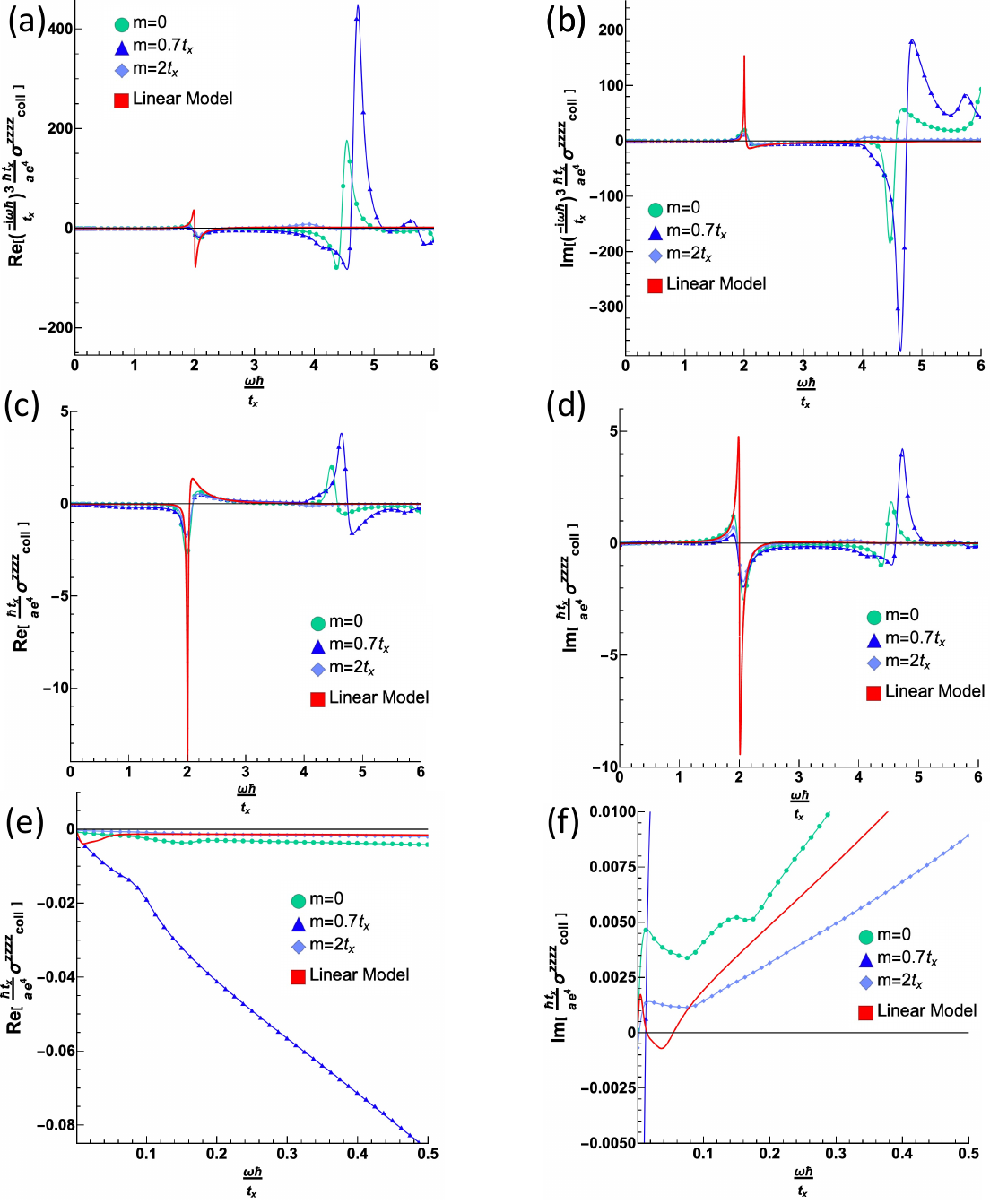}
\caption{Plots of the collective mode contribution to the self-focusing conductivity.  The real and imaginary parts are labeled as (a, c, e) and (b, d, f) respectively.  Plots (a) and (b) are showing $(-i\omega)^3 \sigma^{zzzz}_{\text{coll}}$, plots (c) and (d) are of $ \sigma^{zzzz}_{\text{coll}}$, and plots (e) and (f) capture a zoomed window of (c) and (d).  The parameters $g=t_x$, $t_x = t_y = t_z$, and $\Delta = 0.5 t_x$ were used.  The cutoff for the ideal model was taken at $\Lambda=\pi$ to provide adequate agreement with the lattice model.  The linear model was multiplied by a factor of 1/2 better fit the comparison with that of the lattice model.  The higher energy contributions become dominate for $g$ sufficiently small.}
\label{selfFocusFullPlot}
\end{figure*}

Self-focusing refers to the situation where the magnitudes of the frequencies are all the same, but two of the three frequencies are opposite in sign \cite{MooreDiagrammatic}.  
These opposite frequencies exactly cancel yielding a current with the same frequency as the applied electric field.  
This results in an effective correction to the linear conductivity, since contracting electric fields with opposite frequencies,  $\sigma^{\alpha \beta \gamma \delta}(\omega) E_{\gamma}(\omega) E_{\delta}(-\omega) $, gives an effective response to the third uncontracted electric field.  
As in the harmonic case, $\lim_{\omega \rightarrow 0}\omega^3 \sigma^{zzz}_{\text{coll}}(\omega; -\omega, \omega, \omega) = 0$, and this also holds true for the first three derivatives of this quantity.  
More information on the zero frequency limit can be found in Appendix \ref{zeroFreqIntercept}.

For the collective mode contribution to the conductivity, we will then find contributions from the collective mode at zero frequency, and at frequency $2  \omega$.  
The propagator $D_{++}(0) + e^{2 i \phi}D_{+-}(0)$ for the amplitude mode at $\omega=0$ enters into the diagrammatic calculation when two photons of opposite frequency excite a virtual electron-hole pair, which then recombines to excite a virtual massive collective excitation (see fig.~\ref{ThirdOrderCollectiveDiagrams}). 
Since the massive mode propagator is nonzero at $\omega=0$, it gives a frequency-independent contribution to the conductivity, entering as an overall constant scale factor. 

In contrast, $D_{++}(2  \omega) + e^{2 i \phi}D_{+-}(2  \omega)$ appears in the calculation when the electric fields excite a virtual electron-hole pair at the doubled frequency $2\omega$.  
This enters into the collective conductivity as a frequency-dependent scale factor, which we expect to have a large effect when $2\omega\approx \omega_{\text{res}}$, the resonant frequency of the massive mode.

Similar to the harmonic case, we can explain the behavior of the self-focus conductivity in terms of (virtual) electron-photon processes.  
Several excitation mechanisms are identical to the harmonic processes, such as the two-photon loop with the same frequencies, so only the new features from triangle and loop diagrams are itemized below.
\begin{enumerate}
    \item For two-photon processes where the incident photons have opposite frequencies, the zero-frequency amplitude mode propagator enters the calculation. 
    This provides a constant rescaling of the collective conductivity. 
    For small values of the coupling constant $g$ (i.e. for small values of the resonant frequency $\omega_{\text{res}}$ of the collective mode), this contribution can be quite large.
  This accounts for the large features in Fig.~\ref{selfFocusFullPlot} for large $\omega$ when compared to the response near small $\omega$.
  
   \item Similar to the harmonic case, there exists a resonance in the self-focusing conductivity $2 \omega = \omega_{\text{res}}$ due to the divergence of the massive mode propagator. 
   This is most evident in  Fig.~\ref{selfFocusFullPlot}(e-f). 
   {The massive phonon resonant peaks are smoothed out in comparison to the harmonic case.  
   This is due to symmetrically summing over frequency arguments, which mixes the resonant massive term with $D_{++}(0) + e^{2 i \phi} D_{+-}(0)$, thereby obscuring the resonant peak.  }
    
    \item   The first non-trivial resonance in the self-focusing conductivity comes when one photon is able to excite an electron across the gap with $\omega = 2 (2 \Delta)$.  
    The peaks in Fig.~\ref{selfFocusFullPlot}, which occur at $\omega = 2 t_x$, reflect this. 
    
    \item Similar to item 3, resonant excitation of an electron-hole pair by a single photon with $ \omega = 4| \sqrt{(t_z)^2 + (\Delta)^2} -m |$ is possible at the Brillouin zone edge.  
    This is apparent at $m = 0$ in Fig.~\ref{selfFocusFullPlot}(b, d) where the imaginary part reaches a local minimum near $\omega \approx 4.47 t_x$.
    
\end{enumerate}

The self-focusing conductivity in the ideal model from Fig.~\ref{selfFocusFullPlot} deviates from the lattice model { at larger $\omega$.
This is most notable in comparison of the ideal model with the} $m=0$ model, which agrees reasonably well for the harmonic conductivity in Fig.~\ref{harmonicFullPlot}.  
This discrepancy arises because the idealized linear approximation is valid for low energies, where the electronic states involved are close to the band minimum.  
However, at high frequencies, the lattice model is sensitive to the presence of states at the Brillouin zone boundary, which are not found in the simplistic idealized Weyl model.  
This amplifies features in the high-frequency self-focusing conductivity in the lattice model due to the zero frequency collective mass term, $D_{++}(0) + e^{2 i \phi} D_{+-}(0)$, which is large in the case of $g$ being small (see Appendix~\ref{massiveResonanceSection} for a discussion of the explicit form of the massive mode propagator).  
If $g$ were larger, then higher order perturbations in $g$ would become important to this calculation.  
Hence, for $g$ small enough to require only a single collective mode propagator in the Feynman diagram, the decaying tail of the massive mode propagator will hit the zero frequency intercept at a value inversely proportional to $g$.

\section{Conclusion}
\label{conclusion_section}
\begin{figure}[ht]
      \centering
\begin{minipage}{0.93\hsize}
\centering
\includegraphics[width=0.93\hsize]{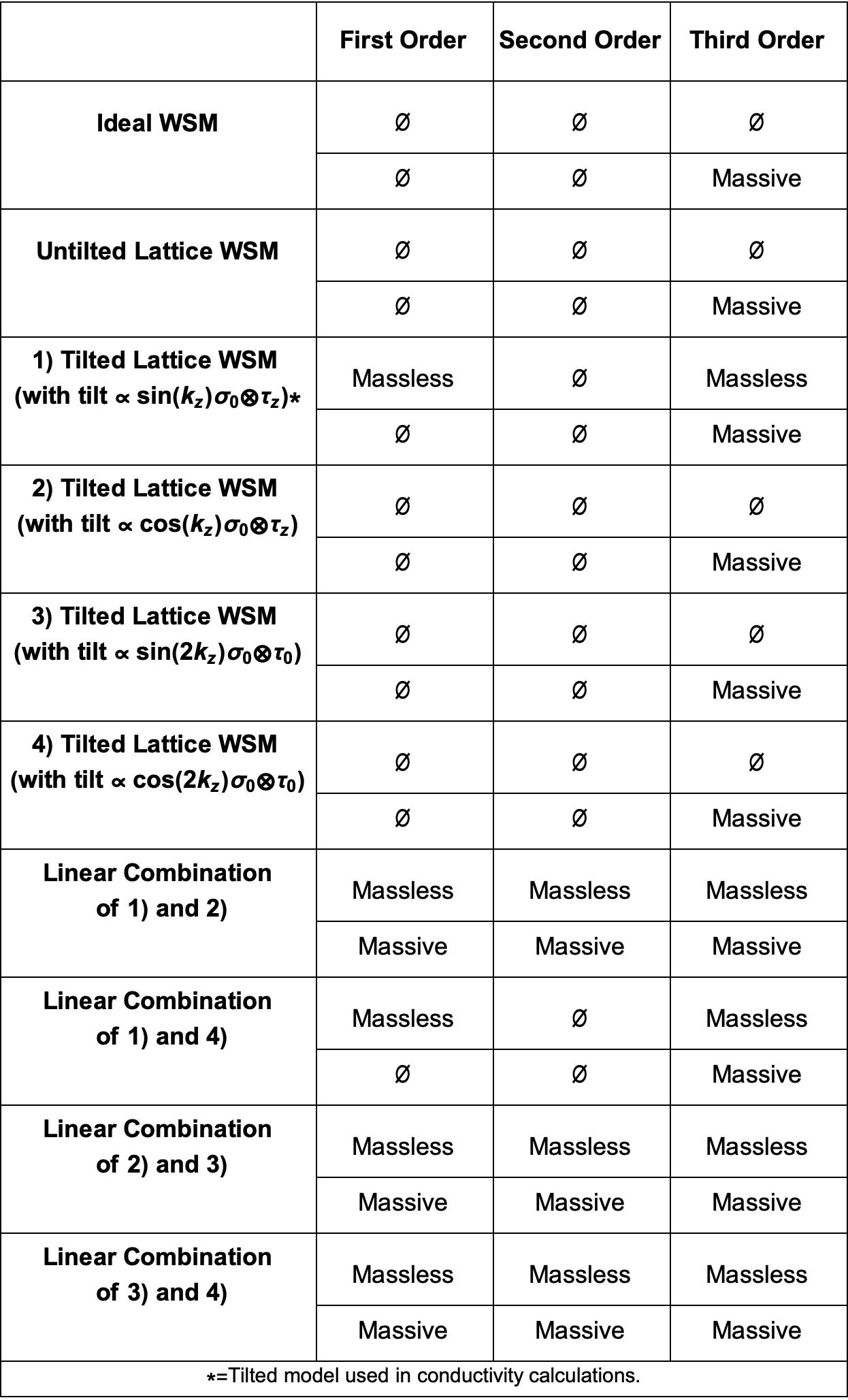}
\end{minipage}
\caption{This table shows how adjusting the Pauli matrix structure and parity of the tilted Weyl Hamiltonian affects the massive and massless collective mode contribution to the conductivity.}
\label{grid}
\end{figure}
In this work, we have examined the nonlinear optical conductivity of both the untilted and tilted Weyl lattice CDW, as well as the ideal linearized Weyl CDW.  
For the linear optical conductivity, we showed the phase mode does not contribute to the current in an untilted model, since the Fermi surface nesting in a multi-band model can couple electrons and holes with the same velocity. 
Furthermore, the amplitude mode also does not contribute to the conductivity due to the parity of the momentum integrand in the perturbative expansion; this is true for both the ideal and lattice Weyl model. 
To recover a DC conductivity from the phase mode, as seen in a single band CDW, we must introduce a tilt in the Weyl system, which allows the CDW to excite electron-hole pairs with nonzero net velocity. 
At higher frequency, we found that the tilted model yields a linear collective conductivity with features marking when the frequency of the electric field is large enough to excite electrons across the single-particle gap as well as features marking when the frequency approaches separation between the band maxima and minima.

Next, for the untilted lattice and linearized models, we showed that the first nontrivial collective contribution to the  conductivity are at the third-order in the electric field, and only the amplitude mode contributes at this order.  
For harmonic generation ($\omega_{\alpha} = \omega_{\beta} = \omega_{\gamma}$), we saw that there is a distinct peak in the optical response near the resonant frequency of the amplitude mode.
We have shown that this feature is independent of how we regularize our low energy model at large momentum, and thus should be a universal feature in Weyl-CDW systems.  
Additionally, when the frequency is such that one, two, or three electrons can be excited across the gap, there is a peak in either the real or imaginary conductivity.  
Other transitions occur when these one, two, or three electrons are able to cross the bandgap at the Brillouin zone boundary.  
When the bands are nondegenerate away the Weyl point, multiple transitions are possible.

We also looked at the third-order self-focusing contribution to the conductivity, ($\omega_{\alpha} = \omega_{\gamma} = -\omega_{\beta}$), which can be viewed as a field-dependent correction to the linear conductivity.  
Two of the incoming frequencies cancel, so this situation is effectively a single photon process.  
Thus, the self-focusing case does not show resonances associated with two or three electron processes.
This contrasts with the harmonic collective conductivity, which exhibits peaks for these multi-particle processes.  
Additionally, because the amplitude mode propagator is finite as $\omega\rightarrow 0$, it provides a background that enhances the peaks in the conductivity at larger frequencies.  
Since the ideal Weyl model is primarily accurate for small frequencies, we see the collective conductivity in the ideal model deviates from the conductivity in the lattice model once states at the Brillouin zone boundary become relevant.

Our results highlight the sensitivity of the collective mode conductivity to symmetries of the Weyl system. 
In particular, we saw that breaking inversion and particle-hole symmetry with the tilting term Eq.~(\ref{tiltedLatticeTerm}) was necessary for the massless mode to contribute to charge transport at linear order. 
We did not, however, endeavor in this work to give a systematic exploration of the effects of symmetry-breaking perturbations on collective mode charge transport; we leave this as a task for future research. 
A first step in this direction is given in Fig.~\ref{grid}, where we summarize the affect of alternative tilting terms on the contributions to the collective mode conductivity. 
Note that we see from the figure that breaking inversion symmetry alone is not sufficient to generate a nonzero second-order collective response, as the correct Pauli matrix structure must also be chosen.  
Linear combinations of different tilt structures may also produce nonvanishing contributions, illustrating the balance between parity and matrix structure.

Our work points out several experimentally relevant features in the collective response of Weyl-CDW systems. 
Going forward, our results also open up several avenues for future research. 
First, our method allows for an extension beyond minimal models of topological CDW materials. 
Extensions to time-reversal invariant systems with multiple Weyl points will allow for a direct application of transport in (TaSe$_4$)$_2$I. 
A similar analysis could be undertaken for spin-density waves and superconducting Weyl systems system. 
This could allow for an exploration of transport signatures of competing phase transitions in these materials\cite{mu_CDW_SC_phase}. 
Recall also that we restricted our analysis here to the regime of small electron-phonon coupling $g$. 
If $g$ is large enough, additional diagrams at higher orders in $g$ may also be included.  
Additionally, other aspects of renormalization may produce interesting effects.  
Examples of such effects include renormalization of the phonon-electron and photon-electron vertices\cite{TakadaIncommensurateCDW}, and an exact analysis of the self-energy.

\begin{acknowledgments}
The authors thank P. Abbamonte, B. Skinner, and F. Mahmood for fruitful discussions. 
The analytical and numerical work of B.B. and R.C.M. was supported by the U.S. DOE, Office of Basic Energy Sciences, Energy Frontier Research Center for Quantum Sensing and Quantum Materials through Grant No. DE-SC0021238. 
B.B. received additional support from the Alfred P. Sloan Foundation, and the National Science Foundation under grant DMR-1945058. 
This work made use of the Illinois Campus Cluster, a computing resource that is operated by the Illinois Campus Cluster Program (ICCP) in conjunction with the National Center for Supercomputing Applications (NCSA) and which is supported by funds from the University of Illinois at Urbana-Champaign.
\end{acknowledgments}

\onecolumngrid
\begin{appendices}
\section{Lattice Model in the Shifted Brillouin Zone}
\label{shiftedZoneSchemeDerivation}
In this section, we review our procedure for redefining the Brillouin zone to shift the Weyl points to the origin analogous to LRA \cite{RLAChargeDensityWaves, Rice_Supplemental_notes_on_orignal_RLA, Rice_Supplemental_notes_on_orignal_RLA, RiceandCrossCDW, FentonCDW}.  
This is a convenient tactic for treating systems with broken momentum conservation.
We start by reorganizing terms in the Fourier expansion of our electron creation operators $c_{\mathbf{R}}$.
\begin{equation}
    c_{\mathbf{R}} = \sum^{\pi, \pi, \pi}_{\mathbf{k}=-\pi, -\pi, -\pi} e^{i \mathbf{k} \cdot \mathbf{R}} c_{\mathbf{k}}= \left(\sum^{\pi, \pi, 0}_{\mathbf{k}=-\pi, -\pi, -\pi}  + \sum^{\pi, \pi, \pi}_{\mathbf{k}=-\pi, -\pi, 0} \right) e^{i \mathbf{k} \cdot \mathbf{R}} c_{\mathbf{k}}.
\end{equation}
From henceforth, the sum over $k_x$ and $k_y$ will be implied, since the only sum being manipulated is $k_z$.  
We also suppress the normalization constant $N$ of our Fourier sums, since all factors of $N$ cancel at the end of the calculation.  
Applying two opposing shifts to each sum by letting $k_z^{\prime} = k_z+ \pi/2$ and $k_z^{\prime \prime} = k_z - \pi/2$, we have
\begin{equation}
\label{equation_for_shifted_conversion}
    c_{\mathbf{R}} = \sum^{ \pi/2}_{k'_z = -\pi / 2}   e^{i (\mathbf{k}^{\prime} - \pi/2 \hat{z} ) \cdot \mathbf{R}} c_{\mathbf{k'} - \pi / 2 \hat{z}} + \sum^{ \pi/2}_{k^{\prime\prime}_z=-\pi / 2}   e^{i (\mathbf{k}^{\prime \prime} + \pi/2 \hat{z} ) \cdot \mathbf{R}} c_{\mathbf{k}^{\prime \prime} + \pi / 2 \hat{z}}.
\end{equation}
Now let $\mathbf{k}^{\prime} = \mathbf{k}$ and $\mathbf{k}^{\prime \prime} = \mathbf{k}$ as a change of dummy indices.  
Consider the interaction term: $ H_2 =2 \sum_{\mathbf{R}} \cos(Q R_z + \phi) c^{\dagger}_{\mathbf{R}}\sigma_z c_{\mathbf{R}}$ at $\mathbf{Q} = \pi$.  
Applying Eq.~\eqref{equation_for_shifted_conversion} yields\cite{RLAChargeDensityWaves, WangAndZhangAxionic}
\begin{equation}
\label{angleManipulationEquation}
\begin{split}
    &\frac{1}{2} \sum_{\mathbf{R}}  \left(e^{i (R_z \pi + \phi )} + e^{-i (R_z \pi + \phi )} \right)  c^{\dagger}_{\mathbf{R}}  c_{\mathbf{R}} 
    \\&= \sum_{\mathbf{R}} \sum_{\mathbf{k}, \mathbf{k}^{\prime \prime}} \frac{1}{2} \left(e^{i (R_z \pi + \phi )} + e^{-i (R_z \pi + \phi )} \right)
     \left( e^{-i (\mathbf{k} - \pi/2 \hat{z} ) \cdot \mathbf{R}} c^{\dagger}_{\mathbf{k} - \pi / 2 \hat{z}} +  e^{-i (\mathbf{k} + \pi/2 \hat{z} ) \cdot \mathbf{R}} c^{\dagger}_{\mathbf{k} + \pi / 2 \hat{z}} \right) \\& \qquad\qquad\qquad \qquad \qquad \qquad \qquad \qquad \times \left( e^{i (\mathbf{k}^{\prime} - \pi/2 \hat{z} ) \cdot \mathbf{R}} c_{\mathbf{k}^{\prime} - \pi / 2 \hat{z}} +  e^{i (\mathbf{k}^{\prime } + \pi/2 \hat{z} ) \cdot \mathbf{R}} c_{\mathbf{k}^{\prime } + \pi / 2 \hat{z}} \right)
     \\&= \frac{1}{2} \sum_{\mathbf{R}} \sum_{\mathbf{k}, \mathbf{k}^{\prime \prime}}  \left(e^{i (R_z \pi + \phi )} + e^{-i (R_z \pi + \phi )} \right)
     \left( e^{-i (\mathbf{k} - \pi/2 \hat{z} ) \cdot \mathbf{R}} c^{\dagger}_{\mathbf{k} - \pi / 2 \hat{z}} +  e^{-i (\mathbf{k} + \pi/2 \hat{z} ) \cdot \mathbf{R}} c^{\dagger}_{\mathbf{k} + \pi / 2 \hat{z}} \right) \\& \qquad\qquad\qquad \qquad \qquad \qquad \qquad \qquad \times \left( e^{i (\mathbf{k}^{\prime} - \pi/2 \hat{z} ) \cdot \mathbf{R}} c_{\mathbf{k}^{\prime} - \pi / 2 \hat{z}} +  e^{i (\mathbf{k}^{\prime } + \pi/2 \hat{z} ) \cdot \mathbf{R}} c_{\mathbf{k}^{\prime } + \pi / 2 \hat{z}} \right)
     \\&= \frac{1}{2}\sum_{\mathbf{R}} \sum_{\mathbf{k}, \mathbf{k}^{\prime \prime}}  \left(e^{i (R_z \pi + \phi )} + e^{-i (R_z \pi + \phi )} \right)
     \left( e^{i (-\mathbf{k} + \mathbf{k}^{\prime}) \cdot \mathbf{R}} c^{\dagger}_{\mathbf{k} - \pi / 2 \hat{z}} c_{\mathbf{k}^{\prime} - \pi /2 \hat{z}} +   e^{i (-\mathbf{k} + \mathbf{k}^{\prime} + \pi \hat{z}) \cdot \mathbf{R}} c^{\dagger}_{\mathbf{k} - \pi / 2 \hat{z}} c_{\mathbf{k}^{\prime} + \pi /2 \hat{z}} \right.
     \\&
     \left.
     \qquad \qquad \qquad\qquad\qquad\qquad\qquad\qquad+ e^{i (-\mathbf{k} + \mathbf{k}^{\prime}) \cdot \mathbf{R}} c^{\dagger}_{\mathbf{k} + \pi / 2 \hat{z}} c_{\mathbf{k}^{\prime} + \pi /2 \hat{z}}  + e^{i (-\mathbf{k} + \mathbf{k}^{\prime} - \pi \hat{z}) \cdot \mathbf{R}} c^{\dagger}_{\mathbf{k} + \pi / 2 \hat{z}} c_{\mathbf{k}^{\prime} - \pi /2 \hat{z}}\right)
      \\&=  \frac{1}{2}\sum_{\mathbf{k}, \mathbf{k}^{\prime \prime}}
     \left(\left( e^{i \phi} \delta_{\mathbf{k}, \mathbf{k}^{\prime} + \pi \hat{z}} + e^{-i \phi} \delta_{\mathbf{k}, \mathbf{k}^{\prime} - \pi \hat{z}} \right) c^{\dagger}_{\mathbf{k} - \pi / 2 \hat{z}} c_{\mathbf{k}^{\prime} - \pi /2 \hat{z}} +   \left( e^{i \phi} \delta_{\mathbf{k}, \mathbf{k}^{\prime} + 2 \pi \hat{z}} + e^{-i \phi} \delta_{\mathbf{k}, \mathbf{k}^{\prime}} \right) c^{\dagger}_{\mathbf{k} - \pi / 2 \hat{z}} c_{\mathbf{k}^{\prime} + \pi /2 \hat{z}} \right.
     \\&
     \left.
     \qquad \qquad \qquad+ \left( e^{i \phi} \delta_{\mathbf{k}, \mathbf{k}^{\prime} + \pi \hat{z}} + e^{-i \phi} \delta_{\mathbf{k}, \mathbf{k}^{\prime} - \pi \hat{z}} \right) c^{\dagger}_{\mathbf{k} + \pi / 2 \hat{z}} c_{\mathbf{k}^{\prime} + \pi /2 \hat{z}}  + \left( e^{i \phi} \delta_{\mathbf{k}, \mathbf{k}^{\prime} } + e^{-i \phi} \delta_{\mathbf{k}, \mathbf{k}^{\prime} - 2 \pi \hat{z}} \right) c^{\dagger}_{\mathbf{k} + \pi / 2 \hat{z}} c_{\mathbf{k}^{\prime} - \pi /2 \hat{z}}\right)
     \\&=  \sum_{\mathbf{k}}
     \left( \cos(\phi)  c^{\dagger}_{\mathbf{k} - \pi / 2 \hat{z}} c_{\mathbf{k} + \pi /2 \hat{z}}  +  \cos(\phi) c^{\dagger}_{\mathbf{k} + \pi / 2 \hat{z}} c_{\mathbf{k} - \pi /2 \hat{z}}\right)
\end{split}
\end{equation}
In the last step, recall the sum is between $-\pi/2$ and $\pi/2$, so only Kronecker deltas of the form $\delta_{-\mathbf{k} + \mathbf{k}^{\prime}}$ and $\delta_{-\mathbf{k} + \mathbf{k}^{\prime} \pm 2 \pi \hat{z}}$ survive.  
Since shifting the Brillouin zone also shifts the periodic boundary, $\mathbf{k} = \mathbf{k}^{\prime} \pm 2 \pi \hat{z}$, which implies that $\delta_{-\mathbf{k} + \mathbf{k}^{\prime} \pm 2 \pi \hat{z}}$ contributes in the sum.

Next we examine how a hopping term like $c^{\dagger}_{\mathbf{R}} c_{\mathbf{R} + \mathbf{n}} + \text{h.c.}$ is modified, where $\mathbf{n}$ is a discrete length $\mathbf{n} = n_1 a \hat{x} + n_2 b \hat{y} + n_3 c \hat{z}$ for $\{n_1, n_2, n_3\}\in \mathds{Z}$. 
We find
\begin{equation}
\label{hopfulEquation}
\begin{split}
       & \sum_{\mathbf{R}} c^{\dagger}_{\mathbf{R}} c_{\mathbf{R} + \mathbf{n}} = \sum_{\mathbf{R}} \sum_{\mathbf{k}, \mathbf{k}^{\prime}} \left( e^{-i (\mathbf{k} - \pi/2 \hat{z} ) \cdot \mathbf{R}} c^{\dagger}_{\mathbf{k} - \pi / 2 \hat{z}} +  e^{-i (\mathbf{k} + \pi/2 \hat{z} ) \cdot \mathbf{R}} c^{\dagger}_{\mathbf{k} + \pi / 2 \hat{z}} \right) 
       \\&
       \qquad\qquad\qquad \qquad  \times \left( e^{i (\mathbf{k}^{\prime} - \pi/2 \hat{z} ) \cdot \left( \mathbf{R} + \mathbf{n} \right)} c_{\mathbf{k}^{\prime} - \pi / 2 \hat{z}} +  e^{i (\mathbf{k}^{\prime } + \pi/2 \hat{z} ) \cdot \left( \mathbf{R} + \mathbf{n} \right)} c_{\mathbf{k}^{\prime } + \pi / 2 \hat{z}} \right)
       \\=& 
       \sum_{\mathbf{R}} \sum_{\mathbf{k}, \mathbf{k}^{\prime}} \left( e^{i (-\mathbf{k} + \mathbf{k}^{\prime}  ) \cdot \mathbf{R}} e^{i (\mathbf{k}^{\prime} - \pi /2 \hat{z})\cdot \mathbf{n}} c^{\dagger}_{\mathbf{k} - \pi / 2 \hat{z}} c_{\mathbf{k}^{\prime} - \pi / 2\hat{z}} +  e^{i (-\mathbf{k} + \mathbf{k}^{\prime} - \pi \hat{z} )\cdot \mathbf{R}} e^{i (\mathbf{k}^{\prime} + \pi /2 \hat{z})\cdot \mathbf{n}}  c^{\dagger}_{\mathbf{k} - \pi / 2 \hat{z}} c_{\mathbf{k}^{\prime} + \pi/2 \hat{z}} \right. 
       \\& 
        \left. \qquad \qquad + e^{i (-\mathbf{k} + \mathbf{k}^{\prime} -\pi \hat{z} ) \cdot \mathbf{R}} e^{i (\mathbf{k}^{\prime} - \pi /2 \hat{z})\cdot \mathbf{n}} c^{\dagger}_{\mathbf{k} + \pi / 2 \hat{z}} c_{\mathbf{k}^{\prime} - \pi / 2\hat{z}} +  e^{i (-\mathbf{k} + \mathbf{k}^{\prime} )\cdot \mathbf{R}} e^{i (\mathbf{k}^{\prime} + \pi /2 \hat{z})\cdot \mathbf{n}}  c^{\dagger}_{\mathbf{k} + \pi / 2 \hat{z}} c_{\mathbf{k}^{\prime} + \pi/2 \hat{z}} \right)
        \\=& 
       \sum_{\mathbf{k}, \mathbf{k}^{\prime}} \left( \delta_{\mathbf{k}, \mathbf{k}^{\prime}} e^{i (\mathbf{k}^{\prime} - \pi /2 \hat{z})\cdot \mathbf{n}} c^{\dagger}_{\mathbf{k} - \pi / 2 \hat{z}} c_{\mathbf{k}^{\prime} - \pi / 2\hat{z}} +  \delta_{ \mathbf{k}, \mathbf{k}^{\prime} - \pi \hat{z} } e^{i (\mathbf{k}^{\prime} + \pi /2 \hat{z})\cdot \mathbf{n}}  c^{\dagger}_{\mathbf{k} - \pi / 2 \hat{z}} c_{\mathbf{k}^{\prime} + \pi/2 \hat{z}} \right. 
       \\& 
        \left. \qquad \qquad + \delta_{ \mathbf{k},  \mathbf{k}^{\prime} -\pi \hat{z} } e^{i (\mathbf{k}^{\prime} - \pi /2 \hat{z})\cdot \mathbf{n}} c^{\dagger}_{\mathbf{k} + \pi / 2 \hat{z}} c_{\mathbf{k}^{\prime} - \pi / 2\hat{z}} +  \delta_{ \mathbf{k} , \mathbf{k}^{\prime} } e^{i (\mathbf{k}^{\prime} + \pi /2 \hat{z})\cdot \mathbf{n}}  c^{\dagger}_{\mathbf{k} + \pi / 2 \hat{z}} c_{\mathbf{k}^{\prime} + \pi/2 \hat{z}} \right)
        \\=& 
       \sum_{\mathbf{k}} \left(  e^{i (\mathbf{k} - \pi /2 \hat{z})\cdot \mathbf{n}} c^{\dagger}_{\mathbf{k} - \pi / 2 \hat{z}} c_{\mathbf{k} - \pi / 2\hat{z}}  +  e^{i (\mathbf{k} + \pi /2 \hat{z})\cdot \mathbf{n}}  c^{\dagger}_{\mathbf{k} + \pi / 2 \hat{z}} c_{\mathbf{k} + \pi/2 \hat{z}}  \right).
\end{split}
\end{equation}
As before, the sum over $k_z$ will restrict the Kronecker deltas that give nonvanishing contributions.  
Notice this equation is bereft of terms mixing the $+\pi / 2$ and $-\pi / 2$ spaces, and introduces a shift of $\pm \pi / 2 \hat{z}$ in $e^{i \mathbf{k} \cdot \mathbf{n}}$ for the respective $\mathbf{k} \pm \pi / 2 \hat{z}$ annihilation/creation subspaces.  
Consider the non-interacting Hamiltonian \cite{BradlynAxionicWeylCDW}, 
\begin{equation}
\begin{split}
H_{0} =& \left( \sum_{\mathbf{R}}\left[i t_{x} c_{R}^{\dagger} \sigma^{x} c_{\mathbf{R}+\hat{x}}+i t_{y} c_{\mathbf{R}}^{\dagger} \sigma^{y} c_{\mathbf{R}+\hat{y}}+t_{z} c_{\mathbf{R}}^{\dagger} \sigma^{z} c_{\mathbf{R}+\hat{z}}\right] +\sum_{\mathbf{R}} \frac{m}{2}\left(c_{\mathbf{R}}^{\dagger} \sigma^{z} c_{\mathbf{R}+\hat{x}}+c_{\mathbf{R}}^{\dagger} \sigma^{z} c_{\mathbf{R}+\hat{y}}-2 c_{\mathbf{R}}^{\dagger} \sigma^{z} c_{\mathbf{R}}\right) 
\right.\\&\left.
-\sum_{\mathbf{R}} t_{z} \cos \frac{Q c}{2} c_{\mathbf{R}}^{\dagger} \sigma^{z} c_{\mathbf{R}}\right)+\mathrm{h.c.} 
\end{split}
\end{equation}
Applying Eq.~\eqref{hopfulEquation}, this Hamiltonian produces Eq.~\eqref{H0Ham}.  
This result relies on $\mathbf{Q} = \pi \hat{z}$, which constitutes breaking the Hamiltonian into two separate chunks and shifting these chunks into a single region.  
However, this procedure is generalizable for any modulation vector $\mathbf{Q} = \ell \pi/n \hat{z}$ for $\ell, n \in \mathds{Z}^+$.  
In this more general case, $n$ indicates into how many regions we need to cut the original Brillouin zone during the shifting process. 
{Note also that the factors of $\cos\phi$ appear in Eq.~(\ref{angleManipulationEquation}) due to the twofold modulation $\mathbf{Q} = \pi / 2\hat{z}$. 
Larger values of $n$, (i.e. higher-fold modulation) would require a careful treatment of how $\phi$ enters into the Kronecker delta constraints in Eq.~\eqref{angleManipulationEquation}.}

\section{Covariance using the Berry Connection}
\label{VelocityOperatorDerivation}
In periodic crystals, matrix elements of the velocity operator contain ``anomalous'' contributions due to the Berry connection \cite{SipeSecondOrderConductivity, MooreDiagrammatic, BlountFormalismToBandTheory}.  
These can be embedded into the covariant version of the derivative through minimal coupling.  
To see this, start from the definition of the density operator \cite{NoltingQuantumMagnetism}:
\begin{equation}
    \rho (r) = \sum^N_{i} \delta(\mathbf{r} - \mathbf{r}_i) = \sum_{\mathbf{k}_1, \mathbf{k}_2, a, b} \langle \mathbf{k}_1, a | \delta(\mathbf{r} - \mathbf{r}_i) | \mathbf{k}_2, b \rangle c^{\dagger}_{\mathbf{k}_1, a} c_{\mathbf{k}_2, b}= \sum_{\mathbf{k}_1, \mathbf{k}_2, a, b} \psi^*_{\mathbf{k}_1, a}(\mathbf{r}) \psi_{\mathbf{k}_2, b}(\mathbf{r}) c^{\dagger}_{\mathbf{k}_1, a} c_{\mathbf{k}_2, b},
\end{equation}
where $\psi_{\mathbf{k}_1, a}(\mathbf{r})$ is the Bloch wavefunction satisfying $\psi_{\mathbf{k}_1 a}(\mathbf{r}) = e^{i \mathbf{k} \cdot \mathbf{r}} u_{\mathbf{k}, a}(\mathbf{r})$, where $u_{\mathbf{k}, a}(\mathbf{r}) = u_{\mathbf{k}, a}(\mathbf{r} + \mathbf{R})$ for Bravais lattice vector $\mathbf{R}$.  
Note the index $a$ is understood to index both orbital degrees of freedom and any valley degrees of freedom arising from zone shifting.  
Recall also that the Bloch wavefunction are orthonormal: $\int [d\mathbf{r}] \psi^{*}_{\mathbf{k}_1, a}(\mathbf{r}) \psi_{\mathbf{k}_2, b}(\mathbf{r}) = \delta(\mathbf{k}_1 - \mathbf{k}_2) \delta_{a b} $.  
Invoking a Fourier transform, we have for the density operator
\begin{equation}
    \rho_{q^{\prime}} = \sum_{k_1, k_2, r }\left( u^{*}_{\mathbf{k}_1, a}(\mathbf{r}) e^{- i \mathbf{k}_1 \cdot \mathbf{r} } e^{ i \mathbf{k}_2 \cdot \mathbf{r} }  u_{\mathbf{k}_2, b}(\mathbf{r}) \right)c^{\dagger}_{\mathbf{k}_1, a} c_{\mathbf{k}_2, b} e^{-i \mathbf{q}^{\prime} \cdot \mathbf{r}}.
\end{equation}
To find the modified current density operator, we want to expand the continuity equation
\begin{equation}
    \partial_t\rho_{\mathbf{q}'} +i\mathbf{q'}\cdot\mathbf{j}_\mathbf{q'} = 0
\end{equation}
as $\mathbf{q}^{\prime} \rightarrow 0$.  
Using the equation of motion $-i\partial_t\rho_{q^\prime}=[H, \rho_{q^{\prime}}]$ yields
\begin{equation}
\label{currentDensityWithBerryCon}
\begin{split}
    &[ \sum_{k, a, b} H_{a b}(\mathbf{k}) c^{\dagger}_{\mathbf{k}, a} c_{\mathbf{k}, b}, \sum_{k_1, k_2, r a^{\prime}, b^{\prime}}\left( u^{*}_{\mathbf{k}_1, a^{\prime}}(\mathbf{r})  u_{\mathbf{k}_2, b^{\prime}}(\mathbf{r})  e^{ i (\mathbf{k}_2 - \mathbf{k}_1 - \mathbf{q}^{\prime}) \cdot \mathbf{r} }  \right)c^{\dagger}_{\mathbf{k}_1, a^{\prime}} c_{\mathbf{k}_2, b^{\prime}}]
    \\&= \sum_{k, k_1, k_2, r, a, b, a^{\prime}, b^{\prime}} H_{a b}(\mathbf{k}) \left( u^{*}_{\mathbf{k}_1, a^{\prime}}(\mathbf{r})  u_{\mathbf{k}_2, b^{\prime}}(\mathbf{r})  e^{ i (\mathbf{k}_2 - \mathbf{k}_1 - \mathbf{q}^{\prime}) \cdot \mathbf{r} }  \right) \left[ c^{\dagger}_{\mathbf{k}, a} c_{\mathbf{k}_2, b^{\prime}} \delta_{\mathbf{k}_1, \mathbf{k}} \delta_{b, a^{\prime}} - c^{\dagger}_{\mathbf{k}_1, a^{\prime}} c_{\mathbf{k}, b} \delta_{\mathbf{k}, \mathbf{k}_2} \delta_{b^{\prime}, a} \right]
    \\&= \sum H_{a a^{\prime}}(\mathbf{k}) \left( u^{*}_{\mathbf{k}, a^{\prime}}(\mathbf{r})  u_{\mathbf{k}_1, b^{\prime}}(\mathbf{r})  e^{ i (\mathbf{k}_1 - \mathbf{k} - \mathbf{q}^{\prime}) \cdot \mathbf{r} }  \right) c^{\dagger}_{\mathbf{k}, a} c_{\mathbf{k}_1, b^{\prime}}  - H_{b^{\prime} b}(\mathbf{k}) \left( u^{*}_{\mathbf{k}_1, a^{\prime}}(\mathbf{r})  u_{\mathbf{k}, b^{\prime}}(\mathbf{r})  e^{ i (\mathbf{k} - \mathbf{k}_1 - \mathbf{q}^{\prime}) \cdot \mathbf{r} }  \right) c^{\dagger}_{\mathbf{k}_1, a^{\prime}} c_{\mathbf{k}, b}.
\end{split}
\end{equation}
The next step is to Taylor expand in orders of $\mathbf{q}^{\prime}$.  
Notice that due to the orthogonality of the Bloch wavefunctions, Eq.~\eqref{currentDensityWithBerryCon} vanishes for $\mathbf{q}^{\prime} = 0$;  the expansion starts at linear order.  
Taking the derivative to compute the first order term gives
\begin{equation}
\begin{split}
    \sum_{r} \partial_{\mathbf{q}^{\prime}} \left. \left( u^{*}_{\mathbf{k}, a^{\prime}}(\mathbf{r})  u_{\mathbf{k}_1, b^{\prime}}(\mathbf{r})  e^{ i (\mathbf{k}_1 - \mathbf{k} - \mathbf{q}^{\prime}) \cdot \mathbf{r} }  \right) \right|_{\mathbf{q}^{\prime} = 0}  & =  -\sum_{r} \left( u^{*}_{\mathbf{k}, a^{\prime}}(\mathbf{r})  u_{\mathbf{k} + \mathbf{q}, b^{\prime}}(\mathbf{r})  \partial_{\mathbf{q}} e^{ i \mathbf{q} \cdot \mathbf{r} }  \right) 
    \\
    & =  -\sum_{r} \partial_{\mathbf{q}} \left( u^{*}_{\mathbf{k}, a^{\prime}}(\mathbf{r})  u_{\mathbf{k} + \mathbf{q}, b^{\prime}}(\mathbf{r})  e^{ i \mathbf{q} \cdot \mathbf{r} }  \right) + \sum_{r} \partial_{\mathbf{q}} \left( u^{*}_{\mathbf{k}, a^{\prime}}(\mathbf{r})  u_{\mathbf{k} + \mathbf{q}, b^{\prime}}(\mathbf{r}) \right) e^{ i \mathbf{q} \cdot \mathbf{r} }  
    \\& =
    - \partial_{\mathbf{q}} \delta(\mathbf{q}) \delta_{a^{\prime}, b^{\prime}} + \sum_{R}  e^{ i \mathbf{q} \cdot \mathbf{R} }  \int_{\text{U.C.}} [d \mathbf{r}]\partial_{\mathbf{q}} \left( u^{*}_{\mathbf{k}, a^{\prime}}(\mathbf{r})  u_{\mathbf{k} + \mathbf{q}, b^{\prime}}(\mathbf{r}) \right)
    \\& =
    - \partial_{\mathbf{q}} \delta(\mathbf{q}) \delta_{a^{\prime}, b^{\prime}} + \delta(\mathbf{q}) \int_{\text{U.C.}} [d \mathbf{r}]  u^{*}_{\mathbf{k}, a^{\prime}}(\mathbf{r}) \partial_{\mathbf{q}} u_{\mathbf{k} + \mathbf{q}, b^{\prime}}(\mathbf{r})
    \\& =
    - \partial_{\mathbf{q}} \delta(\mathbf{q}) \delta_{a^{\prime}, b^{\prime}} - i \delta(\mathbf{q}) \mathcal{A}_{a^{\prime} b^{\prime}}(\mathbf{k}).\label{eq:posoperatormatrix}
\end{split}
\end{equation}
  The previous calculation took advantage of the periodicity in $\mathbf{R}$ . 
  We used the transformation $\mathbf{k}_1 = \mathbf{k} + \mathbf{q}$, and we denote by $\mathcal{A}_{a^{\prime} b^{\prime}}(\mathbf{k}) \equiv i \int_{\text{U.C.}} [d \mathbf{r}] u^*_{\mathbf{k}, a^{\prime}} (\mathbf{r}) \partial_{\mathbf{k}}u_{\mathbf{k}, b^{\prime}} (\mathbf{r})$ the matrix elements of the Berry connection between states indexed by $a^{\prime}$ and $b^{\prime}$.  
  Note that the Brillouin zone shifting procedure in general leads to additional terms proportional to $\delta(\mathbf{q} \pm \mathbf{Q})$ in Eq.~(\ref{eq:posoperatormatrix}).  However, in the limit that $\mathbf{q} \rightarrow 0$ which we consider here, these terms vanish. 
  We emphasize as well that $\mathcal{A}_{a'b'}$ is the Berry connection evaluated in the orbital and valley basis. 
  Because of the Dirac delta $\delta(\mathbf{q})$ in Eq.~\eqref{eq:posoperatormatrix}, it is diagonal in the valley indices. 
  If instead we choose to work in the basis of eigenstates of $H_{ab}(\mathbf{k})$, then we would find $i \mathcal{A}^{\alpha}_{nm} = \int_{\text{U. C.}} [d \mathbf{r}](\partial_{k^{\alpha}} [u^{*}_{\mathbf{k}, a}(\mathbf{r}) U^{\dagger}_{an} (\mathbf{k}))] U_{mb}(\mathbf{k}) u_{\mathbf{k}, b}  (\mathbf{r})$, where the transformation matrix $U$ diagonalizes $H_0$ such that $H^{\prime}_0 = U^{\dagger} H_0 U$\cite{RaoHallViscosity}.
  The simplicity of $\mathcal{A}_{a'b'}$ is one of the advantages of working in the orbital basis mentioned in Sec.~\ref{conductivity_section} of the main text.

  Continuing with our simplification of Eq.~\eqref{currentDensityWithBerryCon}, we have additionally that that $\sum_r \left( u^{*}_{\mathbf{k}_1, a^{\prime}}(\mathbf{r})  u_{\mathbf{k}, b^{\prime}}(\mathbf{r})  e^{ i (\mathbf{k} - \mathbf{k}_1 - \mathbf{q}^{\prime}) \cdot \mathbf{r} }  \right) = - \partial_{\mathbf{q}} \delta(\mathbf{q}) \delta_{a^{\prime}, b^{\prime}} - i \delta(\mathbf{q}) \mathcal{A}_{a^{\prime} b^{\prime}}(\mathbf{k})$ under the transform $\mathbf{k}_1 = \mathbf{k} - \mathbf{q}$.  
  Putting everything together yields
\begin{equation}\label{eq:continuity}
\begin{split}
    &[ \sum_{k, a, b} H_{a b}(\mathbf{k}) c^{\dagger}_{\mathbf{k}, a} c_{\mathbf{k}, b}, \sum_{k_1, k_2, r, a^{\prime}, b^{\prime} }\left( u^{*}_{\mathbf{k}_1, a^{\prime}}(\mathbf{r})  u_{\mathbf{k}_2, b^{\prime}}(\mathbf{r})  e^{ i (\mathbf{k}_2 - \mathbf{k}_1 - \mathbf{q}^{\prime}) \cdot \mathbf{r} }  \right)c^{\dagger}_{\mathbf{k}_1, a^{\prime}} c_{\mathbf{k}_2, b^{\prime}}]
    \\&= 
    \mathbf{q^{\prime}} \cdot \left[ \sum H_{a a^{\prime}}(\mathbf{k}) \left( - \partial_{\mathbf{q}} \delta(\mathbf{q}) \delta_{a^{\prime}, b^{\prime}} - i\delta(\mathbf{q}) \mathcal{A}_{a^{\prime} b^{\prime}}(\mathbf{k})  \right) c^{\dagger}_{\mathbf{k}, a} c_{\mathbf{k} + \mathbf{q}, b^{\prime}}  - H_{b^{\prime} b}(\mathbf{k}) \left(- \partial_{\mathbf{q}} \delta(\mathbf{q}) \delta_{a^{\prime}, b^{\prime}} - i \delta(\mathbf{q}) \mathcal{A}_{a^{\prime} b^{\prime}}(\mathbf{k}) \right) c^{\dagger}_{\mathbf{k} - \mathbf{q}, a^{\prime}} c_{\mathbf{k}, b} \right]
    \\&= \mathbf{q^{\prime}} \cdot \left[ \sum_{k, q, a, b} H_{a b}(\mathbf{k}) \delta(\mathbf{q}) \left( c^{\dagger}_{\mathbf{k}, a} \partial_{\mathbf{q}} c_{\mathbf{k} + \mathbf{q}, b} - \partial_{\mathbf{q}} c^{\dagger}_{\mathbf{k} - \mathbf{q}, a} c_{\mathbf{k}, b} \right) - i \sum_{k, a, b, a^{\prime}, b^{\prime}} c^{\dagger}_{\mathbf{k}, a} H_{a a^{\prime}}(\mathbf{k}) \mathcal{A}_{a^{\prime} b}(\mathbf{k}) c_{\mathbf{k}, b^{\prime}} - c^{\dagger}_{\mathbf{k}, a^{\prime}}  \mathcal{A}_{a^{\prime} b^{\prime}}(\mathbf{k}) H_{b^{\prime} b}(\mathbf{k}) c_{\mathbf{k}, b} \right]
    \\&= \mathbf{q^{\prime}} \cdot \left[ \sum_{k, a, b} H_{a b}(\mathbf{k}) \partial_{\mathbf{k}}(c^{\dagger}_{\mathbf{k}, a} c_{\mathbf{k}, b}) - i \sum_{k, a, b, a^{\prime}} c^{\dagger}_{\mathbf{k}, a} \Big( H_{a a^{\prime}}(\mathbf{k}) \mathcal{A}_{a^{\prime} b}(\mathbf{k}) -  \mathcal{A}_{a a^{\prime}}(\mathbf{k}) H_{a^{\prime} b}(\mathbf{k}) \Big) c_{\mathbf{k}, b} \right]
    \\&= \mathbf{q^{\prime}} \cdot \left[ \sum_{k, a, b} H_{a b}(\mathbf{k}) \partial_{\mathbf{k}}(c^{\dagger}_{\mathbf{k}, a} c_{\mathbf{k}, b}) - i \sum_{k, a, b} c^{\dagger}_{\mathbf{k}, a} [H, \mathcal{A}]_{a b}  c_{\mathbf{k}, b} \right].
\end{split}
\end{equation}
In this form, $[H, \mathcal{A}]_{a b} =  \sum_{a^{\prime}}H_{a, a^{\prime}}(\mathbf{k}) \mathcal{A}_{a^{\prime} b}(\mathbf{k}) -  \mathcal{A}_{a a^{\prime}}(\mathbf{k}) H_{a^{\prime}, b}(\mathbf{k}) $, is just the commutator between the Hamiltonian and Berry connection matrices \cite{MooreDiagrammatic}.  
Applying this logic recursively suggests that every derivative with an expectation value should be covariantly modified to $D^{\mu} \hat{\mathcal{O}}_{a b} = \partial^{\mu} \hat{\mathcal{O}}_{a b} - i [\hat{\mathcal{O}}, \mathcal{A}]_{a b}$ for an arbitrary operator $\hat{\mathcal{O}}$.  
This guarantees expectation values involving derivatives will yield identical answers no matter which basis is chosen.

Finally, we must make a choice as to the Berry connection $\mathcal{A}_{ab}$ in the orbital/valley basis. 
This quantity is model dependent, containing information about the shape and location of the basis orbtials in which our tight-binding model is expressed. 
In particular, we are free to make use of the standard ``tight-binding'' convention, where the basis orbitals are assumed to be compactly localized, centered at the origin of the unit cell, and that the position operator has no off-diagonal elements in terms of the basis orbitals. 
This ensures that $\mathcal{A}_{ab}=0$ within this tight-binding orbital basis. 
We work with this convention throughout the paper.
This choice is appropriate for a lattice model whose primary purpose is to regulate the low-energy theory of an incommensurate syste.  
Note that for a tight-binding model of a \emph{commensurate} CDW, it may not be possible to choose $\mathcal{A}=0$. 
This is because the commensurate CDW distortion enlarges the position space unit cell, resulting in orbitals that are not located at the origin in the enlarged cell. 
The orbital basis Berry connection $\mathcal{A}_{ab}$ then encodes the positions of these orbtials. 
We defer a systematic study of the implications of this subtlety to future work.

There is an additional question: do the off-diagonal elements of the Hamiltonian contribute to the velocity vertex in a nontrivial way? 
To answer this, let us recall that the valley-off-diagonal terms in $H_{ab}(\mathbf{k})$ are given by our mean field expression $H_{+-}(\mathbf{k}) = g\langle b_{\mathbf{Q}} + b^\dag_{-\mathbf{Q}}\rangle = 2\Delta e^{-i\phi}$, which is crucially independent of the crystal momentum $\mathbf{k}$. 
Substituting this into Eq.~\eqref{eq:continuity}, integrating the first term by parts, and using the fact that $\mathcal{A}=0$, we see that $H_{+-}$ does not contribute to the velocity operator. 
Therefore, the standard form $\hat{\mathbf{v}}_{n m} = \langle n |\frac{d \hat{\mathbf{x}}}{d t}| m \rangle$ for the velocity operator is still valid.

\section{Free Phonon Propagator}
\label{FreePhononPropagator}
This section provides the derivation of the free phonon propagator.  
We start with the definition $D_0( \mathscr{T}, \mathbf{q}) = \langle T_{\mathscr{T}} A_{\mathbf{q}, \xi_1}(\mathscr{T}) A^{\dagger}_{\mathbf{q}, \xi_1} \rangle$ for $ A_{\mathbf{q}, \xi_1}(\mathscr{T}) = b_{ \mathbf{q}, \xi_1}(\mathscr{T}) + b^{\dagger}_{-\mathbf{q}, \xi_1}(\mathscr{T})$, where $H_1 = \sum_{q>0, \xi_1} \omega_{\mathbf{q}} (b^{\dagger}_{\mathbf{q}, \xi_1} b_{\mathbf{q}, \xi_1} + b^{\dagger}_{-\mathbf{q}, \xi_1} b_{-\mathbf{q}, \xi_1})$.  To proceed, notice that $[H_1, b_{\mathbf{q}, \xi_1}] = - \omega_{\mathbf{q}} b_{\mathbf{q}, \xi_1}$ and $[H_1, b^{\dagger}_{-\mathbf{q}, \xi_1}] =  \omega_{\mathbf{q}} b^{\dagger}_{-\mathbf{q}, \xi_1}$.  
Using these commutation relations and the Hadamard lemma, which produces $e^{H_1 \mathscr{T}} b_{\mathbf{q}, \xi_1} e^{-H_1 \mathscr{T}} = e^{- \omega_{\mathbf{q}} \mathscr{T}} b_{\mathbf{q}, \xi_1}$ and $e^{H_1 \mathscr{T}} b^{\dagger}_{-\mathbf{q},\xi_1} e^{-H_1 \mathscr{T}} = e^{\omega_{\mathbf{q}} \mathscr{T}} b^{\dagger}_{-\mathbf{q}, \xi_1}$, the phonon propagator becomes \cite{BruusAndFlensberg}
\begin{equation}
\begin{split}
    D_0( \mathscr{T}, \mathbf{q})  = & -\left\langle \Big( b_{ \mathbf{q}, \xi_1}(\mathscr{T}) + b^{\dagger}_{-\mathbf{q}, \xi_1}(\mathscr{T}) \Big) \Big( b^{\dagger}_{\mathbf{q}, \xi_1} + b_{-\mathbf{q}, \xi_1} \Big) \right \rangle \theta(\mathscr{T})  -\left\langle  \Big( b^{\dagger}_{\mathbf{q}, \xi_1} + b_{-\mathbf{q}, \xi_1} \Big) \Big( b_{\mathbf{q}, \xi_1}(\mathscr{T}) + b^{\dagger}_{-\mathbf{q}, \xi_1}(\mathscr{T}) \Big) \right\rangle \theta(-\mathscr{T})
    \\ = &
    -\left \langle e^{H_1 \mathscr{T}} \Big( b_{\mathbf{q}, \xi_1} + b^{\dagger}_{-\mathbf{q}, \xi_1}\Big) e^{-H_1 \mathscr{T}} \Big( b^{\dagger}_{\mathbf{q}, \xi_1} + b_{-\mathbf{q}, \xi_1} \Big) \right \rangle \theta(\mathscr{T})  -\left \langle  \Big( b^{\dagger}_{\mathbf{q}, \xi_1} + b_{-\mathbf{q}, \xi_1} \Big) e^{H_1 \mathscr{T}} \Big( b_{\mathbf{q}, \xi_1} + b^{\dagger}_{-\mathbf{q}, \xi_1} \Big) e^{-H_1 \mathscr{T}} \right\rangle \theta(-\mathscr{T})
    \\ = &
    -\Big( e^{-\omega_{\mathbf{q}} \mathscr{T}} \langle  b_{\mathbf{q}, \xi_1} b^{\dagger}_{\mathbf{q}, \xi_1} \rangle + e^{\omega_{\mathbf{q}} \mathscr{T}} \langle b^{\dagger}_{-\mathbf{q}, \xi_1} b_{-\mathbf{q}, \xi_1} \rangle \Big) \theta(\mathscr{T})  - \Big( e^{\omega_{\mathbf{q}} \mathscr{T}} \langle b^{\dagger}_{\mathbf{q}, \xi_1} b_{\mathbf{q}, \xi_1} \rangle + e^{- \omega_{\mathbf{q}} \mathscr{T}} \langle b_{-\mathbf{q}, \xi_1} b^{\dagger}_{-\mathbf{q}, \xi_1} \rangle  \Big) \theta(-\mathscr{T}).
\end{split}
\end{equation}
We next transform to Matsubara frequencies $\omega = 2 n \pi / \beta$ with $n \epsilon \mathds{Z}$.  
In addition, the boson distribution takes the place of the expectation values, $\langle b_{ \mathbf{q}, \xi_1} b^{\dagger}_{ \mathbf{q}, \xi_1} \rangle = 1+n_B(\omega_{\mathbf{q}}) $ and $\langle b^{\dagger}_{ \mathbf{q}, \xi_1} b_{ \mathbf{q}, \xi_1} \rangle = n_B(\omega_{\mathbf{q}}) $. 
This yields 
\begin{equation}
    D_0(i\omega, \mathbf{q}) = -\frac{ 2 \omega_{\mathbf{q}}}{\omega^2 + \omega^2_{\mathbf{q}}}.
\end{equation}

\section{Intercept of Massless Linear Collective Conductivity}
\label{zeroFreqIntercept}

Because the massless collective mode propagator diverges as $\omega\rightarrow 0$, it leads to a divergent contribution to the conductivity as $\omega \rightarrow 0$.  
To see how this arises, we will analyze the asymptotics of the massless collective contribution to the linear conductivity. 
First we note that the massless mode propagator, $D_{++}(\omega)- e^{2 i \phi} D_{+-}(\omega)$ diverges as $~\frac{1}{\omega^2}$. 
Additionally, note that the $d_1$-th component of the loop contribution from Fig.~\ref{LinearCollectiveDiagrams}, $\left(G^{\text{loop}}\right)^{\alpha}_{d_1}$, behaves like $~\frac{1}{\omega}$ for large frequency.  
To find the value of $\lim_{\omega\rightarrow 0} \sigma_{zz}(\omega)$, consider the general form of $\left(G^{\text{loop}}\right)^{\alpha}_{d_1}(\omega)$ for any arbitrary Hamiltonian,
\begin{equation}
    \left(G^{\text{loop}}\right)^{\alpha}_{d_1}(\omega) = \int [d \mathbf{k}] \frac{f^{\alpha}_0(\mathbf{k}) +f^{\alpha}_1(\mathbf{k}) \omega +  \cdot \cdot \cdot + f^{\alpha}_{n-1}(\mathbf{k}) \omega^{n-1}}{g^{\alpha}_0(\mathbf{k}) +g^{\alpha}_1(\mathbf{k}) \omega +  \cdot \cdot \cdot + g^{\alpha}_{n}(\mathbf{k}) \omega^n}.
\end{equation}
The functions $f^{\alpha}_{i}(\mathbf{k})$ and $g^{\alpha}_i(\mathbf{k})$ multiply $\omega^i$ in the rational function representation of the loop diagram.  
For example, in the tilted lattice model from Eq.~\eqref{tiltedLatticeTerm}, the degree of polynomial is $n=10$.  
Another important property of the lattice model is that $f^{\alpha}_j(\mathbf{k}) = 0 $ for $\{j= 2 i   | i \in \mathds{N}\}$ and $g^{\alpha}_j(\mathbf{k}) = 0 $ for $\{j= 2 i - 1 | i \in \mathds{N}\}$ due to parity constraints.  
Correspondingly, the massless mode propagator may be written as
\begin{equation}
    D_{++}(\omega)- e^{2 i \phi} D_{+-}(\omega)= \frac{1}{\frac{\omega^2_Q - \omega^2}{2 \omega_\mathbf{Q}} - \int[\mathbf{k}] \frac{m_{1, 0}(\mathbf{k}) + m_{1, 1}(\mathbf{k}) \omega + \cdot \cdot \cdot m_{1, \ell-1}(\mathbf{k}) \omega^{\ell-1}}{m_{2, 0}(\mathbf{k}) + m_{2, 1}(\mathbf{k}) \omega + \cdot \cdot \cdot m_{2, \ell}(\mathbf{k}) \omega^{\ell}}}.
\end{equation}
Now, the functions $m_{1, i}(\mathbf{k})$ and $m_{2, i}(\mathbf{k})$ describe the coefficients of the rational function in the denominator, which are due to our recursive treatment of the electron-phonon interactions.  
Also, $\ell$ gives the degree of the polynomials, where $\ell = 8$ in the tilted lattice model.  
The characteristic property of the massless mode propagator $D_{++}(\omega)- e^{2 i \phi} D_{+-}(\omega)$ is that $\frac{\omega_\mathbf{Q}}{2} = \int[\mathbf{k}] \frac{m_{1, 0}(\mathbf{k})}{m_{2, 0}(\mathbf{k})}$, due to the gap equation.  
Taking into account the symmetries of the tilted lattice model, it follows that $m_{1, j}(\mathbf{k}) = 0 $ for $\{j= 2 i -1   | i \in \mathds{N}\}$ and $m_{2, j}(\mathbf{k}) = 0   $ for $\{j= 2 i -1 | i \in \mathds{N}\}$. 

Returning to the loop diagram, recall that the components obey $\left(G^{\text{loop}}\right)^{z}_{1}(\omega)=- \left(G^{\text{loop}}\right)^{z}_{2}(\omega)$.
Therefore, since the relevant conductivity is $\omega \sigma^{zz}(\omega) = - 2i\frac{e^2}{\hbar} \left[ \left(G^{\text{loop}}\right)^{z}_{1}(\omega) \right]^2 \left( D_{++}(\omega)- e^{2 i \phi} D_{+-}(\omega) \right) $, L'Hôpital's rule must be applied twice as $\omega \rightarrow 0$ in the case of the tilted lattice model.  
Simplifying, this gives
\begin{equation}
   \lim_{\omega \rightarrow 0} \omega \sigma^{zz}(\omega) = -i\int [d \mathbf{k}_1] [d \mathbf{k}_2] \frac{2 f^z_1(\mathbf{k}_1) f^z_1(\mathbf{k}_2) }{g^z_{0}(\mathbf{k_1}) g^z_{0}(\mathbf{k_2}) } \frac{1}{\omega_\mathbf{Q} + \int [d \mathbf{k}_3] \frac{2 m_{1, 2} (\mathbf{k}_3)}{m_{2, 0} (\mathbf{k}_3)} }.
\end{equation}
In the tilted lattice model, these intercept values are small (see Fig.~\ref{linearCollectiveConductivityPlotWithTilt}(i)), which is characteristically different from the cosine dispersion in Appendix \ref{cosineSection}.  
Based on our arguments in Sec.~\ref{LinearCollectiveSection}, this is because the massless collective contribution to the conductivity depends perturbatively on the tilt. 
The result is small tilts will produce only small changes in the conductivity, such as the intercept at $\omega = 0$. 

\section{Mass of the Amplitude Mode Propagator}
\label{massiveResonanceSection}
\begin{figure*}[h]
      \centering
\centering
\includegraphics[width=0.8\hsize]{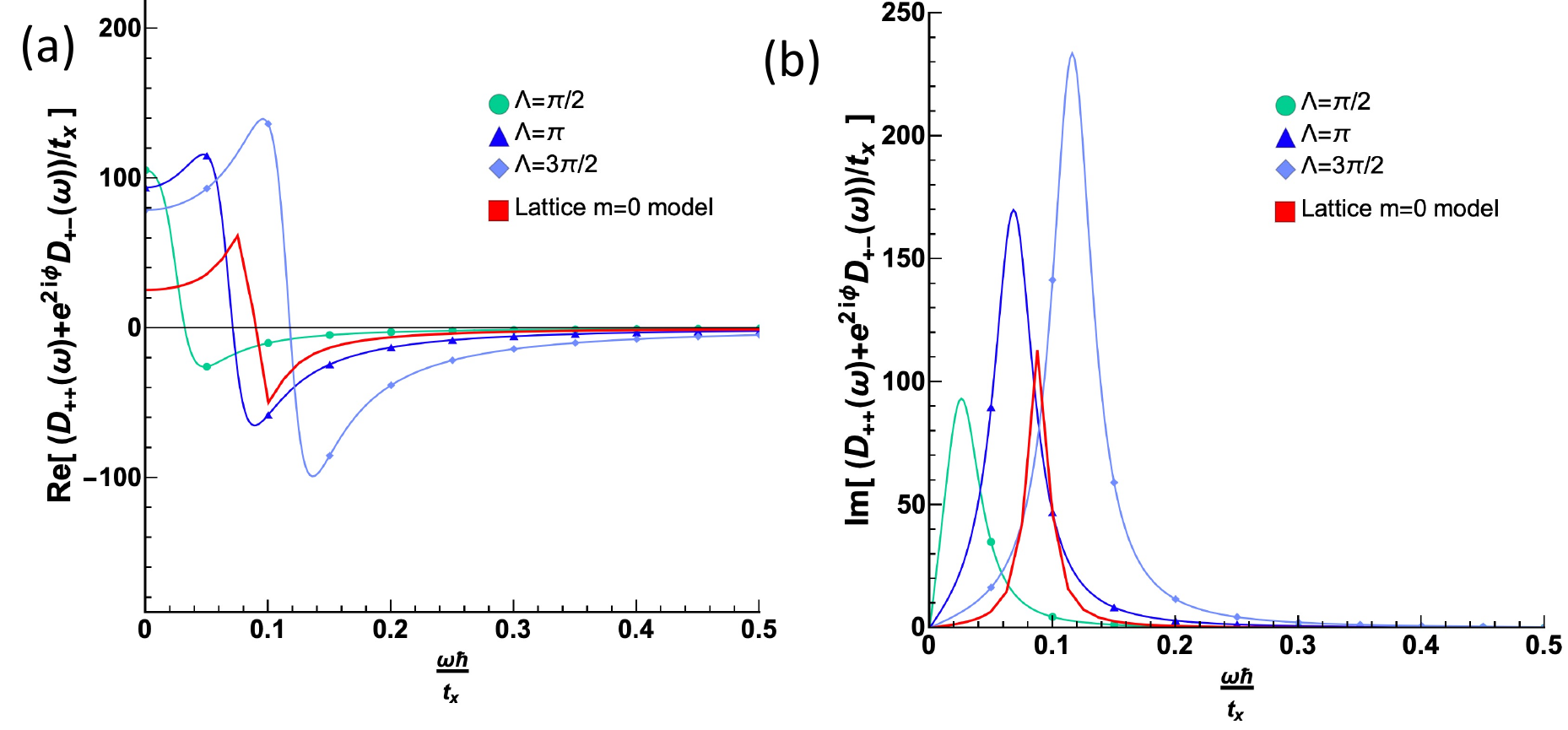}
\caption{Real (a) and imaginary (b) part of the massive collective mode propagator $D_{++}(\omega) + e^{2 i \phi} D_{+-}(\omega)$ for the ideal linearized model.  The cutoff $\Lambda$ is important in determining the resonant frequency, as well as the intercept at $\omega = 0$.  In the main text, a cutoff of $\Lambda = \pi $ was used.}
\label{massiveAgainstCutoff}
\end{figure*}
As stated in the main text at Eq.~\eqref{MassiveMode}, the massive phonon mode propagator (when Wick rotated) is 
\begin{equation}
\label{massivePhononProp}
    D_{++}(\omega) + e^{2 i \phi} D_{+-}(\omega) =
    \frac{1}{\frac{ -\omega^2}{2 \omega_{\mathbf{Q}}}   + g^2\int[d \mathbf{k}] \frac{\omega^2 - 4 (2 \Delta)^2}{E_{\mathbf{k}} (\omega^2 -4 E_{\mathbf{k}}^2 )}}.
\end{equation}

The mass of the collective mode is determined by the resonant frequency, $\omega_{\text{res}}$, at which this propagator diverges.  
Setting the denominator of Eq.~(\ref{massivePhononProp}) equal to zero, we obtain a transcendental equation.  
However, a solution may be estimated in the ideal model by expanding perturbatively when $\Delta \gg \omega_{\text{res}}$.  
Substituting the gap equation Eq~(\ref{idealGapEquation}), the squared resonant frequency is{
\begin{align}
\label{resonantFreq}
    \omega_{\text{res}}^2 =& \frac{\int [d \mathbf{k}] \frac{(2 \Delta)^2}{E^3_{\mathbf{k}}} }{  \frac{1}{4g^4\int[d \mathbf{k}] E^{-1}_{\mathbf{k}}} + \int [d \mathbf{k}] \frac{\epsilon^2_{\mathbf{k}}}{4 E^5_{\mathbf{k}}}}
    \\ = &
    \frac{(g \Delta)^4 \beta_1}{(t_x t_y t_z)^2 \beta_2 + g^4 \Delta^2 \beta_3},
\end{align}
where $\beta_{1, 2, 3}$ are numbers defined by unitless integrals $\beta_1 = \int \frac{du dv dw}{(2 \pi)^3} \frac{1}{2}(u^2 + v^2 + w^2 + 1)^{-3/2}$, $\beta_2 = \left[ \int \frac{du dv dw}{(2 \pi)^3} 2 (u^2 + v^2 + w^2 + 1)^{-1/2} \right]^{-1}$, and $\beta_3 =  \int \frac{du dv dw}{(2 \pi)^3} \frac{1}{8} (u^2 + v^2 + w^2)(u^2 + v^2 + w^2 + 1)^{-5/2} $.}  
Near this frequency, the propagator $D_{++}(\omega) + e^{2 i \phi} D_{+-}(\omega)$ diverges.  
In the weak coupling limit $g\ll (t_xt_yt_z)^{1/3}$, we see that
\begin{equation}
\omega_{\text{res}}\rightarrow \sqrt{\frac{\beta_1}{(t_xt_yt_z)^2\beta_2}}(g\Delta)^2.
\end{equation}
This is generically  less than the single-particle gap $2\Delta$.  
As shown in Fig.~\ref{massiveAgainstCutoff}, the unitless integrals, $\beta_1$, $\beta_2$, and $\beta_3$ parametrize the cutoff dependence of the integrals. 
The cutoff $\Lambda$ is important to consider when dealing with the ideal Weyl model.  
Another subtlety is the dependence of the propagator on the phonon self energy, $i \eta$.  
A sufficiently large $\eta$ can make the decaying tail of the real massive propagator intercept the axis at different points. 
The cutoff also influences this intercept.
Although we focus on the disorder-free $\eta\rightarrow 0$ limit in the text, we must still keep in mind the role of $\eta$ when we interpret our numerical calculations.

\section{Vanishing of the Massless Mode Propagator in the One-Photon Loop}
\label{MasslessModeVanishingDerivation}
This section provides mathematical rationale for why the massless mode propagator does not contribute to the conductivity in untilted Weyl semimetal models.  
Mathematically, the massless mode, $D_{++}(\omega) - e^{2 i \phi} D_{+-}(\omega)$, is multiplied by $\omega^2$ to ensure the singularity in $D_{++}(\omega) - e^{2 i \phi} D_{+-}(\omega)$ cancels through a series of L'Hôpital's Rules.  
Eq.~\eqref{LinearCollectiveFullCalc} indicates that in the expression for the collective conductivity, a factor of $\omega^2$ will appear in the numerator {due to the Green's function in Eq.~\eqref{greenFunctions}} for the diagonal elements, $G_{n_1, \xi, n_2, \xi}(i(\nu + \omega), \mathbf{k}) \propto i(\nu + \omega) \delta_{n_1 n_2}/ (E^2_{\mathbf{k}} + \nu^2)$ for all choices of node index $\xi$.  
{In this section, we will show that the Pauli matrix algebra will force these terms to give zero contribution to the conductivity, unlike LRA's quadratic dispersion model.} 
Although there are other contributions to the conductivity that go as $\omega^2$ upon a Taylor expansion of the denominator of the Green's functions, those terms vanish due to integration parity. Note first that the velocity vertex is block diagonal in the $\xi$ subspace. 
This implies that $h^{\mu}_{n_1 \xi_1, n_2 \xi_2} \propto \delta_{\xi_1 \xi_2} (\sigma^{\mu})_{n_1, n_2}$ {up to an overall sign in accord with Eq.~\eqref{linearizedVelocityVertexWithOneIndex}}.  
Next, the vertex $P_{n_1 \xi_1, n_2 \xi_2; d_1}$, for any chosen value of $d_1$, is non-zero only for $\xi_1 = - \xi_2$, {as demonstrated in Eqs.~\eqref{phononElectronVertexIn} and \eqref{phononElectronVertexOut}}.  
Therefore, the contributions to the conductivity that do not trivially integrate to zero from a single photon collective loop (see the single loop part from  Eq.~\eqref{LinearCollectiveFullCalc} and from Fig.~\ref{LinearCollectiveFullCalc}) are proportional to
\begin{equation}
\label{VanishingMassless1}
\text{tr}\left(G_{-\xi, \xi}(i\nu_1, \mathbf{k}_1) h^{\mu}_{\xi, \xi} G_{\xi, \xi}(i\nu_1 + i\omega, \mathbf{k}_1) P_{\xi, -\xi; \xi}\right).
\end{equation}
Since the $\xi$ index is the only important one at this point, the $n$ indices were suppressed in Eq.~\eqref{VanishingMassless1}.  
The Pauli matrix structure for either choice $\xi = \pm$ is identical, {up to an overall sign}.  
Substituting the respective Pauli matrices into \eqref{VanishingMassless1}, the trace becomes
\begin{equation}
\label{vanishingCondition2}
\text{tr}\left(G_{-\xi, \xi}(i\nu_1, \mathbf{k}_1) h^{\mu}_{\xi, \xi} G_{\xi, \xi}(i\nu_1 + i\omega, \mathbf{k}_1) P_{\xi, -\xi; \xi}\right) \propto \text{Tr}\left( \sigma_z \sigma_{\mu} \sigma_0 \sigma_z \right) \propto \delta_{\mu 0} \delta_{zz}.
\end{equation}
In the last step we used the following trace identity\cite{KoshelevFourPauliTrace}: \begin{equation*}
\operatorname{tr}\left(\sigma_{\alpha} \sigma_{\beta} \sigma_{\gamma} \sigma_{\mu}\right)=2\left(\delta_{\alpha \beta} \delta_{\gamma \mu}-\delta_{\alpha \gamma} \delta_{\beta \mu}+\delta_{\alpha \mu} \delta_{\beta \gamma}\right)+4\left(\delta_{\alpha \gamma} \delta_{0 \beta} \delta_{0 \mu}+\delta_{\beta \mu} \delta_{0 \alpha} \delta_{0 \gamma}\right)-8 \delta_{0 \alpha} \delta_{0 \beta} \delta_{0 \gamma} \delta_{0 \mu}+2 i \sum_{(\alpha \beta \gamma \mu)} \varepsilon_{0 \alpha \beta \gamma} \delta_{0 \mu}.
\end{equation*}
Therefore the massless mode propagator will not contribute to the conductivity, since the velocity index only runs over $\mu \: \in \: \{x, y, z\}$.  
This vanishing occurs even before integration or summing over $\xi$.  
This stems from the even number of subbands and the symmetries internal to each Weyl node.  
{As a result, the Pauli matrix algebra between bands is important in determining the conductivity.
This contrasts with the 1D single band case, where the loop contribution is the trace of a $1 \times 1$ matrix. (See. Appendix~\ref{cosineSection} below).}  
Furthermore, along the direction of the CDW $\hat{z}$, the velocities on each band are opposite and so cancel in the collective masssless term.  
These opposing on-node velocities can be interpreted as  LRA\cite{RLAChargeDensityWaves} and anti-LRA CDWs canceling out, which is further discussed in Section \ref{LinearCollectiveSection}.
A tilting term in the Weyl model will allow for the appearance of a term proprotional to $\sigma_0$ in the velocity operator, allowing for Eq.~\eqref{vanishingCondition2} to be nonzero.

\section{Collective Conductivity in a 1D CDW}
\label{cosineSection}

\begin{figure}[ht]
\begin{center}
\includegraphics[width=0.93\hsize]{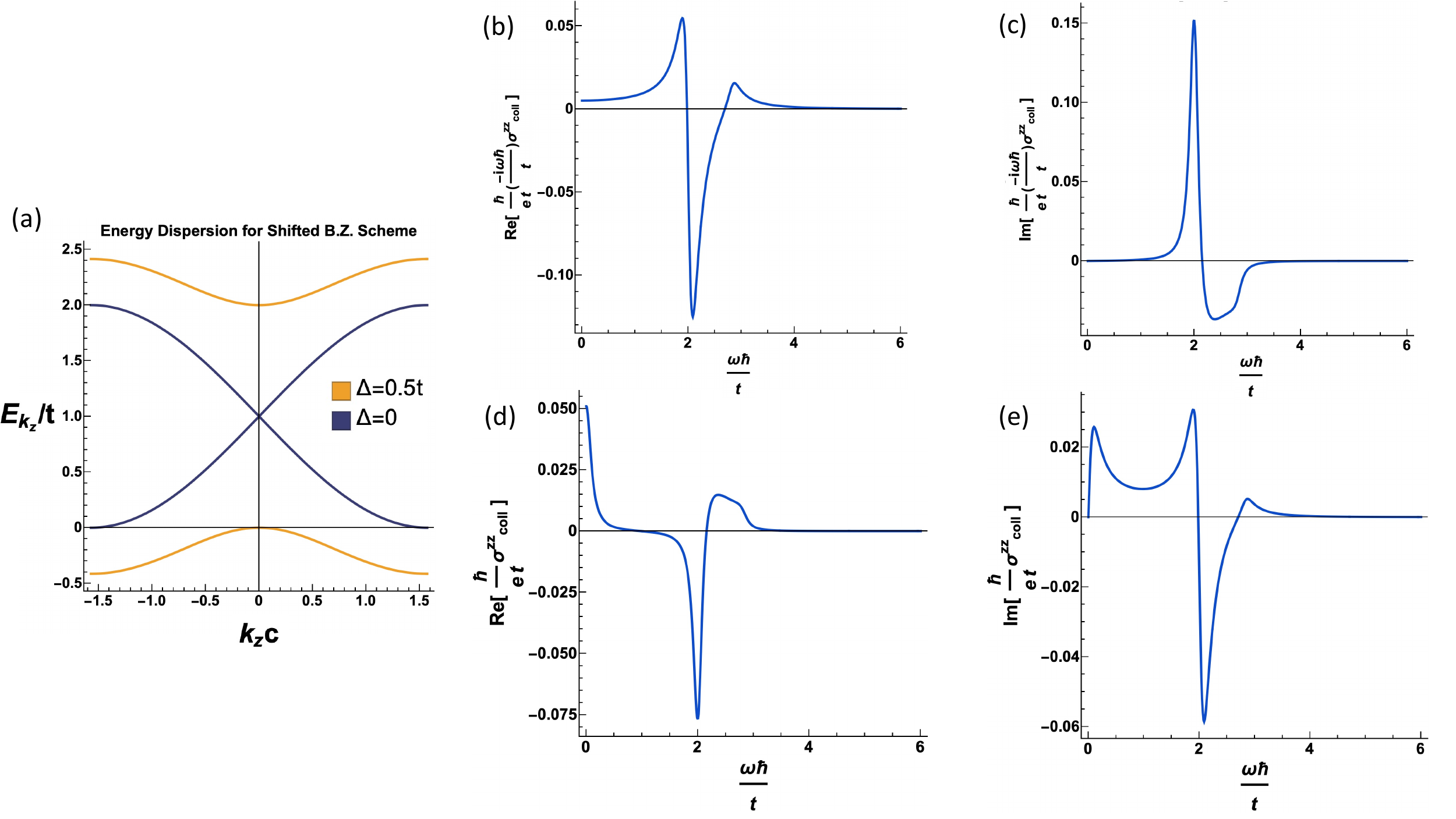}
\caption{Spectrum and conductivity for the tight-binding model of a $1$D CDW. The spectrum is shown in (a), with both $\Delta = 0$ and $\Delta = 0.5 t$.  The plots of $i\omega \sigma^{zz}_{\text{coll}}$ ((b) and (c)) and $\sigma^{zz}_{\text{coll}}$ ((d) and (e)) are each shown in natural units.  The real ((b) and (d)) and imaginary ((c) and (e)) parts are provided for the parameters $\eta = 0.1 t/\hbar$ and a gap $\Delta = 0.5 t$, where the small self-energy $\omega \rightarrow \omega + i \eta$ serves to numerically resolve the plot. }
\label{cosineFullPlot}
\end{center}
\end{figure}
This section will examine the collective conductivity from a toy 1D model to compare to the tilted Weyl semimetal results.  
Consider a simple 1D cosine dispersion given by the tight-binding Hamiltonian $H_{\text{1D}}= t \sum_{\mathbf{R}} -c^{\dagger}_{\mathbf{R}} c_{\mathbf{R}+\hat{z}} + c^{\dagger}_{\mathbf{R}} c_{\mathbf{R}}  + \text{h.c.}$, where $\mathbf{R} = R \hat{z}$.  
Taking a Fourier transform, $c_{\mathbf{R}} = \sum^{\pi}_{\mathbf{k} = -\pi} e^{i \mathbf{k} \cdot \mathbf{R}} c_{\mathbf{k}}$, the Hamiltonian in the full Brillouin zone scheme is $H_{\text{1D}} =   t\sum_{\mathbf{k}}\left(1- \cos(\mathbf{k}) \right) c^{\dagger}_{\mathbf{k}} c_{\mathbf{k} }$. 
We consider the model at half filling, so that the Fermi wavevector is $\mathbf{Q} = \pi \hat{z}$. 
We will then introduce a charge-density wave distortion coupling the two Fermi points (ignoring commensuration effects as in the main text).  
This model was chosen since it can be compared to the LRA results at quadratic order \cite{RLAChargeDensityWaves, FentonCDW}.  
Applying the same shifted Brillouin zone scheme as the models in the main text (see Appendix \ref{shiftedZoneSchemeDerivation}), we rewrite the Hamiltonian as
\begin{equation}
\label{cosine1D}
H_{\text{1D}}= \sum^{\pi / 2}_{k_z = -\pi/2} \vec{c}^{\dagger \: \prime}_{k_z} \left[t \tau_0 + t \sin(k_z) \tau_z \right] \vec{c}^{\: \prime}_{k_z},
\end{equation}
where recall $\vec{c}^{\dagger \: \prime}_{k_z}$ denotes creation in the shifted subspace from momentum $\pm Q/2$.  
The spectrum of $H_{\text{1D}}$ in the shifted Brillouin zone scheme with the gap term from Eq.~\eqref{gapAndPhonon} is plotted in Fig.~\ref{cosineFullPlot}.  

We now move on to compute the propagators for this model.  
{As compared with the main text, note that the electron-phonon vertex is different, now given by as $P_{1, \xi_1, 1 \xi_2; d_1} = g \delta_{d_1 \xi_1} \left( \tau_{x} \right)_{\xi_1, \xi_2} $.}  
The diagram from Fig.~\ref{LinearCollectiveDiagrams} and its corresponding Eq.~\eqref{LinearCollectiveFullCalc} are still applicable.  
The only difference is the integral is one-dimensional, spanning just $\int^{\pi/2}_{-\pi/2}[d  k_z]$.  The Green's function matrix for the electrons is
\begin{equation}
    G(i\nu, k_z) = \begin{bmatrix}
\frac{-i \nu + t(1+\sin(k_z))}{(2 \Delta)^2 + (t \sin(k_z))^2-(t - i \nu)^2 } & \frac{- (2 \Delta) e^{-i \phi}}{(2 \Delta)^2 + (t \sin(k_z))^2-(t - i \nu)^2 }\\
\frac{- (2 \Delta) e^{i \phi}}{(2 \Delta)^2 + (t \sin(k_z))^2-(t - i \nu)^2 }  & \frac{-i \nu + t(1-\sin(k_z))}{(2 \Delta)^2 + (t \sin(k_z))^2-(t - i \nu)^2 }
\end{bmatrix}.
\end{equation}
Evaluating Eq.~(\ref{GeneralD}) to compute the collective conductivity, we find that the only the massless mode contributes, due to the even parity and single band of the initial cosine dispersion.  
The massless mode propagator is
\begin{equation}
    D_{++}(\omega) -  e^{2 i \phi}D_{+-}(\omega) = \frac{1}{\frac{-\omega^2 + \omega^2_Q}{2 \omega_\mathbf{Q}} - 2 g^2 \int [d k_z] \frac{(2 \Delta)^2 + (t \sin(k_z))^2}{\sqrt{(2 \Delta)^2 +(t \sin(k_z))^2} \left[ 4 \left( (2 \Delta)^2 + (t \sin(k_z))^2 \right) - \omega^2 \right]} },
\end{equation}
and, from the general gap Eq.~(\ref{gapGeneral}),
\begin{equation}
    \omega_{Q} = g^2 \int [d k_z] \frac{1}{\sqrt{(2 \Delta)^2 +(t \sin(k_z))^2}}.
\end{equation}
{For completeness, the single photon electron loop is also defined by:
\begin{equation}
    G_{\text{loop in}}(\omega) = \int [d \mathbf{k}] \frac{2 \Delta g t \cos(k_z)}{\sqrt{(2 \Delta)^2 + (t \sin(k_z))^2} (4(2 \Delta)^2 + 4 (t \sin(k_z))^2 - \omega^2)} \begin{bmatrix}
e^{i \phi} (\omega + 2 \sin(k_z)) \\
e^{-i \phi} (-\omega + 2 \sin(k_z))
\end{bmatrix},
\end{equation}
and $G_{\text{loop out}}(\omega) = G^{*}_{\text{loop in}}(\omega)$. 
The collective conductivity is then given by $\sigma_{\text{coll}}(\omega) = \frac{-e^2}{i \omega} G_{\text{loop in}}(\omega) D(\omega) G_{\text{loop out}}(\omega)$.}

{The definition of the phase and amplitude mode propagators used in this paper, $D_{++}(\omega) \pm e^{2 i \phi} D_{+-}(\omega)$, are different than the definition used in LRA, $D^{\text{LRA}}_{++}(\omega) \pm D^{\text{LRA}}_{+-}(\omega)$.  
The reason for this is that LRA implements a different convention for the mean field decomposition: those authors take $b_\mathbf{Q}(t) = (\Delta/g + \delta b)e^{i\delta\theta}$, where the phase $\phi$ is absorbed into a shift in the origin of the dynamical phase $\delta\theta$. 
As a result, the factors of $e^{i \phi}$ are distributed differently among the collective phonon propagator and the loop diagrams.  
However, LRA's method and this paper's method will produce the same outcome, as all factors of $e^{i \phi}$ cancel in the final result, $\sigma_{\text{coll}}(\omega)$.} 

The collective linear conductivity is illustrated in Fig.~\ref{cosineFullPlot}.  
We see that, due to the contribution of the massless mode, $\text{Re} \left\{ -i\omega \sigma^{zz}_{\text{coll}} (\omega) \right\}$ is finite as $\omega \rightarrow 0$. 
This corresponds to a divergence in the real part of the DC conductivity at zero frequency.  
Another feature similar to the tilted Weyl collective conductivity is the presence of a peak in the conductivity that occurs for frequencies larger than the single-particle gap, which is at {$\omega = 2 (2 \Delta)$} ($\omega = 2 t $ in Fig.~\ref{cosineFullPlot}). 
The frequency at which the electric field is able to excite electrons from the bottom band to the highest point in the upper band is given by {$\omega = t + \sqrt{(2 \Delta)^2 + t^2}$ ($ \omega = 2.414 t $ in Fig.~\ref{cosineFullPlot})}. 
It is marked by a subtle change is slope (real) or a maximum (imaginary) in the conductivity.  
The final characteristic occurs at {$\omega = 2 \sqrt{(2 \Delta)^2 + t^2}$ ($ \omega = 2.828 t / \hbar $ in Fig.~\ref{cosineFullPlot})} where the lowest part of the energy dispersion in the bottom band excites past the highest energy in the upper band.  
This is marked by a decrease in the conductivity towards zero.

\begin{figure}[ht]
\begin{center}
\includegraphics[width=0.93\hsize]{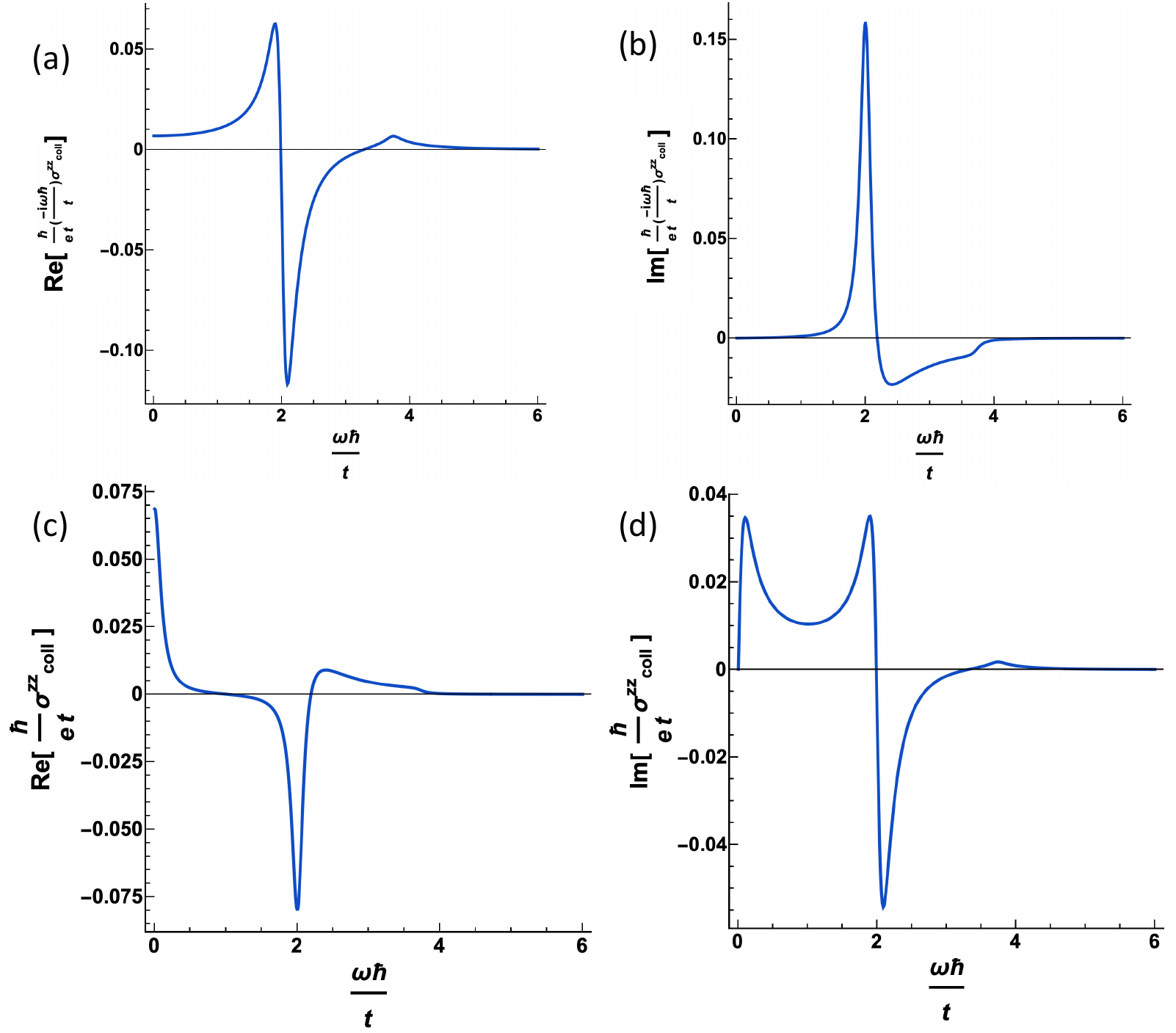}
\caption{Conducivity in the linearized model of a $1$D CDW, as first calculated by LRA.  Plots (a) and (c) are the real part, while plots (b) and (d) are the imaginary part.  The parameters $\Delta = 0.5 t$ and $\omega \rightarrow \omega + i \eta$ with $\eta = 0.1 t /\hbar $ were used. } 
\label{quadradicPlots}
\end{center}
\end{figure}

The Lee-Rice-Anderson result for the linearized model may be recovered by approximating our Hamiltonian to linear order in $k_z$.  
Making this approximation in Eq.~\eqref{cosine1D} yields Fig.~\ref{quadradicPlots} for the conductivity, which agrees with the results of Ref.~\onlinecite{FentonCDW}.  
For convenience, the cutoff momentum is taken at $\Lambda = \pi / 2$. 
Note that for small frequencies, the conductivity in the linearized model Fig.~\ref{quadradicPlots} is qualitatively similar to the lattice result in Fig.~\ref{cosineFullPlot}. 
We will now show that in this low energy limit, we can recover the analytic results of Refs.~\cite{RLAChargeDensityWaves,FentonCDW}.  
To facilitate this, the cutoff momentum is taken as $\Lambda \rightarrow \infty$, which is possible in 1D, but yields a divergence answer in 3D.  
Then the collective conductivity may be described simply.  
Defining the unitless function
\begin{equation}
f\left( \frac{\omega} {4 \Delta} \right) \equiv\frac{1}{2 \pi} \frac{2}{\frac{\omega}{4 \Delta}\sqrt{1 - \frac{\omega^2}{(4 \Delta)^2}}} \arctan\left( \frac{\frac{\omega}{4 \Delta}}{\sqrt{1 - \frac{\omega^2}{(4 \Delta)^2}}}  \right),
\end{equation}
the conductivity becomes 
\begin{equation}\sigma_{\text{coll}}(\omega) = \frac{-e^2}{i \omega}\left( \frac{2^5 t_z \omega_\mathbf{Q} g^2 f^2(\frac{\omega}{ 4 \Delta})}{t_z (4 \Delta)^4 + \omega_\mathbf{Q} g^2 (4 \Delta)^2 f(\frac{\omega}{4 \Delta})} \right),
\end{equation}
in agreement with Ref.~\cite{FentonCDW}

\section{Third-Order Collective Conductivity Near Zero Frequency}
\label{collectiveInterceptVsSelfEnergy}
\begin{figure}[ht]
      \centering
\begin{minipage}{0.8\hsize}
\centering
\includegraphics[width=0.4\hsize]{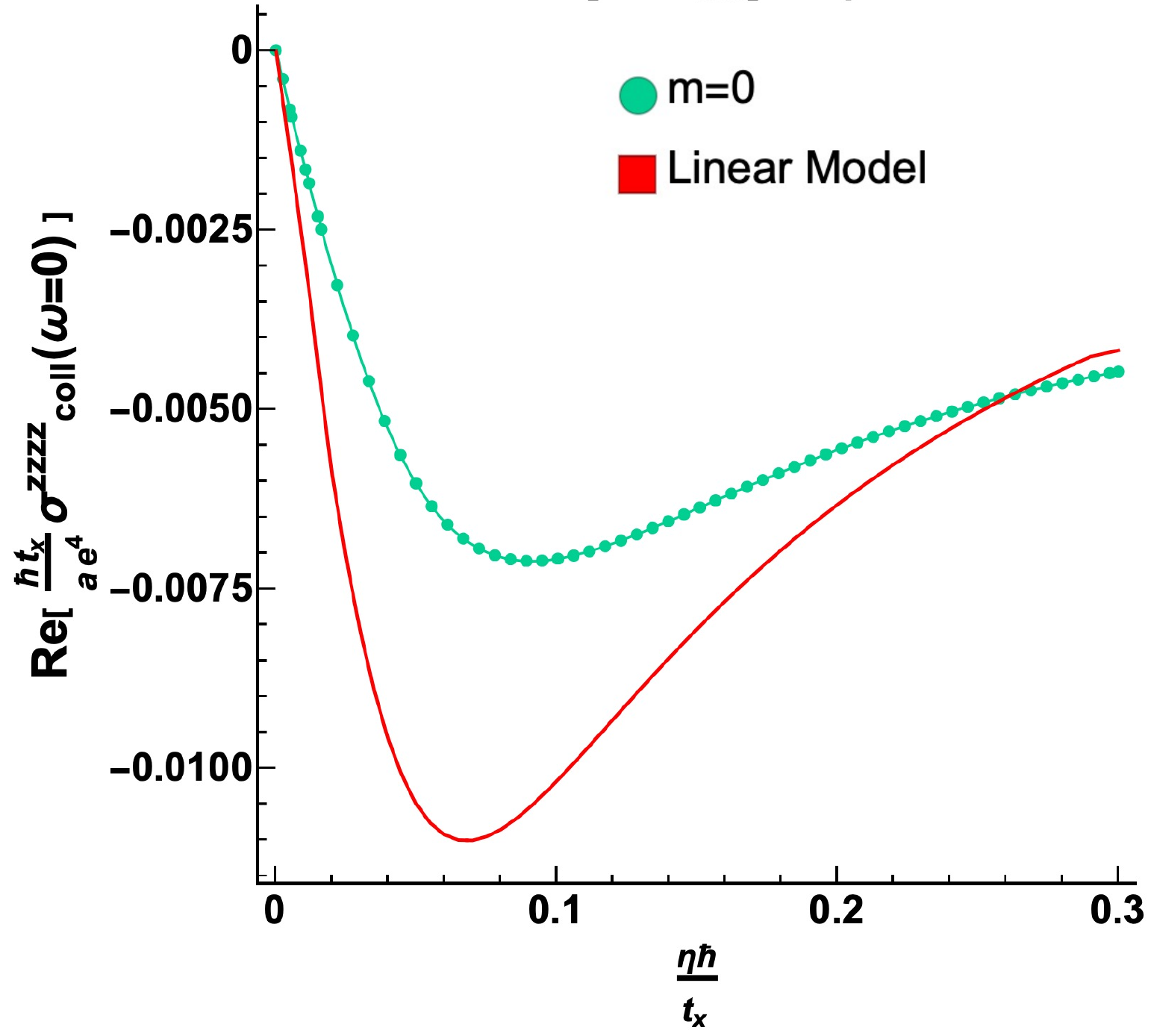}
\end{minipage}
\caption{Real part of the third-order collective conductivity at $\omega = 0$ as a function of the self-energy parameter, $\eta$.  The idealized Weyl model and the $m = 0$ lattice model are sufficient to generalize to the other models throughout the paper.}
\label{ConvergenceOfCollectiveConductivityvsEta}
\end{figure}

The harmonic generation and self-focus plots shown in Fig.~\ref{harmonicFullPlot} and \ref{selfFocusFullPlot} seem to have nonzero intercepts at $\omega = 0$.  
Since gapping out the Weyl system should yield an insulator, the reader might expect both real and imaginary parts of the collective conductivity to vanish in the zero frequency limit.  
This nonzero intercept is a relic of introducing a small imaginary self-energy to the frequency, $\omega \rightarrow \omega + i \eta$.  
The self-energy dictates how fast the resonant frequency decays, giving rise to this nonzero intercept.  
To illustrate that $\eta$ is the culprit, we have plotted $\text{Re} \left[ \lim_{\omega_{\beta}, \omega_{\gamma}, \omega_{\delta}  \rightarrow 0} \sigma^{zzzz}_{\text{coll}}(\omega_{\beta \gamma \delta}; \omega_{\beta}, \omega_{\gamma}, \omega_{\delta}) \right]$ as a function of $\eta$ in Fig.~\ref{ConvergenceOfCollectiveConductivityvsEta}.  
We see that as the small imaginary frequency vanishes, we recover our zero intercept of the collective conductivity.  
Ergo, one should be cautious about numerically taking the full, complex frequency, $\omega + i \eta$ to zero when comparing to an insulator.  
We show this for the idealized Weyl model and the $m=0$ lattice model in the Figure, but the argument will generalize to all values of $m$.

The imaginary part of the collective conductivity, on the other hand, does go to zero as frequency vanishes.  
Furthermore, $\text{Im} \left[ \lim_{\omega_{\beta}, \omega_{\gamma}, \omega_{\delta}  \rightarrow 0} \sigma^{zzzz}_{\text{coll}}(\omega_{\beta \gamma \delta}; \omega_{\beta}, \omega_{\gamma}, \omega_{\delta}) \right]$ is independent of $\eta$, and so would give a horizontal line through zero on a plot like Fig.~\ref{ConvergenceOfCollectiveConductivityvsEta}.

\end{appendices}

\twocolumngrid
\bibliography{CDWFormalNotes_ref}

\end{document}